\documentclass[%
reprint,
superscriptaddress,
amsmath,amssymb,
aps,
]{revtex4-2}

\usepackage{zref-xr} % 替代 xr/xr-hyper
\zxrsetup{toltxlabel}

\usepackage{lipsum}
\usepackage{bm}
\usepackage{amssymb}
\usepackage{amstext}
\usepackage{amsmath}
\usepackage{mathrsfs}
\usepackage{graphicx}
\usepackage{color}
\usepackage{amsthm}
\usepackage{floatrow}
\usepackage{array}
\renewcommand\arraystretch{2}
\usepackage{listings}
\usepackage{booktabs}
\usepackage{changepage}
\usepackage{xurl}
\usepackage{hyperref}% add hypertext capabilities
\begin{document}
	\preprint{APS/123-QED}
	\title{Geometry- and topology-controlled synchronization phase transition on manifolds}% Force line breaks with \\
	\thanks{Correspondence should be addressed to Yang Tian.}%

\author{Yang Tian}
\email{tyanyang04@gmail.com \& yang.tian@infplane.com}
\affiliation{Infplane Computing Technologies Ltd, Beijing, 100080, China}

	%\affiliation{
	% Lunar Base
	%\affiliation{%
	
	%\collaboration{CLEO Collaboration}%%\noaffiliation
	
	
\begin{abstract}
In this work, we explore how the geometry and topology of the underlying
manifold shape the synchronization phase transition of a system. To do so, we extend the Kuramoto-Sakaguchi model
from spheres to compact, connected, orientable, and homogeneous Riemannian
manifolds of arbitrary dimension. Starting from the mean-field kinetic equation on the manifold, we
derive a local response equation for the order parameter near the incoherent
state and separate the geometric and topological contributions to the phase
transition out of the incoherent state. The manifold geometry determines the
averaged projection factor $\kappa\left(M\right)$, which directly controls the
coupling strength required to destabilize the incoherent state. The critical
coupling is determined jointly by this geometric factor and the
response of the intrinsic drift fields. The manifold topology affects the
phase transition through the Euler characteristic $\chi\left(M\right)$: the
Poincar\'e-Hopf relation fixes the net defect charge of the incipient ordered
texture, and the local reduction and sign conditions stated below allow the
same Euler-characteristic data to constrain the cubic coefficient of the
reduced response equation.
In that conditional class, a non-zero Euler characteristic gives the cubic sign
used in the reduced normal form, and an additional stabilization condition
gives a discontinuous local transition. When
$\chi\left(M\right)=0$, the local branch is determined by the normal-form
coefficients rather than by the Euler characteristic alone.
We evaluate these geometric and topological indicators on representative families including hyperspheres, sphere products, complex Grassmannians, complex projective spaces, flat
tori, real Stiefel manifolds, rotation groups, and unitary groups. Our framework recovers the topological part of the classical hyperspherical parity distinction and extends it to a
broad class of non-spherical state spaces. It therefore identifies the
geometric factor that controls the coupling strength required for the loss of
stability of the incoherent state, the unconditional topological defect
constraint, and the conditional topological selection of the local phase
transition scenario in diverse synchronization phenomena.
\end{abstract}
\maketitle
\section{Introduction }The Kuramoto model serves as a prototypical framework for investigating low-dimensional synchronization phenomena \cite{Kuramoto1975,Rodrigues2016PhysRep,Pietras2019PhysRep,Bick2020JMathNeurosci}. Its generalized form, the Kuramoto-Sakaguchi model, also plays a crucial role in studying both synchronization and asynchronization in low-dimensional systems \cite{SakaguchiKuramoto1986,Rodrigues2016PhysRep,Pietras2019PhysRep}. However, to accurately describe a broader range of biological, chemical, and physical systems \cite{Breakspear2010FrontHumNeurosci,Lu2016Chaos,Witthaut2022RMP}, it is necessary to consider high-dimensional settings ($D\geq 3$), in which the original low-dimensional formulations ($D\leq 2$) of the Kuramoto model and the Kuramoto-Sakaguchi model become inadequate \cite{Chandra2019PRX,Lipton2021Chaos,Zou2023PRL}. Moreover, the phase space of a high-dimensional system need not be a standard $D$-dimensional sphere. In that case, high-dimensional extensions built specifically on the hypersphere may fail to represent the underlying state space \cite{Markdahl2019Automatica,Tron2012CDC,GolseHa2019ARMA,Buzanello2022Chaos,DeAguiar2023PRE}.

In this study, we extend the Kuramoto-Sakaguchi model to compact, connected, orientable, and homogeneous $D$-dimensional Riemannian manifolds, which may possess non-trivial topology or an arbitrary number of holes \cite{Arvanitoyeorgos2003Homogeneous}. This class already contains the principal geometric examples relevant to synchronization, including spheres, flat tori, Stiefel manifolds, rotation groups, and unitary groups \cite{Chandra2019PRX,Markdahl2019Automatica,Tron2012CDC,GolseHa2019ARMA,Fiori2018DCDSB,Fiori2020SIAM,Montenbruck2015SCL,Fiori2018NonlinearDyn,EdelmanAriasSmith1998SIAM}. Related Lohe-type and unitary-group models provide a complementary matrix-valued route to manifold synchronization \cite{GolseHa2019ARMA,HaKoRyoo2017JSP,HaKoRyoo2018JSP,KimKim2023KRM,KimKim2025EJAM,HaPark2020SIADS,HaKangPark2021JMP,HaKangPark2021CPAA,HaPark2021JSP,Ryoo2025CMS,Lohe2019JMP,HaKimPark2020PhysicaD}. The original Kuramoto model, the original Kuramoto-Sakaguchi model, and the generalized Kuramoto model on the hypersphere all serve as special cases of our extended model.

\begin{table*}[t]
  \caption{Logical structure of the present work. The table follows the route
  used in the paper: we first formulate the Kuramoto-Sakaguchi model on a
  general state manifold, then derive the kinetic equation, separate the
  linear and nonlinear response terms near the incoherent state, and finally
  identify the geometric and topological information that enters the
  synchronization phase transition.}
  \label{Table-framework}
  \centering
  \begingroup
  \small
  \newcommand{\FrameworkTableRow}[3]{%
  \noindent
  \begin{minipage}[t]{0.18\textwidth}\raggedright #1\end{minipage}\hfill
  \begin{minipage}[t]{0.38\textwidth}\raggedright #2\end{minipage}\hfill
  \begin{minipage}[t]{0.38\textwidth}\raggedright #3\end{minipage}\par
  \vspace{0.5em}\hrule\vspace{0.5em}}
  \hrule\vspace{0.5em}
  \FrameworkTableRow{\textbf{Section and step}}
  {\textbf{What we derive}}
  {\textbf{How the result is used}}
  \FrameworkTableRow{In Sec. II, manifold Kuramoto-Sakaguchi model}
  {We place the oscillator state on a compact homogeneous manifold \(M\), fix
  an embedding \(F:M\hookrightarrow\mathbb{R}^{D_a}\), and project the coupling
  term onto the tangent space.}
  {This gives a common dynamical model for spherical and non-spherical state
  spaces and specifies where geometry can enter the coupling.}
  \FrameworkTableRow{In Sec. II D, mean-field kinetic equation}
  {We derive the density equation on the tangent bundle and expand it near the
  incoherent state.}
  {The expansion separates the response into a linear spectral problem and a
  nonlinear local response equation.}
  \FrameworkTableRow{In Sec. III B, linear response and geometry}
  {We linearize the kinetic equation and evaluate the coupling-induced source
  term in the chosen order parameter mode.}
  {The averaged tangent-projection factor \(\kappa(M)\) gives the geometric
  contribution to the critical coupling \(K_c\), together with the
  response of the intrinsic drift fields.}
  \FrameworkTableRow{In Secs. III C--E, nonlinear response and topology}
  {We reduce the local order parameter response to a one-mode normal form and
  use the Poincar\'e-Hopf relation for the critical tangent field.}
  {The Euler characteristic gives an unconditional defect-charge constraint.
  Under the locality and sign conditions, it also gives the cubic sign used in
  the reduced local response.}
  \FrameworkTableRow{In Sec. IV, representative manifolds and numerical tests}
  {We compute \(\kappa(M)\), \(\chi(M)\), defect constraints, and local
  transition indicators on representative spherical and non-spherical
  manifolds.}
  {These calculations and simulations test how geometry determines \(K_c\) and
  how topology constrains defects and, conditionally, the local transition
  scenario.}
  \vspace{-0.5em}\hrule
  \endgroup
\end{table*}

Our analysis yields three main conclusions and Table \ref{Table-framework} summarizes how the theoretical framework is built. First, the manifold geometry determines the averaged tangent-projection factor $\kappa\left(M\right)$. This factor multiplies the coupling-induced source term in the linearized kinetic equation. The critical coupling $K_{c}$ is then determined jointly by this geometric factor and by the response of the intrinsic drift fields. In scalar reductions where the transport response factorizes as $s_{M}\left(\lambda;h\right)=\kappa\left(M\right)\Xi\left(\lambda;h\right)$, the scalar factor $\Xi\left(\lambda;h\right)$ is the intrinsic-drift response and the critical coupling is $K_{c}=1/\left[\kappa\left(M\right)\Xi\left(0^{+};h\right)\right]$ for $\alpha=0$. Second, the manifold topology gives an unconditional defect constraint and a conditional constraint on local phase transition scenarios. The unconditional statement is the Poincar\'e-Hopf relation: the total index of the zeros of the critical tangent field equals the Euler characteristic \cite{Milnor1997DifferentiableViewpoint}, so a non-zero Euler characteristic imposes a non-zero net defect charge on the critical texture. The phase transition statement additionally uses the locality conditions of Appendix \ref{Appendix-D} and the sign condition stated in Appendix \ref{Appendix-E}. Under those assumptions, a non-zero Euler characteristic gives the cubic sign used in the one-mode normal form; if the stabilizing condition $\Lambda_{5}>0$ also holds, the local transition is discontinuous. When $\chi\left(M\right)=0$, the local branch is determined by the normal-form coefficients rather than by the Euler characteristic alone. Here the sign condition means that the contribution of neighborhoods of the zeros weighted by their local indices has a definite sign and dominates the remaining part of the cubic density that is not weighted by the indices. Third, these geometric, spectral, and topological ingredients control physical properties that are directly relevant to statistical physics. They determine the critical coupling for the loss of stability of the incoherent state, the conditional phase transition type, the possible presence of hysteresis, and the defect content of the incipient ordered texture, and we compute these indicators on representative manifolds, including hyperspheres, sphere products, complex Grassmannians, complex projective spaces, flat tori, real Stiefel manifolds, rotation groups, and unitary groups \cite{EdelmanAriasSmith1998SIAM,Harris1992AlgebraicGeometry}.

The present framework also provides a basis for several further statistical-physics problems on high-dimensional state spaces. It can be used to study finite-size critical fluctuations near the loss of stability of the incoherent state, defect kinetics and charge balance in the emerging ordered texture, metastability and noise-induced switching on the discontinuous side, and nonequilibrium hysteresis under slow or fast parameter ramps. It also opens a route to systematic comparisons among multi-angle oscillators, matrix synchronization models, and other manifold-valued oscillator systems within a common geometric and topological language. In this sense, the theory identifies which part of the loss of stability of the incoherent state is determined by geometry, which part of the nonlinear response is constrained by topology, and which physical properties remain to be selected by the analytic structure of the underlying model.

\section{General model on the manifold}
\subsection{Geometric setting}

In this work, we consider a smooth, connected, compact, orientable, and homogeneous Riemannian manifold
$\left(M,g\right)$ of dimension $D$ without boundary. The state of every oscillator $i$ is defined on this manifold, $\sigma_{i}\in M$. The Riemannian metric
$g$ induces the Levi-Civita connection $\nabla_{M}$ and the volume
measure $\mu$ \cite{Lee2018Riemannian,Arvanitoyeorgos2003Homogeneous}.
For any $\sigma_{i}\in M$, we denote by $T_{\sigma_{i}} M$ the tangent space at
$\sigma_{i}$, equipped with the inner product
$\langle\cdot,\cdot\rangle_{g}$ and the norm $\|\cdot\|_{g}$ \cite{Lee2018Riemannian}.

According to Nash's embedding theorem \cite{Nash1956AnnMath,Lee2012Smooth}, there exists an isometric embedding
$F_{0}:\left(M,g\right)\hookrightarrow\left(\mathbb{R}^{D_{0}},\langle\cdot,\cdot\rangle\right)$.
In this work, we define a smooth embedding
\begin{align}
    F : M \hookrightarrow \mathbb{R}^{D_{a}} ,\;\;\;D_{a}\ge D_{0} \label{II-A-EQ001}
\end{align}
and identify $M$ with its image $F(M)\subset\mathbb{R}^{D_{a}}$.
The differential $\mathsf{d}F_{\sigma_{i}}:T_{\sigma_{i}} M\rightarrow\mathbb{R}^{D_{a}}$ is injective for
each $\sigma_{i}\in M$. In some special cases (e.g., the hypersphere $M=S^{D}$),
we additionally set $\vert F\left(\sigma_{i}\right)\vert=1$ for all $\sigma_{i}\in M$ such that
$F\left(M\right)$ lies on the unit sphere.

Through the embedding, we realize each tangent space $T_{\sigma_{i}} M$ as a
linear subspace of $\mathbb{R}^{D_{a}}$. We denote
\begin{align}
    \mathsf{P}_{\sigma_{i}}^\perp : \mathbb{R}^{D_{a}} \to T_{\sigma_{i}} M\label{II-A-EQ002}
\end{align}
as the orthogonal projection with respect to the Euclidean inner product \cite{Lee2018Riemannian}.
This projection ensures that all vector fields defining the dynamics
remain tangent to the manifold $M$. The transport formulation introduced below still makes sense on a general compact orientable manifold, while the homogeneous assumption is used when we reduce the averaged geometric operator to a scalar factor in Sec. \ref{Linear-response-subsection}.

\begin{table*}[t]
  \caption{Representative special cases of the general Kuramoto-Sakaguchi model Eqs. (\ref{II-B-EQ001}, \ref{II-B-EQ003}). In all the cases, the state of oscillator $i$ is denoted by $\sigma_{i}\in M$, where $M$ is
a smooth compact Riemannian manifold. $F:M\hookrightarrow\mathbb{R}^{D_{a}}$
is a fixed smooth embedding into an ambient Euclidean space
$\mathbb{R}^{D_{a}}$. The coupling term is
$\Gamma\left(\sigma_{i},\sigma_{j};\alpha\right)
 = \mathsf{P}^{\perp}_{\sigma_{i}}
    \left[\left(\mathbf{I}-\alpha A\right)F\left(\sigma_{j}\right)\right]$,
where $\mathsf{P}^{\perp}_{\sigma_{i}}$ denotes the orthogonal
projection from $\mathbb{R}^{D_{a}}$ onto the tangent space
$T_{\sigma_{i}}M$, $A\in\mathfrak{so}\left(D_{a}\right)$ is a fixed
skew-symmetric matrix representing a linearized generator of rotations
in the embedding space, and $\alpha\in\mathbb{R}$ is a scalar
phase-lag parameter. In the table, each row corresponds to a
particular choice of $M$, $F$ and $A$ (e.g., $M=S^{1}$ with
$F\left(\sigma\right)=\left(\cos\sigma,\sin\sigma\right)$ and
$A=0$ in the first row, or $M=S^{D-1}$ with $F\left(\sigma_{i}\right)=\sigma_{i}$ and
$A\in\mathfrak{so}\left(D_{a}\right)$ in the fourth row), which reproduces a special
Kuramoto-type model proposed in previous studies \cite{Rodrigues2016PhysRep,Pietras2019PhysRep,Chandra2019PRX,Lipton2021Chaos,Zou2023PRL,Buzanello2022Chaos,DeAguiar2023PRE,Markdahl2019Automatica,Markdahl2017TAC,Tron2012CDC,Lohe2019JMP,HaKimPark2020PhysicaD,Boccaletti2016PhysRep,SkardalArenas2020CommunPhys}. Please see Appendix \ref{Appendix-B} for the detailed definitions of all these special cases. Their associated mathematical notions are not elaborated in the main text since they are irrelevant to our subsequent analysis.}
  \label{Table-1}
  \begin{ruledtabular}
  \begin{tabular}{llll}
    Model &
    $M$ &
    $F$ &
    Phase-lag \\ \hline
    Kuramoto on $S^{1}$ &
    $S^{1}$ &
    $(\cos\sigma,\sin\sigma)$ &
    $A=0$, $\alpha=0$ \\[0.2em] \hline
    Kuramoto-Sakaguchi on $S^{1}$ &
    $S^{1}$ &
    $\left(\cos\sigma,\sin\sigma\right)$ &
    $A=J$, small $|\alpha|$ \\[0.2em] \hline
    Kuramoto on $S^{D-1}$ &
    $S^{D-1}$ &
    Inclusion $F\left(\sigma\right)=\sigma$ &
    $A=0$, $\alpha=0$ \\[0.2em] \hline
    Frustrated Kuramoto on $S^{D-1}$ &
    $S^{D-1}$ &
    $F\left(\sigma_{i}\right)=\sigma_{i}$ &
    $A\in\mathfrak{so}(D)$ \\[0.2em] \hline
    Vector-phase Kuramoto on $\mathbb{T}^{d}$ &
    $\mathbb{T}^{d}$ &
    Product of $S^{1}$ embeddings &
    Block-diagonal $A$ \\[0.2em] \hline
    Kuramoto on Stiefel/rotation manifolds &
    $\mathrm{St}\left(p,n\right)$, $\mathrm{SO}\left(n\right)$ &
    Canonical matrix embedding &
    $A$ acts by $BX$ or $XB$ \\[0.2em] \hline
    Lohe/unitary Kuramoto &
    $\mathrm{U}\left(d\right)$, $\mathrm{SU}\left(d\right)$ &
    $F\left(U\right)=U$ in $\mathbb{C}^{d\times d}$ &
    $A$ in $\mathfrak{u}\left(d\right)$ \\[0.2em] \hline
    Network-distributed Kuramoto &
    Any of the above $M$ &
    Corresponding $F$ &
    $K\to K_{ij}$, same $A$
  \end{tabular}
  \end{ruledtabular}
\end{table*}

\subsection{General model in discrete form }Based on the definitions presented above, we first introduce our general model in its discrete form. The Kuramoto–Sakaguchi model on the manifold is written as
\begin{align}
    \frac{\mathsf{d}}{\mathsf{d}t}\sigma_{i}=V_{i}\left(\sigma_{i}\right)+\frac{K}{N}\sum_{j=1}^{N}\Gamma\left(\sigma_{i},\sigma_{j};\alpha\right),\label{II-B-EQ001}
\end{align}
where $K$ denotes the coupling strength, $N$ is the number of oscillators, and $\alpha$ denotes the phase-lag effect. Each $V_{i}$ is a smooth tangent vector field of oscillator $i$ that satisfies
\begin{align}
    \nabla_{M}\cdot V_{i}=0,\label{II-B-EQ002}
\end{align}
which is the natural condition ensuring that the spatially uniform incoherent state used below remains stationary under the intrinsic drift. We define the function $\Gamma$ as
\begin{align}
   \Gamma\left(\sigma_{i},\sigma_{j};\alpha\right)=\mathsf{P}_{\sigma_{i}}^{\perp}\left[\left(\mathbf{I}-\alpha A\right)F\left(\sigma_{j}\right)\right],\label{II-B-EQ003}
\end{align}
in which $\mathsf{P}_{\sigma_{i}}^{\perp}$ maps any vector in the ambient space $\mathbb{R}^{D_{a}}$ to the tangent space of $M$ with respect to $\sigma_{i}$. The function $F$ isometrically embeds oscillator states on $M$ into the ambient space $\mathbb{R}^{D_{a}}$ to support the synchronization analysis in an Euclidean space. The anti-symmetric operator $A$ is used to describe the non-reciprocal coupling caused by phase lags, whose strength is measured by $\alpha$. We interpret $\mathbf{I}-\alpha A$ as a linearized phase-lag operator. If an exact finite phase lag is represented by the orthogonal operator $R_{\alpha}=\exp\left(-\alpha A\right)$, then $\mathbf{I}-\alpha A$ is its first-order truncation in $\alpha$. We use this truncation to keep the coupling linear in the embedded order parameter. The operator $\mathbf{I}-\alpha A$ itself is not required to map $F\left(M\right)$ into $F\left(M\right)$ or to generate an exact motion on $M$; the projection $\mathsf{P}_{\sigma_{i}}^{\perp}$ enforces tangency of the velocity at the receiving state $\sigma_{i}$. Thus, for every $i$, the right-hand side of Eq. (\ref{II-B-EQ001}) lies in $T_{\sigma_i}M$. The phase-lag term breaks the reciprocal structure when $\vert\alpha\vert>0$, in analogy with the classic Kuramoto-Sakaguchi coupling on the circle \cite{SakaguchiKuramoto1986}.

In Eq. (\ref{II-B-EQ003}), $\mathsf{P}_{\sigma_{i}}^{\perp}F\left(\sigma_{j}\right)$ enforces synchronization among oscillators while $-\alpha \mathsf{P}_{\sigma_{i}}^{\perp}AF\left(\sigma_{j}\right)$ imposes asynchronization. The action of $\mathsf{P}_{\sigma_{i}}^{\perp}$ ensures that the change in oscillator state $\sigma_{i}$ remains confined to the tangent space, irrespective of whether the evolution of $\sigma_{i}$ is governed by synchronizing or desynchronizing interactions. To analyze the global state of the system, the order parameter of the system is defined as
\begin{align}
r\left(t\right)=\frac{1}{N}\sum_{i=1}^{N}F\left(\sigma_{i}\right),\label{II-B-EQ004}
\end{align}
which is the empirical mean of the embedded states $F\left(\sigma_{i}\right)$ in the ambient Euclidean space \cite{Kuramoto1975,SakaguchiKuramoto1986,Rodrigues2016PhysRep,Pietras2019PhysRep}. The particular choice of embedding $F$
determines which geometric features of the manifold are probed by the
order parameter. In the classical examples, $F$ is an isometric
embedding into the unit sphere, so that $\vert r\left(t\right)\vert\in\left[0,1\right]$ quantifies
the degree of synchronization. As $\vert r\vert$ approaches $0$, the system is in an asynchronization phase. As $\vert r\vert$ approaches $1$, the system is in a synchronization phase.

\subsection{Connection to existing models }For better conceptual clarity, we now explain the connection between the abstract model presented above and existing concrete models.

In a special case where $\left(M,g\right)$ reduces to a standard sphere $S^{D-1}\subset \mathbb{R}^{D}$, we actually require $D_{a}=D$, $F\left(\sigma_{i}\right)=\sigma_{i}$, and $\Vert \sigma_{i}\Vert=1$ (i.e., on the unit sphere). Meanwhile, the projection operator reduces to $\mathsf{P}_{\sigma_{i}}^{\perp}=\mathbf{I}-\sigma_{i}\sigma_{i}^{\top}$ due to changes on the Riemannian metric. Moreover, the natural frequency field $V_{i}$ reduces to the generator of rotation $\Omega_{i}$ such that $V_{i}\left(\sigma_{i}\right)\rightarrow \Omega_{i}\sigma_{i}$. These changes enable us to rewrite Eqs. (\ref{II-B-EQ001}, \ref{II-B-EQ003}) as
\begin{align}
    &\frac{\mathsf{d}}{\mathsf{d}t}\sigma_{i}=\Omega_{i}\sigma_{i}+\notag\\&\frac{K}{N}\sum_{j=1}^{N}\left[\sigma_{j}-\left(\sigma_{i}\cdot\sigma_{j}\right)\sigma_{i}-\alpha\left(A\sigma_{j}-\left(\sigma_{i}\cdot A\sigma_{j}\right)\sigma_{i}\right)\right],\label{II-C-EQ001}
\end{align}
which is the extended Kuramoto-Sakaguchi model on the $D$-dimensional hypersphere. If we further set $\alpha=0$, there is no phase-lag effect and we obtain the extended Kuramoto model on the $D$-dimensional hypersphere
\begin{align}
    \frac{\mathsf{d}}{\mathsf{d}t}\sigma_{i}=\Omega_{i}\sigma_{i}+\frac{K}{N}\sum_{j=1}^{N}\left[\sigma_{j}-\left(\sigma_{i}\cdot\sigma_{j}\right)\sigma_{i}\right],\label{II-C-EQ002}
\end{align}
which contains only synchronization forces. The gradient flow structure under this condition is elaborated in Appendix \ref{Appendix-A}. When $D=2$, Eqs. (\ref{II-C-EQ001}-\ref{II-C-EQ002}) reduce to the original Kuramoto-Sakaguchi model and the original Kuramoto model, respectively.

In more general cases, the manifold formulation can recover different
Kuramoto-type models as special cases. Different choices of the
manifold $M$ and the embedding $F$ correspond to different
state spaces and interaction geometries, while the different value of the phase-lag
parameter $\alpha$ introduces different Kuramoto-Sakaguchi type
frustration. A few representative examples are
summarized in Table \ref{Table-1}. Their definitions and relations with Eqs. (\ref{II-B-EQ001}, \ref{II-B-EQ003}) are presented in Appendix \ref{Appendix-B}.

\subsection{General model in continuous form }Building on the established correspondence with previous models, we proceed to develop the continuous counterpart of our discrete formulation. The continuum formulation arises in the mean-field limit $N\to\infty$ of Eqs. (\ref{II-B-EQ001}, \ref{II-B-EQ003}), whose detailed derivation is presented in Appendix \ref{Appendix-C} \cite{Sznitman1991PropagationChaos,GolseHa2019ARMA}. In general, we consider an empirical measure on $M$ and relate it with the limit
$N\to\infty$ in the weak formulation of the dynamics. We separate the distribution of intrinsic drift fields from the conditional spatial density and write the limiting measure on the tangent bundle as $\rho\left(\sigma,V,t\right)h\left(V\right)\mathsf{d}\mu\left(\sigma\right)\mathsf{d}V$, where $\rho\left(\sigma,V,t\right)$ denotes the conditional density of oscillators at time $t$, in state $\sigma\in M$, for a fixed intrinsic field $V$, and $h\left(V\right)$ is the time-independent probability density of $V$. The conditional density satisfies
\begin{align}
    \int_{M}\rho\left(\sigma,V,t\right)\mathsf{d}\mu\left(\sigma\right)=1\label{II-D-EQ001}
\end{align}
for every fixed $V$, where $TM$ is the tangent bundle of the manifold (i.e., the union of all the tangent spaces), $\mu$ denotes a volume measure, and $\mathsf{d}\mu\left(\sigma\right)=\sqrt{\det g}\mathsf{d}^{D}x$ is the Riemannian volume element \cite{Lee2012Smooth,Lee2018Riemannian}.

In the continuous setting, the order parameter of the system can be expressed as
\begin{align}
    r\left(t\right)=\int_{TM}\int_{M}F\left(\sigma\right)\rho\left(\sigma,V,t\right)h\left(V\right)\mathsf{d}\mu\left(\sigma\right)\mathsf{d}V,\label{II-D-EQ002}
\end{align}
which is the expectation of $F\left(\sigma\right)$ with respect to the probability measure $\rho\left(\sigma,V,t\right)\mathsf{d}\mu\left(\sigma\right)h\left(V\right)\mathsf{d}V$ (here $h\left(V\right)$ is the probability density of $V$ at each position). To reflect the influence of the global coupling force, represented by the order parameter, on the evolution of oscillator states, we propose the velocity field equation of the oscillator states
\begin{align}
\mathbf{v}\left(\sigma,V,t\right)=V\left(\sigma\right)+K\mathsf{P}_{\sigma}^{\perp}\left(r\left(t\right)-\alpha A r\left(t\right)\right).\label{II-D-EQ003}
\end{align}
In this equation, $V\left(\sigma\right)$ describes the intrinsic velocity that depends on the local position $\sigma$ and reflects heterogeneity. For simplicity, we assume $\mathbb{E}\left[V\left(\sigma\right)\right]=0$ and $\mathbb{E}\left[V\left(\sigma\right)\otimes V\left(\sigma\right)\right]=\Delta^{2}\mathbf{I}$, where $\Delta$ measures the variance of intrinsic velocity and $\mathbf{I}$ is the identity tensor on the tangent space. $K\mathsf{P}_{\sigma}^{\perp}r\left(t\right)$ drives the system toward synchronization given the feedback effects of the order parameter. The non-reciprocal coupling term $-K\mathsf{P}_{\sigma}^{\perp}\alpha A r\left(t\right)$ models the reaction delays inherent in real systems and disrupts synchronization. The conservation law below uses only the fact that $\mathbf{v}\left(\sigma,V,t\right)$ is tangent to $M$, which is guaranteed by the projection $\mathsf{P}_{\sigma}^{\perp}$. It does not require $\mathbf{I}-\alpha A$ to be an exact rotation on the embedded manifold or to be divergence-free. The divergence-free assumption in Eq. (\ref{II-B-EQ002}) is imposed on the intrinsic drift field $V_i$ so that the incoherent uniform density remains stationary under the intrinsic transport.

The structure of the velocity field in Eq. (\ref{II-D-EQ003}) is consistent with the mean-field velocity obtained in the continuum limit. In Appendix \ref{Appendix-C} we show that, at the level of the limiting
density, the drift term entering the continuity equation is given by
an intrinsic part $V\left(\sigma\right)$ plus an interaction part
obtained by averaging the discrete coupling
$\Gamma\left(\sigma_{i},\sigma_{j};\alpha\right)$ over the
distribution of oscillator states. When the interaction is expressed
in terms of the order parameter $r\left(t\right)$ as in
Eq. (\ref{II-D-EQ002}), this averaged coupling reduces exactly to the
projection $\mathsf{P}_{\sigma}^{\perp}
\left(r\left(t\right)-\alpha A r\left(t\right)\right)$ in
Eq. (\ref{II-D-EQ003}).

The velocity field defined above effectively transports the probability density over the manifold to create the flow of a ``probability fluid". This probability flow describes how the statistical distribution of the oscillator system evolves over time in the phase space. In general, the evolution equation of this flow is written as
\begin{align}
\frac{\partial }{\partial t}\rho+\nabla_{M}\cdot\mathbf{J}=0,\label{II-D-EQ004}
\end{align}
where $\nabla_{M}$ is the previously defined covariant divergence operator and $\mathbf{J}=\rho \mathbf{v}$ is a probability density flow tensor \cite{Lee2018Riemannian}. For any $\sigma\in M$, $\mathbf{J}\left(\sigma\right)$ indicates the direction of probability flow and $\vert\mathbf{J}\left(\sigma\right)\vert$ quantifies the flow rate. The divergence $\nabla_{M}\cdot\mathbf{J}$ measures the net outflow per unit time and unit volume. There is a net outflow of probability density from the local volume element when $\nabla_{M}\cdot\mathbf{J}>0$ and there is a net inflow of probability density when $\nabla_{M}\cdot\mathbf{J}<0$. More precisely, the above equation can be rewritten in an explicit form under a local coordinate chart $\{x^{1},\ldots,x^{D}\}$ by explictly calculating the covariant divergence
\begin{align}
\frac{\partial }{\partial t}\rho+\frac{1}{\sqrt{\det g}}\frac{\partial}{\partial x^{k}}\left(\sqrt{\det g}\rho v^{k}\right)=0,\label{II-D-EQ005}
\end{align}
in which $v^{k}$ is the contravariant component of the velocity field (i.e., $\mathbf{v}=v^{k}\partial_{k}$) and $\rho v^{k}=\mathbf{J}^{k}$ is the contravariant component of the probability flow density. In Eq. (\ref{II-D-EQ005}), $\sqrt{\det g}\rho v^{k}$ measures the invariant probability flux density through a coordinate surface (i.e., the amount of probability crossing a unit coordinate area per unit time). $\sqrt{\det g}$ encodes the influence of the manifold curvature on the probability flow since $\det g$ varies sharply in regions of high curvature. When $\sqrt{\det g}$ enlarges, probability flow deceleration and density dilution occur, making curvature equivalent to a repulsive potential. When $\sqrt{\det g}$ reduces, probability flow accelerates and probability density concentrates, making curvature equivalent to an attractive potential. The normalization coefficient $\frac{1}{\sqrt{\det g}}$ normalizes this flux variation to unit volume.

Eqs. (\ref{II-D-EQ004}-\ref{II-D-EQ005}) interpret the probability flow as an incompressible fluid over the manifold and describe its local conservation law for every fixed intrinsic field $V$. The corresponding global conservation law is
\begin{align}
\frac{\partial }{\partial t}\int_{TM}\int_{M}\rho\left(\sigma,V,t\right)h\left(V\right)\mathsf{d}\mu\left(\sigma\right)\mathsf{d}V=0.\label{II-D-EQ006}
\end{align}
The corresponding boundary condition is either that the manifold $M$ is compact, or that a zero-flux condition (i.e., the Neumann-type condition) is imposed at the boundary. Subject to these constraints, the total probability density is conserved over time, with only local flows occurring.

\section{Geometry- and topology-governed phase transition}
\subsection{Analysis framework }After extending the Kuramoto-Sakaguchi model to compact, connected, orientable, and homogeneous $D$-dimensional Riemannian manifolds, we present the derivation of our main theorem in this section. We separate the result into two parts. First, for the chosen embedding, we use the averaged tangent projection to define the coefficient $\kappa\left(M\right)$. This coefficient multiplies the coupling-induced source term in the linearized kinetic equation. We obtain the actual critical coupling from the full density-level spectral problem after the response of the intrinsic drift fields is evaluated, so the threshold is determined jointly by geometry and by the distribution of intrinsic drift fields. Second, we use the Euler characteristic to obtain the unconditional Poincar\'e-Hopf defect constraint on the critical tangent field. We then use the additional locality and sign conditions of Appendices \ref{Appendix-D} and \ref{Appendix-E} to obtain conditional statements about the cubic coefficient and the local phase transition scenario. In that conditional class, a non-zero Euler characteristic gives the cubic sign used by the one-mode normal form, and the further stabilizing condition $\Lambda_{5}>0$ gives a discontinuous local transition. When $\chi\left(M\right)=0$, the local branch is determined by the normal-form coefficients rather than by the Euler characteristic alone. Below, we use the continuous form of the generalized model to elaborate our derivation. We first formulate the linear kinetic response of the order parameter under perturbations and then establish its connection with the geometric and topological characteristics of the manifold.

To implement this analysis, we consider a situation in which the system is initialized at a ground state
\begin{align}
\rho_{0}\left(\sigma,V\right)=\frac{1}{\textsf{Vol}\left(M\right)}\label{III-A-EQ001}
\end{align}
for every fixed $V$ (i.e., the spatial distribution is uniform over the manifold conditional on each intrinsic drift field, here $\textsf{Vol}\left(M\right)$ measures the total volume of $M$). This uniform distribution lies far from the synchronized state, leading to a near-zero value of the order parameter $r_{0}\approx 0$. We add small perturbations on $\rho_{0}$ to affect $r_{0}$
\begin{align}
\rho=\rho_{0}+\delta_{\rho},\;\;\;\;\;r=r_{0}+\delta_{r}.\label{III-A-EQ002}
\end{align}
These small perturbations can be used to model the influence of interactions between oscillators on the probability flow described in Eq. (\ref{II-D-EQ004}). Note that $\rho_{0}$ is a constant at every point on the manifold for each fixed $V$.

Under small perturbations, we first determine the linear stability of the incoherent state at the level of the full density perturbation. Once we identify a simple critical eigenvalue, we reduce the local nonlinear dynamics of the order parameter to a response equation of the form
\begin{align}
\frac{\partial}{\partial t}r=\mathcal{L}\left(r\right)+\mathcal{N}\left(r\right),\label{III-A-EQ003}
\end{align}
where $\mathcal{L}\left(r\right)$ is the linear operator inherited from the kinetic spectral problem and $\mathcal{N}\left(r\right)$ is the non-linear term. In the following analysis, we identify the geometric source coefficient multiplying the coupling-induced source term, define the corresponding spectral threshold, and then relate the nonlinear coefficient to the topology of the manifold under the local reduction assumptions.

\begin{table*}[t]
  \caption{Geometric contribution to the loss of stability of the incoherent
  state for a fixed embedding. The coefficient
  $\kappa\left(M\right)$ is the isotropic averaged tangent-projection factor.
  It measures how strongly a mean-field perturbation is converted into a
  tangent coupling force on the manifold. The critical coupling is then
  determined by this geometric factor together with the intrinsic-drift
  response.}
  \label{Table-2}
  \begin{ruledtabular}
  \renewcommand{\arraystretch}{1.22}
  \begin{tabular}{@{}ll@{}}
    \begin{minipage}[t]{0.22\textwidth}
    Geometric role
    \end{minipage} &
    \begin{minipage}[t]{0.75\textwidth}
    Detailed effect on the loss of stability of the incoherent state
    \end{minipage} \\ \hline
    \begin{minipage}[t]{0.22\textwidth}\raggedright
    Direct role in the linearized kinetic equation
    \end{minipage} &
    \begin{minipage}[t]{0.75\textwidth}\raggedright
    Geometry determines how much of a mean-field perturbation becomes a tangent
    force on $M$. In an isotropic response mode, $\kappa\left(M\right)$ sets the
    geometric strength of the destabilizing coupling feedback:
    $\overline{\mathsf{P}}=\textsf{Vol}\left(M\right)^{-1}
    \int_{M}\mathsf{P}_{\sigma}^{\perp}\mathsf{d}\mu\left(\sigma\right)$, and
    $\overline{\mathsf{P}}=\kappa\left(M\right)\mathbf{I}$ in the scalar
    isotropic reduction. The geometric coupling factor is then
    $K\kappa\left(M\right)$.
    \end{minipage} \\[0.9em] \hline
    \begin{minipage}[t]{0.22\textwidth}\raggedright
    Role in the critical coupling
    \end{minipage} &
    \begin{minipage}[t]{0.75\textwidth}\raggedright
    Geometry directly affects the coupling strength required to destabilize the
    incoherent state, but it does not determine the critical coupling alone.
    The critical coupling is fixed jointly by $\kappa\left(M\right)$ and the
    $h$-dependent susceptibility $\mathcal{S}\left(\lambda;h\right)$ through
    $\det\!\left[\mathbf{I}-K\mathcal{S}\left(\lambda;h\right)
    \left(\mathbf{I}-\alpha A\right)\right]=0$. If
    $s_{M}\left(0^{+};h\right)=\kappa\left(M\right)\Xi\left(0^{+};h\right)$,
    then the scalar intrinsic-drift response is $\Xi\left(0^{+};h\right)$ and
    $K_{c}=1/\left[\kappa\left(M\right)\Xi\left(0^{+};h\right)\right]$.
    \end{minipage} \\[0.4em]
  \end{tabular}
  \end{ruledtabular}
\end{table*}

\subsection{Linearized kinetic response and spectral threshold}\label{Linear-response-subsection}

\paragraph{Density-level linearization.-}We first determine the linear response at the level of the full density perturbation. From Eq. (\ref{II-D-EQ003}), the perturbation of the velocity field is
\begin{align}
\delta_{\mathbf{v}}
=
K\mathsf{P}_{\sigma}^{\perp}
\left(\mathbf{I}-\alpha A\right)\delta_{r}.
\label{III-B-EQ001}
\end{align}
We insert $\rho=\rho_{0}+\delta_{\rho}$ and
$\mathbf{v}=V+\delta_{\mathbf{v}}$ into the continuity equation
Eq. (\ref{II-D-EQ004}) and keep only linear terms. This gives
\begin{align}
\frac{\partial}{\partial t}\delta_{\rho}
=
\mathcal{A}_{V}\left[\delta_{\rho}\right]
-K\rho_{0}\nabla_{M}\cdot
\left[
\mathsf{P}_{\sigma}^{\perp}
\left(\mathbf{I}-\alpha A\right)\delta_{r}
\right],
\label{III-B-EQ002}
\end{align}
where
\begin{align}
\mathcal{A}_{V}\left[\varphi\right]
=
-\nabla_{M}\cdot\left(\varphi V\right)
\label{III-B-EQ003}
\end{align}
is the transport generator associated with the intrinsic drift field. The order parameter perturbation is the first embedded moment
\begin{align}
\delta_{r}
=
\int_{TM}\int_{M}
F\left(\sigma\right)\delta_{\rho}\left(\sigma,V,t\right)
h\left(V\right)\mathsf{d}\mu\left(\sigma\right)\mathsf{d}V.
\label{III-B-EQ004}
\end{align}
Eqs. (\ref{III-B-EQ002}, \ref{III-B-EQ004}) form the correct linearized problem. They are not closed as an autonomous equation for $\delta_{r}$, because the transport term $\mathcal{A}_{V}\left[\delta_{\rho}\right]$ depends on the full density perturbation, not only on its first embedded moment.

\paragraph{Geometric source term.-}Although the moment equation is not closed, we can still isolate the geometric part of the coupling-induced source term in Eq. (\ref{III-B-EQ002}). For every tangent vector field $\mathbf{X}$ on the compact manifold,
\begin{align}
\int_{M}F\left(\sigma\right)\nabla_{M}\cdot\mathbf{X}\mathsf{d}\mu\left(\sigma\right)
=
-\int_{M}\mathsf{d}F_{\sigma}\left[\mathbf{X}\left(\sigma\right)\right]\mathsf{d}\mu\left(\sigma\right).
\label{III-B-EQ005}
\end{align}
Therefore, with
$\mathbf{Y}=\left(\mathbf{I}-\alpha A\right)\delta_{r}$,
we obtain the direct moment of the coupling-induced source term as
\begin{align}
&-K\rho_{0}\int_{TM}\int_{M}
F\left(\sigma\right)
\nabla_{M}\cdot
\left[\mathsf{P}_{\sigma}^{\perp}\mathbf{Y}\right]
h\left(V\right)\mathsf{d}\mu\left(\sigma\right)\mathsf{d}V\notag\\
&=
\frac{K}{\textsf{Vol}\left(M\right)}
\int_{M}\mathsf{P}_{\sigma}^{\perp}\mathbf{Y}\mathsf{d}\mu\left(\sigma\right).
\label{III-B-EQ006}
\end{align}
Up to this point, we have not used homogeneity. If the homogeneous embedding subspace probed by the order parameter is isotropic, then the averaged tangent projection acts as a scalar multiple of the identity on that subspace:
\begin{align}
\frac{1}{\textsf{Vol}\left(M\right)}
\int_{M}\mathsf{P}_{\sigma}^{\perp}\mathbf{Y}\mathsf{d}\mu\left(\sigma\right)
=
\kappa\left(M\right)\mathbf{Y}.
\label{III-B-EQ007}
\end{align}
We therefore identify $\kappa\left(M\right)$ as the averaged tangent-projection factor multiplying the coupling-induced source term,
\begin{align}
\frac{K}{\textsf{Vol}\left(M\right)}
\int_{M}\mathsf{P}_{\sigma}^{\perp}\mathbf{Y}\mathsf{d}\mu\left(\sigma\right)
=
K\kappa\left(M\right)\mathbf{Y}.
\label{III-B-EQ008}
\end{align}
This identity specifies the coupling-induced source term rather than a closed growth equation for $\delta_{r}$; the intrinsic-drift resolvent is evaluated separately.

\paragraph{Transport resolvent and self-consistency.-}We note that the operator $\mathcal{A}_{V}$ is skew-adjoint on $L^{2}\left(M,\mu\right)$ when $\nabla_{M}\cdot V=0$:
\begin{align}
\int_{M}f\,\mathcal{A}_{V}\left[g\right]\mathsf{d}\mu
=
-\int_{M}g\,\mathcal{A}_{V}\left[f\right]\mathsf{d}\mu.
\label{III-B-EQ009}
\end{align}
However, this skew-adjointness does not by itself determine the real part of the spectrum after the rank-finite mean-field coupling is added. We must obtain the loss of stability from the coupled density-moment problem. We therefore use a normal-mode ansatz
$\delta_{\rho}=e^{\lambda t}\eta\left(\sigma,V\right)$ and
$\delta_{r}=e^{\lambda t}q$. For $\operatorname{Re}\lambda>0$,
Eq. (\ref{III-B-EQ002}) gives
\begin{align}
\left(\lambda-\mathcal{A}_{V}\right)\eta
=
-K\rho_{0}\nabla_{M}\cdot
\left[
\mathsf{P}_{\sigma}^{\perp}
\left(\mathbf{I}-\alpha A\right)q
\right].
\label{III-B-EQ010}
\end{align}
We apply the resolvent of the transport generator and impose the moment condition. This gives
\begin{align}
q
=
K\mathcal{S}\left(\lambda;h\right)
\left(\mathbf{I}-\alpha A\right)q,
\label{III-B-EQ011}
\end{align}
where the linear susceptibility operator is
\begin{align}
\mathcal{S}\left(\lambda;h\right)z
=&
-\frac{1}{\textsf{Vol}\left(M\right)}
\int_{TM}\int_{M}
F\left(\sigma\right)
\left(\lambda-\mathcal{A}_{V}\right)^{-1}\notag\\
&\times
\nabla_{M}\cdot
\left[
\mathsf{P}_{\sigma}^{\perp}z
\right]
h\left(V\right)\mathsf{d}\mu\left(\sigma\right)\mathsf{d}V.
\label{III-B-EQ012}
\end{align}
We therefore determine the critical coupling from the spectral self-consistency condition
\begin{align}
\det
\left[
\mathbf{I}
-K\mathcal{S}\left(\lambda;h\right)
\left(\mathbf{I}-\alpha A\right)
\right]
=0,
\qquad
\operatorname{Re}\lambda=0.
\label{III-B-EQ013}
\end{align}
Equivalently, we may approach the neutral spectrum by analytic continuation from $\operatorname{Re}\lambda>0$. In this formulation, the averaged projection factor $\kappa\left(M\right)$ is contained in the geometric source term in Eq. (\ref{III-B-EQ008}), whereas the actual threshold also depends on the intrinsic-drift distribution through the resolvent in Eq. (\ref{III-B-EQ012}). We define the intrinsic-drift response as the $h$-dependent susceptibility $\mathcal{S}\left(\lambda;h\right)$ in Eq. (\ref{III-B-EQ012}); when the scalar factorization $s_{M}\left(\lambda;h\right)=\kappa\left(M\right)\Xi\left(\lambda;h\right)$ holds, the corresponding scalar intrinsic-drift response is $\Xi\left(\lambda;h\right)$.

\paragraph{Scalar isotropic reduction.-}We use Eq. (\ref{III-B-EQ013}) as the general threshold condition. We now extract a closed-form consequence in the common isotropic case. Suppose that the order parameter subspace is an irreducible representation of the transitive symmetry group of $M$, that the drift ensemble $h$ is invariant under this symmetry, and that the phase-lag generator $A$ preserves the same subspace. Then $\mathcal{S}\left(\lambda;h\right)$ commutes with the symmetry action and Schur's lemma gives
\begin{align}
\mathcal{S}\left(\lambda;h\right)\big|_{\mathcal{E}}
=
s_{M}\left(\lambda;h\right)\mathbf{I}_{\mathcal{E}},
\label{III-B-EQ014}
\end{align}
on the order parameter subspace $\mathcal{E}$. We then determine the loss of
linear stability from the scalar mode equations
\begin{align}
1
=
K\,s_{M}\left(\lambda;h\right)\beta_{\ell},
\qquad
\beta_{\ell}\in
\operatorname{spec}
\left[
\left(\mathbf{I}-\alpha A\right)\big|_{\mathcal{E}}
\right].
\label{III-B-EQ015}
\end{align}
For a stationary loss of stability with $\alpha=0$ and $s_{M}\left(0^{+};h\right)>0$, Eq. (\ref{III-B-EQ015}) reduces to the explicit threshold formula
\begin{align}
K_{c}
=
\frac{1}{s_{M}\left(0^{+};h\right)}.
\label{III-B-EQ016}
\end{align}
If, in addition, the transport response factorizes on this subspace as
$s_{M}\left(\lambda;h\right)=\kappa\left(M\right)\Xi\left(\lambda;h\right)$, then
\begin{align}
K_{c}
=
\frac{1}{\kappa\left(M\right)\Xi\left(0^{+};h\right)}.
\label{III-B-EQ017}
\end{align}
Eq. (\ref{III-B-EQ017}) applies for $\alpha=0$. In this factorized scalar reduction, $K_{c}$ is inversely proportional to both the averaged tangent-projection factor $\kappa\left(M\right)$ and the drift susceptibility $\Xi\left(0^{+};h\right)$. At fixed $\Xi$, increasing $\kappa\left(M\right)$ lowers the critical coupling required for the incoherent state to lose stability, while changes in the intrinsic heterogeneity modify $\Xi\left(0^{+};h\right)$. Table \ref{Table-2} summarizes how the averaged tangent projection contributes to the coupling-induced source term and, after the intrinsic-drift response is evaluated, to the threshold condition.

\paragraph{Spherical reductions.-}We first check the density-level formulation by recovering the standard Kuramoto threshold in the one-dimensional limit. For $M=S^{1}$, we write the two real embedding coordinates in complex notation as $F\left(\theta\right)=e^{i\theta}$, take $V=\omega\partial_{\theta}$, set $h\left(V\right)\mathsf{d}V=g\left(\omega\right)\mathsf{d}\omega$, and set $\alpha=0$. The first Fourier mode of Eq. (\ref{III-B-EQ011}) gives the usual scalar self-consistency relation
\begin{align}
1
=
\frac{K}{2}
\int_{\mathbb{R}}
\frac{g\left(\omega\right)}{\lambda+i\omega}
\mathsf{d}\omega.
\label{III-B-EQ018}
\end{align}
For an even unimodal frequency density and a stationary neutral mode, we take $\lambda\to0^{+}$ and obtain
\begin{align}
K_{c}
=
\frac{2}{\pi g\left(0\right)},
\label{III-B-EQ019}
\end{align}
with the conventional Kuramoto normalization \cite{Kuramoto1975,Rodrigues2016PhysRep,Pietras2019PhysRep}. This calculation recovers the full threshold formula in the circle limit and retains the intrinsic-drift distribution in the critical coupling.

For higher-dimensional hyperspheres, we keep \(D\) as the manifold dimension and write the standard case as \(M=S^{D}\subset\mathbb{R}^{D+1}\). The same framework recovers the averaged tangent-projection coefficient \(\kappa\left(S^{D}\right)=D/\left(D+1\right)\) in the coupling-induced source term. The threshold is then obtained from the susceptibility condition in Eq. (\ref{III-B-EQ013}) after one specifies the distribution of intrinsic rotation fields. We now specialize this condition to the standard hyperspherical rotation fields. Let \(M=S^{D}\subset\mathbb{R}^{D+1}\), let
$V_{\Omega}\left(\sigma\right)=\Omega\sigma$ with
$\Omega\in\mathfrak{so}\left(D+1\right)$, and write the rotation-field distribution as
$h\left(\Omega\right)\mathsf{d}\Omega$. For
$f_{z}\left(\sigma\right)=\sigma^{\top}z$, direct spherical calculus gives
\begin{align}
\nabla_{S^{D}}f_{z}
&=
\mathsf{P}_{\sigma}^{\perp}z,
\notag\\
\nabla_{S^{D}}\cdot
\left(\mathsf{P}_{\sigma}^{\perp}z\right)
&=
-D f_{z},
\notag\\
\mathcal{A}_{V_{\Omega}}f_{z}
&=
f_{\Omega z}.
\label{III-B-EQ020}
\end{align}
Thus the first spherical-harmonic subspace is invariant under the intrinsic
transport, and Eq. (\ref{III-B-EQ012}) reduces to
\begin{align}
\mathcal{S}_{S^{D}}\left(\lambda;h\right)z
=&
\frac{D}{D+1}
\int_{\mathfrak{so}\left(D+1\right)}
\left(\lambda\mathbf{I}_{D+1}-\Omega\right)^{-1}z
\notag\\
&\times
h\left(\Omega\right)\mathsf{d}\Omega .
\label{III-B-EQ021}
\end{align}
Eq. (\ref{III-B-EQ021}) is the hyperspherical susceptibility associated with the generalized rotation field \(V_{\Omega}\). In this expression, the universal projection factor \(D/\left(D+1\right)\) multiplies the drift resolvent, and the rotation-field distribution is included in the same resolvent average.

We obtain a scalar closed-form threshold by choosing a solvable rotation-field ensemble. Let \(D+1=2m\), let \(J\) be a fixed complex structure with \(J^{\top}=-J\) and \(J^{2}=-\mathbf{I}_{D+1}\), and choose
$\Omega=\omega J$ with
$h\left(\Omega\right)\mathsf{d}\Omega=g\left(\omega\right)\mathsf{d}\omega$.
Then Eq. (\ref{III-B-EQ021}) gives
\begin{align}
\mathcal{S}_{S^{D}}\left(\lambda;g\right)z
=&
\frac{D}{D+1}
\int_{\mathbb{R}}
\left(\lambda\mathbf{I}_{D+1}-\omega J\right)^{-1}z
g\left(\omega\right)\mathsf{d}\omega .
\label{III-B-EQ022}
\end{align}
If $g$ is even, the term proportional to $J$ cancels and
\begin{align}
s_{S^{D}}\left(\lambda;g\right)
=
\frac{D}{D+1}
\int_{\mathbb{R}}
\frac{\lambda g\left(\omega\right)}
{\lambda^{2}+\omega^{2}}
\mathsf{d}\omega .
\label{III-B-EQ023}
\end{align}
For an even unimodal density and a stationary neutral mode,
$\lambda\to0^{+}$ gives
\begin{align}
K_{c}
=
\frac{D+1}{D\pi g\left(0\right)},
\qquad
D+1=2m .
\label{III-B-EQ024}
\end{align}
The case \(D=1\) reduces to Eq. (\ref{III-B-EQ019}). For \(D>1\), Eq. (\ref{III-B-EQ024}) gives the threshold for this specified Hopf-rotation ensemble. Other rotation-field distributions are evaluated from Eq. (\ref{III-B-EQ021}). This solvable specialization displays how the averaged projection factor \(D/\left(D+1\right)\) and the drift susceptibility enter the same threshold condition.

\begin{figure*}[!t]
\includegraphics[width=\textwidth]{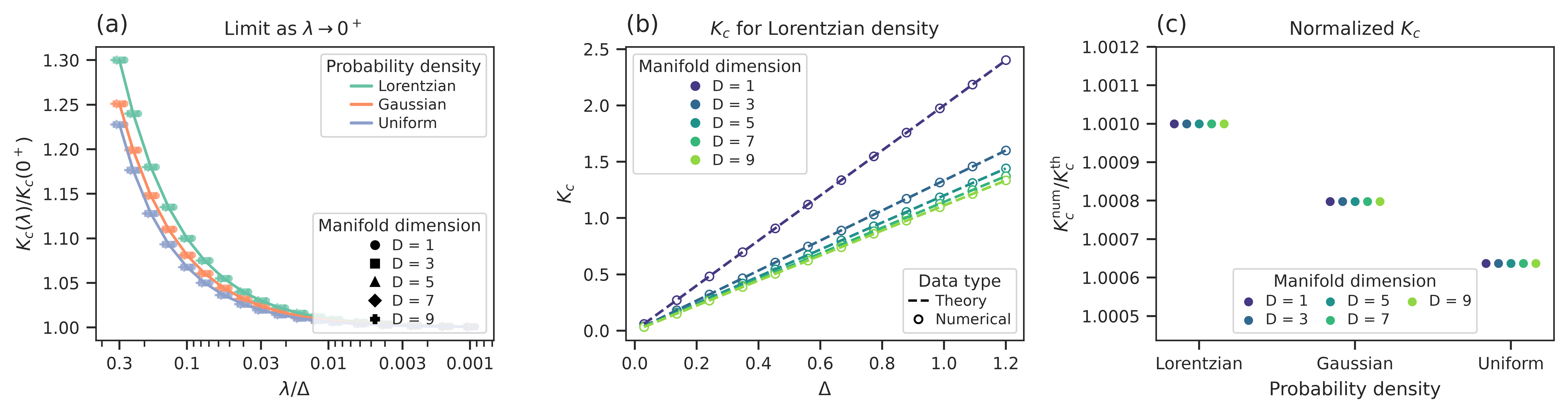}
\caption[Numerical verification of the critical coupling on hyperspheres]{Numerical verification of \(K_{c}\) on hyperspheres for the
Hopf-rotation ensemble. The symbol \(D\) denotes the manifold dimension, so the
tested manifolds are \(S^{D}\subset\mathbb{R}^{D+1}\) with
\(D=1,3,5,7,9\). The parameter \(\Delta\) is the width parameter of the
intrinsic rotation-rate probability density \(g_{\Delta}\), and \(\lambda>0\)
is the parameter in Eq. (\ref{III-B-EQ023}) that is taken to \(0^{+}\) when
computing \(K_{c}\). (a) The plotted quantity is
\(K_{c}\left(\lambda;D,\Delta\right)/
K_{c}\left(0^{+};D,\Delta\right)\), where
\(K_{c}\left(\lambda;D,\Delta\right)=
1/s_{S^{D}}\left(\lambda;g_{\Delta}\right)\). The ratio approaches one as
\(\lambda/\Delta\to0^{+}\). (b) For the Lorentzian density, the computed value
of \(K_{c}\left(\lambda;D,\Delta\right)\) at \(\lambda/\Delta=10^{-3}\) follows
the lines \(K_{c}^{\mathrm{th}}=\left(D+1\right)\Delta/D\). The different
slopes show the effect of the geometric projection factor \(D/\left(D+1\right)\).
(c) After normalization by \(K_{c}^{\mathrm{th}}\), the computed values for
Lorentzian, Gaussian, and uniform probability densities remain close to one for
every tested manifold dimension. This shows that the geometric factor and the
intrinsic-drift probability density enter the same formula for \(K_{c}\).}
\label{Fig-linear-threshold}
\end{figure*}

For the numerical check below, we write the scalar rotation-frequency density as
\(g_{\Delta}\), where \(\Delta\) is the width parameter of the intrinsic
rotation-rate probability density. In the Lorentzian case,
\(g_{\Delta}\left(\omega\right)=
\Delta/\left[\pi\left(\omega^{2}+\Delta^{2}\right)\right]\), so
\(\pi g_{\Delta}\left(0\right)=1/\Delta\) and Eq. (\ref{III-B-EQ024})
becomes
\begin{align}
K_{c}^{\mathrm{th}}\left(D,\Delta\right)
=
\frac{D+1}{D}\Delta .
\label{III-B-EQ025}
\end{align}
For the Gaussian and uniform cases we use the same neutral formula
\(K_{c}^{\mathrm{th}}=\left(D+1\right)/\left[D\pi g_{\Delta}\left(0\right)\right]\)
with the corresponding value of \(g_{\Delta}\left(0\right)\). We also evaluate
the finite-\(\lambda\) susceptibility in Eq. (\ref{III-B-EQ023}) and define
\begin{align}
K_{c}\left(\lambda;D,\Delta\right)
=
\frac{1}{s_{S^{D}}\left(\lambda;g_{\Delta}\right)} .
\label{III-B-EQ026}
\end{align}
The limit \(\lambda\to0^{+}\) is the neutral spectral limit used to compute
\(K_{c}\). Thus Figure \ref{Fig-linear-threshold} tests the susceptibility
calculation in Eq. (\ref{III-B-EQ023}) and the resulting formula for \(K_{c}\).
We next examine how a finite empirical sample of the geometric and
intrinsic-drift distributions approaches the same calculation.

Figure \ref{Fig-linear-threshold} gives a numerical check of
Eq. (\ref{III-B-EQ024}) on \(S^{D}\). We use three probability densities to separate
two effects: the dimension-dependent geometric factor \(D/\left(D+1\right)\)
and the drift response through \(\pi g_{\Delta}\left(0\right)\). Figure
\ref{Fig-linear-threshold}(a) confirms that the susceptibility
\(s_{S^{D}}\left(\lambda;g_{\Delta}\right)\) has the required limit as
\(\lambda\to0^{+}\). Figure \ref{Fig-linear-threshold}(b) then tests the
Lorentzian case, where Eq. (\ref{III-B-EQ025}) gives a straight-line relation
between \(K_{c}^{\mathrm{th}}\) and \(\Delta\). Figure
\ref{Fig-linear-threshold}(c) shows that the same formula for \(K_{c}\) also
organizes Gaussian and uniform probability densities once each result is
normalized by its own \(K_{c}^{\mathrm{th}}\). Therefore the numerics test the
specific conclusion of the linearized kinetic equation: geometry determines the
strength factor multiplying the coupling-induced source term, while the
intrinsic drift distribution supplies the susceptibility factor in the critical
coupling.

We further test the same threshold condition by replacing the density-level
averages in Eq. (\ref{III-B-EQ023}) by finite empirical averages. For a sample
\(\left\{x_{j},\omega_{j}\right\}_{j=1}^{N}\) on \(S^{D}\), with the
order parameter perturbation taken along a fixed embedding direction, we use
\begin{align}
\kappa_{N}
=
\frac{1}{N}
\sum_{j=1}^{N}
\left(1-x_{j,1}^{2}\right)
\label{III-B-EQ027}
\end{align}
as the empirical tangent-projection factor. For the Lorentzian intrinsic
rotation-rate density and a small spectral regularization
\(\epsilon=10^{-2}\Delta\), we then compute
\begin{align}
K_{c,N}^{\left(\epsilon\right)}
=
\left[
\kappa_{N}
\frac{1}{N}
\sum_{j=1}^{N}
\frac{\epsilon}{\epsilon^{2}+\omega_{j}^{2}}
\right]^{-1}.
\label{III-B-EQ028}
\end{align}
The corresponding density-level quantity \(K_{c}^{\left(\epsilon\right)}\) is
obtained by replacing the empirical averages in Eq. (\ref{III-B-EQ028}) by
their integrals. The comparison therefore checks the finite-sample convergence
of the two factors in Eq. (\ref{III-B-EQ028}).

\begin{figure*}[!t]
\includegraphics[width=\textwidth]{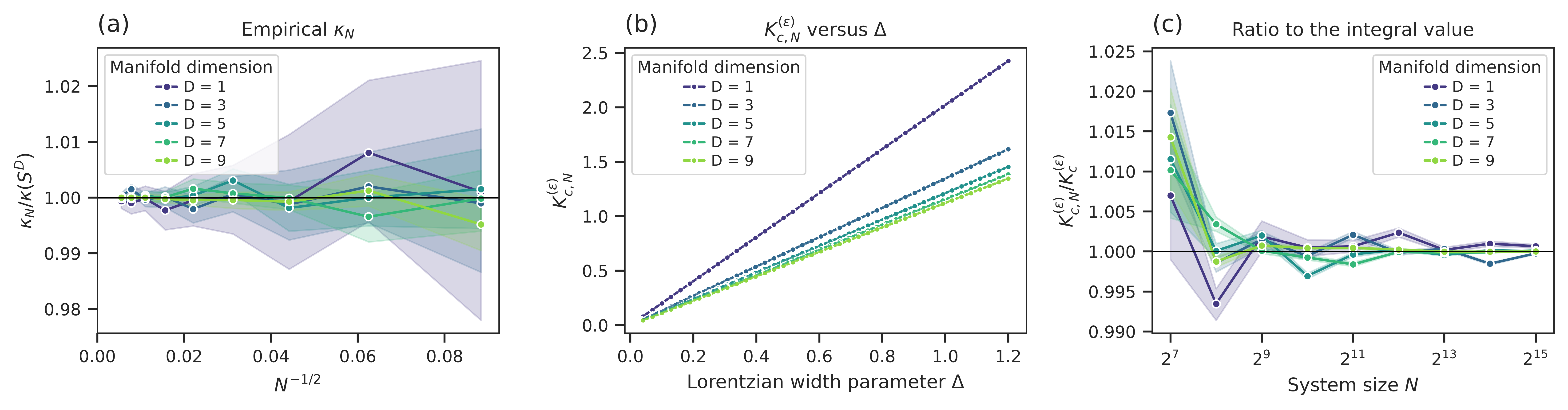}
\caption[Finite-sample verification of the critical coupling]{Finite-sample verification of the critical coupling computed from
Eq. (\ref{III-B-EQ028}).
The symbol \(D\) denotes the manifold dimension, and the tested manifolds are
\(S^{D}\subset\mathbb{R}^{D+1}\) with \(D=1,3,5,7,9\). The intrinsic
rotation-rate probability density is Lorentzian with width parameter
\(\Delta\), and the spectral regularization is fixed at
\(\epsilon=10^{-2}\Delta\). (a) The empirical geometric factor
\(\kappa_{N}\), normalized by the theoretical value \(\kappa\left(S^{D}\right)\),
converges to unity as the sample size increases. This verifies the finite-sample
recovery of the averaged tangent projection. (b) For the largest sample size
\(N=32768\), the empirical value \(K_{c,N}^{\left(\epsilon\right)}\) follows the
curves obtained by replacing the sums in Eq. (\ref{III-B-EQ028}) by integrals.
The markers and bands show empirical seed averages and confidence intervals,
and the thin dashed curves show the corresponding integral values. (c) The ratio
\(K_{c,N}^{\left(\epsilon\right)}/K_{c}^{\left(\epsilon\right)}\), averaged over
all tested \(\Delta\) values and random seeds, approaches one with increasing
\(N\). This confirms that the finite empirical approximation converges to the
same calculation of \(K_{c}\) used in Figure \ref{Fig-linear-threshold}.}
\label{Fig-finite-sample-threshold}
\end{figure*}

Figure \ref{Fig-finite-sample-threshold} complements Figure
\ref{Fig-linear-threshold}. The first figure verifies the closed-form
density-level susceptibility. The second figure verifies that the two empirical
averages in Eq. (\ref{III-B-EQ028}), namely the tangent-projection average and
the intrinsic-drift resolvent average, converge to the corresponding integral
values. The numerical evidence therefore supports the interpretation that
geometry fixes the coupling strength factor, whereas the intrinsic-drift
probability density fixes the response factor that appears together with it in
the critical coupling.

\begin{table*}[t]
  \caption{Topological contribution to the nonlinear response analysis. The
  defect constraint is unconditional. The statements about $\Lambda_{3}$ and
  the local phase transition scenario require the locality condition of
  Appendix \ref{Appendix-D}, the sign condition of Appendix \ref{Appendix-E},
  and, for a stabilized discontinuous branch, the additional condition
  $\Lambda_{5}>0$.}
  \label{Table-3}
  \begin{ruledtabular}
  \renewcommand{\arraystretch}{1.22}
  \begin{tabular}{@{}ll@{}}
    \begin{minipage}[t]{0.23\textwidth}
    Topological role
    \end{minipage} &
    \begin{minipage}[t]{0.74\textwidth}
    Detailed effect on the nonlinear response
    \end{minipage} \\ \hline
    \begin{minipage}[t]{0.23\textwidth}\raggedright
    Topology constrains the critical texture
    \end{minipage} &
    \begin{minipage}[t]{0.74\textwidth}\raggedright
    We apply topology to the critical tangent field
    \(u_{c}\left(\sigma\right)=\mathsf{P}_{\sigma}^{\perp}
    \left(r_{c}-\alpha A r_{c}\right)\), not through the ambient
    order parameter vector alone. The zeros of \(u_c\) are defect cores, and
    Poincar\'e-Hopf gives
    \(\sum_{\ell}\operatorname{ind}_{p_{\ell}}\left(u_{c}\right)
    =\chi\left(M\right)\). If the zeros are simple, then
    \(N_{\mathrm{def}}\geq\left|\chi\left(M\right)\right|\). Thus a
    non-zero Euler characteristic imposes a non-zero net defect charge on the
    critical texture. This statement does not determine the sign of
    \(\Lambda_{3}\).
    \end{minipage} \\[0.9em] \hline
    \begin{minipage}[t]{0.23\textwidth}\raggedright
    Topology can affect \(\Lambda_{3}\) under stated assumptions
    \end{minipage} &
    \begin{minipage}[t]{0.74\textwidth}\raggedright
    Under the locality assumptions of Appendix \ref{Appendix-D} and the signed
    dominance condition of Appendix \ref{Appendix-E}, the contribution of
    neighborhoods of the zeros weighted by their local indices controls the
    cubic coefficient. In that conditional class,
    the topological defect data can give \(\Lambda_{3}<0\). Without these
    dynamical assumptions, \(\chi\left(M\right)\neq0\) alone is not a sign
    theorem for \(\Lambda_{3}\).
    \end{minipage} \\[0.9em] \hline
    \begin{minipage}[t]{0.23\textwidth}\raggedright
    Topology can affect the local transition under stated assumptions
    \end{minipage} &
    \begin{minipage}[t]{0.74\textwidth}\raggedright
    If the conditional sign analysis gives \(\Lambda_{3}<0\) and the quintic
    coefficient satisfies \(\Lambda_{5}>0\), the one-mode normal form has a
    stabilized subcritical branch and a discontinuous local transition. If
    \(\chi\left(M\right)=0\), the local branch is determined by the normal-form
    coefficients rather than by topology.
    \end{minipage} \\[0.4em]
  \end{tabular}
  \end{ruledtabular}
\end{table*}

\subsection{Non-linear response term of the order parameter response equation}\label{Nonlinear-response-subsection}

In the previous section, we formulated the linearized kinetic spectral problem
and identified the geometric source coefficient $\kappa\left(M\right)$ that multiplies the
coupling-induced source term in the linearized kinetic equation [see Eqs. (\ref{III-B-EQ008}-\ref{III-B-EQ013})].
We now focus on the nonlinear coefficient that determines the local
bifurcation type near the incoherent state. We separate two levels of the
argument. The Poincar\'e-Hopf index relation for the critical tangent field is
topological and unconditional. The conclusions about the sign of the cubic
coefficient and about the local phase transition scenario are conditional
results of a one-mode local reduction. The technical reduction is collected in
Appendix \ref{Appendix-D}; here we retain only the ingredients needed to state
these conditions. We assume that a simple critical eigenvalue selects the
critical collective mode. We also assume transversality, so this mode changes
stability at a definite critical coupling. We further impose the vanishing of
the quadratic coefficient, which rules out a preferred sign of the order
parameter at the incoherent state. Appendix \ref{Appendix-D} gives the precise
form used here.

\paragraph{Reduced amplitude equation.-}We assume that the steady response
problem admits a smooth local reduction near the critical coupling
$K=K_{c}$. We assume that the critical eigenvalue of the linearized
order parameter operator is simple. We assume that this eigenvalue crosses
zero transversely. We also assume that the quadratic normal-form coefficient
vanishes. Appendix \ref{Appendix-D} then yields a one-dimensional scalar
amplitude equation for the critical collective coordinate. This equation is
the one-mode reduction of the full response equation
$\frac{\partial}{\partial t}r=\mathcal{L}\left(r\right)+\mathcal{N}\left(r\right)$
in Eq. (\ref{III-A-EQ003}). Its linear term is the reduced action of
$\mathcal{L}$ near $K=K_{c}$, whereas its cubic term is the leading nonlinear
contribution generated by $\mathcal{N}$. The reduced equation reads
\begin{align}
\Psi\left(a;K\right)
&=
\lambda\left(K\right)a+\Lambda_{3}a^{3}\notag\\
&\quad
+O\left(a^{4}\right)
+O\left(\left|K-K_{c}\right|a^{2}\right)
=0,\label{III-C-EQ001}
\end{align}
together with the critical response branch
$r\left(a\right)=a\,r_{c}+O\left(a^{2}\right)$, where $r_{c}$ is the critical
order parameter direction. The coefficient $\Lambda_{3}$ admits the explicit
formula derived in Appendix \ref{Appendix-D}. This reduction does not change
the model. It only separates the critical collective coordinate from the
stable modes that are determined by it near threshold.

\paragraph{Topological relation.-}We connect $\Lambda_{3}$ with the topology of
$M$ by associating $r_{c}$ with the induced critical tangent field
\begin{align}
u_{c}\left(\sigma\right)
&=
\mathsf{P}_{\sigma}^{\perp}
\left(r_{c}-\alpha A r_{c}\right).\label{III-C-EQ002}
\end{align}
This field is the natural geometric object in our model. The microscopic
transport law in Eq. (\ref{II-D-EQ003}) depends on the order parameter through
the projected tangent field
$\mathsf{P}_{\sigma}^{\perp}\left(r-\alpha A r\right)$. For this reason, the
topological information is carried by the induced critical tangent field
$u_{c}$ in the nonlinear response. It is not determined by $r_{c}$ alone.
Under the locality and symmetry conditions stated in Appendix
\ref{Appendix-D}, the cubic coefficient can be written as
\begin{align}
\Lambda_{3}
=
\int_{M}T\left(\sigma\right)\!\left[u_{c}\left(\sigma\right),u_{c}\left(\sigma\right),u_{c}\left(\sigma\right)\right]
\,\mathsf{d}\mu\left(\sigma\right),\label{III-C-EQ003}
\end{align}
where $T\left(\sigma\right)$ is a smooth symmetric trilinear tensor field.
We use these conditions to express the locality and symmetry structure of the
model. The coupling term acts pointwise on the manifold, and the homogeneous
structure organizes the relevant tensor fields into symmetry components. Appendix
\ref{Appendix-D} states the condition under which the elimination of stable
modes preserves this structure.
The Poincar\'e-Hopf theorem \cite{Milnor1997DifferentiableViewpoint} then implies that, when the Euler characteristic
$\chi\left(M\right)$ of $M$ is non-zero,
the field $u_{c}$ must have zeros. This topologically forced zero set does not
by itself determine the sign of $\Lambda_{3}$. To obtain a definite sign
statement, we impose the sign condition stated in Appendix \ref{Appendix-E}.
This condition concerns the local cubic density
$q\left(\sigma\right)=T\left(\sigma\right)\!\left[
u_{c}\left(\sigma\right),u_{c}\left(\sigma\right),u_{c}\left(\sigma\right)
\right]$. It requires the contributions from neighborhoods of the zeros to
admit the indexed decomposition in Eq. (\ref{EEQ3}), and it requires the
weighted indexed sum $\Gamma\left(u_{c}\right)$ to be positive and to dominate
the remaining contribution that is not weighted by the indices, as in
Eq. (\ref{EEQ6}).
Within this conditional class, $\chi\left(M\right)\neq 0$ implies
$\Lambda_{3}<0$. When $\chi\left(M\right)=0$, the Poincar\'e-Hopf relation
does not supply a sign for $\Lambda_{3}$, and the sign is decided by the
analytic structure of the reduced nonlinearity.

\paragraph{Bifurcation consequence.-}We also use Appendix \ref{Appendix-D} to write
$\lambda\left(K\right)=\lambda^{\prime}\!\left(K_{c}\right)\left(K-K_{c}\right)+
O\left(\left(K-K_{c}\right)^{2}\right)$ with
$\lambda^{\prime}\!\left(K_{c}\right)\neq 0$, such that any nontrivial
small-amplitude branch with $\Lambda_{3}\neq 0$ satisfies
\begin{align}
\left|a\right|
\sim
\sqrt{-\frac{\lambda\left(K\right)}{\Lambda_{3}}}.\label{III-C-EQ004}
\end{align}
Therefore, if $\lambda^{\prime}\!\left(K_{c}\right)>0$, then
$\Lambda_{3}>0$ yields a supercritical continuous branch on the linearly
unstable side, while $\Lambda_{3}<0$ yields a subcritical branch on the
linearly stable side and therefore a discontinuous transition after the branch
is stabilized by higher-order terms. We combine this analytic bifurcation
criterion with the conditional sign statement from Appendix \ref{Appendix-E}.
Thus, within the locality and sign assumptions of Appendices
\ref{Appendix-D} and \ref{Appendix-E}, a non-zero Euler characteristic removes
the supercritical small-amplitude synchronization branch from the one-mode
local normal form. A zero Euler characteristic supplies no such sign
restriction and leaves both continuous and discontinuous local scenarios to be
decided by the analytic coefficients.
The condition $\Lambda_{3}\neq 0$ is also standard. This condition only states
that the cubic term is the first nonlinear term that controls the local branch.
When symmetry leaves a higher-dimensional critical eigenspace, we apply the same local
conclusions branch by branch after selecting a symmetry-broken critical
direction. The present one-mode reduction should be read in that branchwise
sense for such cases.
Subsecs. \ref{Defect-structure-subsection} and
\ref{Scaling-scenarios-subsection}, together with Appendices
\ref{Appendix-F}-\ref{Appendix-I}, record the corresponding defect structure,
the weak-field scaling laws near a continuous branch, and the quintic scales of
the subcritical regime.
We summarize the topological origin of the nonlinear response term in Table \ref{Table-3}.

For the hypersphere $M=S^{D-1}$, we have
$\chi\left(S^{D-1}\right)=1+(-1)^{D-1}$, so $\chi=0$ precisely when $D$ is
even. The topological part of the present framework therefore identifies the
same even/odd division that appears in high-dimensional Kuramoto models. The
actual local branch on either side of this division is still determined by the
conditional normal-form coefficients described above.

\begin{table*}[t]
  \caption{Conditional local phase transition scenarios in the one-mode
  normal form. Topology alone supplies the Poincar\'e-Hopf defect constraint.
  The first row additionally uses the locality and sign conditions of
  Appendices \ref{Appendix-D} and \ref{Appendix-E}; the discontinuous branch
  also requires the stabilizing condition $\Lambda_{5}>0$. For
  $\chi\left(M\right)=0$, the local branch is determined by the normal-form
  coefficients rather than by the Euler characteristic alone.}
  \label{Table-4}
  \begin{ruledtabular}
  \begin{tabular}{@{}llll@{}}
    Euler characteristic &
    Additional coefficient condition &
    Local phase transition scenario &
    Leading local law \\ \hline
    $\chi\left(M\right)\neq 0$ &
    \begin{tabular}[t]{@{}l@{}}
    Appendices \ref{Appendix-D} and \ref{Appendix-E} give \\
    $\Lambda_{3}<0$; $\Lambda_{5}>0$ stabilizes \\
    the subcritical branch.
    \end{tabular} &
    \begin{tabular}[t]{@{}l@{}}
    Conditional discontinuous transition \\
    with hysteresis.
    \end{tabular} &
    \begin{tabular}[t]{@{}l@{}}
    $\Delta K_{\mathrm{hys}}
    \sim
    \Lambda_{3}^{2}/\left(4\lambda^{\prime}\!\left(K_{c}\right)\Lambda_{5}\right)$, \\
    $\left|a_{\mathrm{jump}}\right|
    =
    \sqrt{-\Lambda_{3}/\Lambda_{5}}$
    \end{tabular} \\[0.2em] \hline
    $\chi\left(M\right)=0$ &
    $\Lambda_{3}>0$ &
    \begin{tabular}[t]{@{}l@{}}
    Continuous transition selected by the \\
    analytic normal-form coefficient, not \\
    by topology.
    \end{tabular} &
    \begin{tabular}[t]{@{}l@{}}
    $\beta=\frac{1}{2}$, $\gamma=1$, \\
    $\delta=3$
    \end{tabular} \\[0.2em] \hline
    $\chi\left(M\right)=0$ &
    $\Lambda_{3}<0<\Lambda_{5}$ &
    \begin{tabular}[t]{@{}l@{}}
    Discontinuous transition selected by \\
    the analytic normal-form coefficients, \\
    not by topology.
    \end{tabular} &
    \begin{tabular}[t]{@{}l@{}}
    $\Delta K_{\mathrm{hys}}
    \sim
    \Lambda_{3}^{2}/\left(4\lambda^{\prime}\!\left(K_{c}\right)\Lambda_{5}\right)$, \\
    $\left|a_{\mathrm{jump}}\right|
    =
    \sqrt{-\Lambda_{3}/\Lambda_{5}}$
    \end{tabular} \\[0.2em] \hline
    $\chi\left(M\right)=0$ &
    $\Lambda_{3}=0<\Lambda_{5}$ &
    \begin{tabular}[t]{@{}l@{}}
    Tricritical continuous transition after \\
    analytic tuning by the normal-form \\
    coefficients.
    \end{tabular} &
    \begin{tabular}[t]{@{}l@{}}
    $\left|a_{\ast}\right|
    \sim
    \left|K-K_{c}\right|^{1/4}$, \\
    $\left|a\right|
    \sim
    \left|h\right|^{1/5}$
    \end{tabular}
  \end{tabular}
  \end{ruledtabular}
\end{table*}

\subsection{Critical defect structure of the critical mode}\label{Defect-structure-subsection}

We now give the spatial interpretation of the unconditional topological
constraint. The induced tangent field
$u_{c}\left(\sigma\right)=\mathsf{P}_{\sigma}^{\perp}
\left(r_{c}-\alpha A r_{c}\right)$ gives the local direction in which the
synchronized component first develops on $M$. In the language of statistical
physics, we regard $u_{c}$ as the ordered texture selected at the threshold
\cite{SarkarGupte2021PRE,RouzaireLevis2022FrontPhys,Rouzaire2021PRL}. An
isolated zero of $u_{c}$ is then a topological defect of that texture, and the
zero itself is the defect core. At such a point, the local ordered direction
cannot be extended smoothly, because the critical mode does not select any
tangent direction there.

We assume that all zeros of $u_{c}$ are isolated. Then the Poincar\'e-Hopf theorem
gives
\begin{align}
\sum_{\ell=1}^{N_{\mathrm{def}}}\operatorname{ind}_{p_{\ell}}\left(u_{c}\right)
=
\chi\left(M\right),\label{III-D-EQ001}
\end{align}
where $N_{\mathrm{def}}$ denotes the number of defect cores in the critical
texture. Under the generic nondegeneracy condition stated in Appendix
\ref{Appendix-H}, each isolated zero has index $\pm 1$. Therefore
\begin{align}
N_{\mathrm{def}}
\geq
\left|\chi\left(M\right)\right|.\label{III-D-EQ002}
\end{align}
We identify the integer $\operatorname{ind}_{p_{\ell}}\left(u_{c}\right)$ as the
topological charge carried by the $\ell$th defect core.
Eq. (\ref{III-D-EQ001}) states that the net defect charge of the critical texture is
fixed by the Euler characteristic of $M$.

This conclusion is independent of the sign of $\Lambda_{3}$. A non-zero Euler
characteristic prevents a smooth defect-free critical texture. In the generic
case, it forces the ordered texture that first appears at threshold to contain at least
$\left|\chi\left(M\right)\right|$ simple defect cores. The critical pattern
therefore carries a finite background topological charge from the start.

When $\chi\left(M\right)=0$, topology allows a smooth defect-free critical
texture. In that sense, the Poincar\'e-Hopf constraint no longer imposes a
defect-count constraint. If critical defects are present and all of them are
simple, then their positive and negative topological charges must balance. The
critical texture may still contain local defects, but topology does not impose a
nonzero background charge. Appendix \ref{Appendix-H} records this counting
argument in a precise form.

\subsection{Conditional local phase transition scenarios}\label{Scaling-scenarios-subsection}

We finally use the local reductions of Appendices \ref{Appendix-F} and
\ref{Appendix-G} to classify the local statistical-physics scenario determined
by the cubic and quintic coefficients near the loss of stability of the
incoherent state. The important point is not that topology changes the
numerical values of the generic mean-field exponents in this one-dimensional
local theory. The important point is that, after the locality and sign
conditions have been imposed, the Euler characteristic can restrict which
normal-form branch is available.

If $\chi\left(M\right)\neq 0$ and the locality and sign conditions in
Appendices \ref{Appendix-D} and \ref{Appendix-E} hold, then
$\Lambda_{3}<0$. Within this conditional class, the generic supercritical
continuous branch in Appendix \ref{Appendix-F} is absent from the one-mode
normal form. The tricritical continuous branch discussed in Appendix
\ref{Appendix-G} is also absent within the same class, because it would require
$\Lambda_{3}=0$. Under the additional analytic condition $\Lambda_{5}>0$, the
local transition is discontinuous, and the local hysteresis width and jump
amplitude are then determined by Appendix \ref{Appendix-G}.

If $\chi\left(M\right)=0$, topology imposes no sign constraint on
$\Lambda_{3}$. The continuous branch is obtained only when the analytic
coefficient satisfies $\Lambda_{3}>0$. A discontinuous branch is obtained when
$\Lambda_{3}<0<\Lambda_{5}$. A tricritical continuous branch requires the
additional tuning $\Lambda_{3}=0$ with $\Lambda_{5}>0$. Thus the zero Euler
characteristic condition is nonselective in the local normal form: it removes the
Poincar\'e-Hopf defect constraint, but the realized local scenario is still
chosen by the analytic coefficients of the reduced equation. The local
exponents are fixed by the reduced normal form itself.

We summarize these conditional local phase transition scenarios in Table
\ref{Table-4}.

\begin{figure*}[!t]
\includegraphics[width=\textwidth]{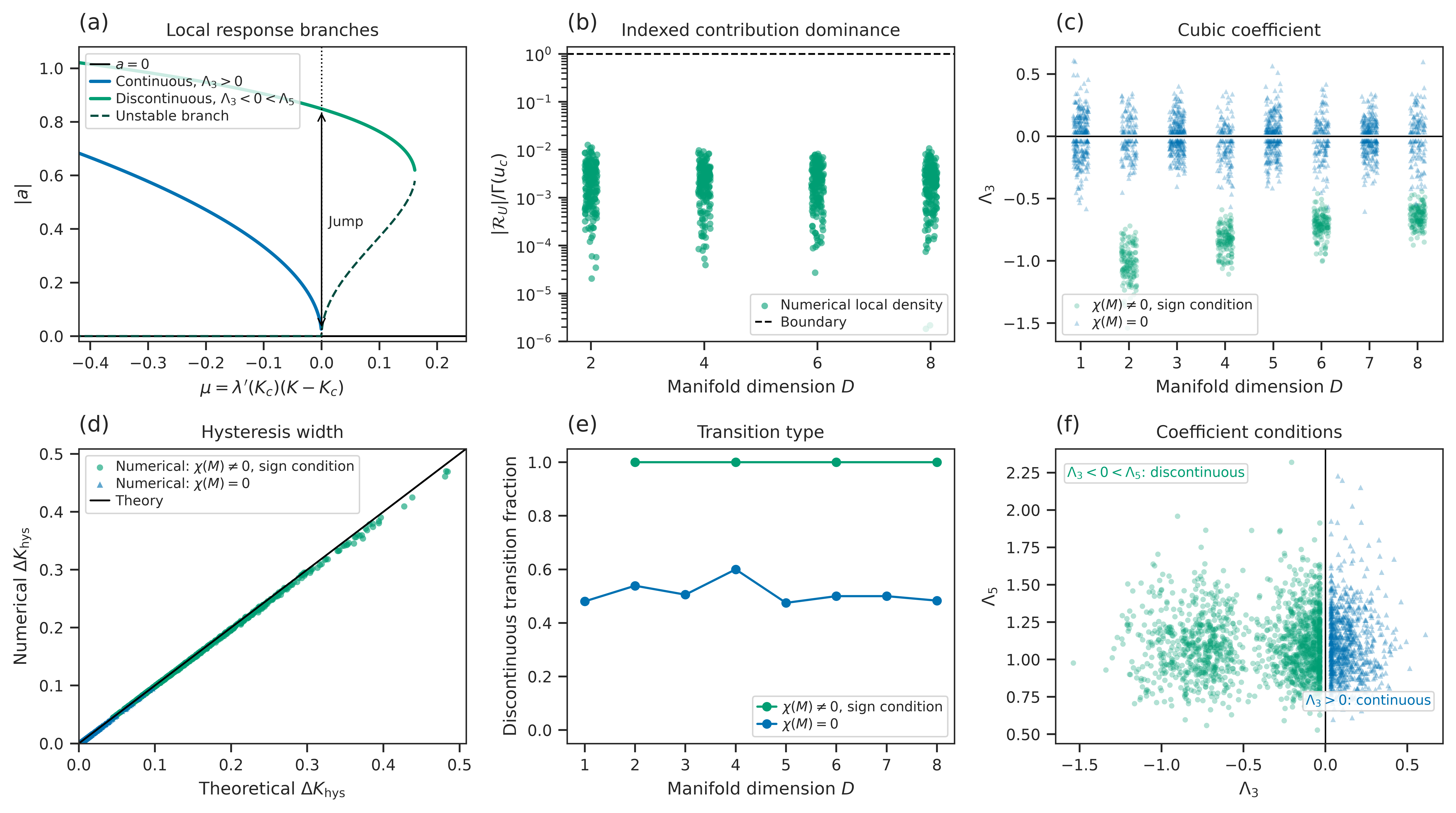}
\caption[Numerical test of the topological contribution to local transition type]{Numerical test of the topological contribution to the local transition
type on \(S^{D}\) and \(\mathbb{T}^{D}\). We write
\(\mu=\lambda^{\prime}\!\left(K_{c}\right)\left(K-K_{c}\right)\), and the
computed coefficients are inserted into Eq. (\ref{TEQ1}). (a) Local response
branches of the reduced equation. When \(\Lambda_{3}>0\), the nonzero branch
satisfies \(a^{2}=-\mu/\Lambda_{3}\), so the continuous branch exists for
\(\mu\le0\) and reaches \(a=0\) at \(\mu=0\). When
\(\Lambda_{3}<0<\Lambda_{5}\), the finite-amplitude branch persists to the
saddle-node point at positive \(\mu\), and its value at \(\mu=0\) gives the
jump amplitude of the discontinuous transition. The dashed curve is the
unstable branch. (b) Numerical evaluation of the sign condition for the
nonzero Euler characteristic sphere cases. Only \(D=2,4,6,8\) appear in this
panel because \(\chi\left(S^{D}\right)=2\) for even \(D\), whereas
\(\chi\left(S^{D}\right)=0\) for odd \(D\) and
\(\chi\left(\mathbb{T}^{D}\right)=0\) for all \(D\). Here
\(\Gamma\left(u_{c}\right)\) is the contribution of the zero neighborhoods
weighted by their local indices and \(\mathcal{R}_{U}\) is the background
contribution not weighted by the indices. The sampled
ratios satisfy
\(\left|\mathcal{R}_{U}\right|/\Gamma\left(u_{c}\right)<1\), which gives
\(\Lambda_{3}=-\Gamma\left(u_{c}\right)+\mathcal{R}_{U}<0\). (c) Cubic
coefficient computed from the same local density. The sign condition in the
nonzero Euler characteristic class gives \(\Lambda_{3}<0\), whereas the zero
Euler characteristic class contains both signs. (d) Hysteresis width of the
local response equation. The horizontal coordinate is Eq. (\ref{GEQ15}), the
vertical coordinate is obtained by numerical continuation of Eq. (\ref{TEQ1}),
and the black line is the theoretical diagonal. (e) Transition type over
\(D=1,\ldots,8\). The class with \(\chi\left(M\right)\neq0\) and the sign
condition gives discontinuous local transitions throughout the tested
dimensions for which that class is present. The class with
\(\chi\left(M\right)=0\) contains both transition types. (f) Classification in
the \(\left(\Lambda_{3},\Lambda_{5}\right)\) plane. The computed coefficients
fall into the continuous condition \(\Lambda_{3}>0\) or the discontinuous
condition \(\Lambda_{3}<0<\Lambda_{5}\).}
\label{Fig-topological-transition}
\end{figure*}

We next test the local consequence of the topological classification for the
phase transition. The numerical experiment is performed on the reduced
response equation derived above, with coefficients computed from explicit
critical tangent fields and local cubic densities. The oscillator state spaces
are \(S^{D}\) and \(\mathbb{T}^{D}\), with \(D=1,\ldots,8\). For \(S^{D}\), we
use the projected constant field
\(u_{c}\left(x\right)=e_{D+1}-\left(e_{D+1}\cdot x\right)x\). For
\(\mathbb{T}^{D}\), we use
\(u_{c}\left(\theta\right)=\left(\sin\theta_{1},\ldots,\sin\theta_{D}\right)\).
We compute the local defect indices from finite difference Jacobians at the
zeros of these fields. We then construct local cubic densities, compute
\(\Lambda_{3}\) from the contribution of the zero neighborhoods weighted by
their local indices and from the background contribution not weighted by the
indices, and numerically continue
\begin{align}
\Psi\left(a;K\right)
&=
\lambda^{\prime}\left(K_{c}\right)\left(K-K_{c}\right)a
+\Lambda_{3}a^{3}
+\Lambda_{5}a^{5}
\notag\\
&\quad
+\Lambda_{7}a^{7}
+\rho\,\lambda^{\prime}\left(K_{c}\right)\left(K-K_{c}\right)a^{3}.
\label{TEQ1}
\end{align}
In the simulations shown in Figure \ref{Fig-topological-transition}, we set
\(\lambda^{\prime}\left(K_{c}\right)=1\). The transition type is assigned after
the coefficients are computed and the response equation is continued.

Figure \ref{Fig-topological-transition} tests the phase transition consequence
of the indexed local-density calculation. Figure
\ref{Fig-topological-transition}(a) shows how the computed sign of
\(\Lambda_{3}\) changes the local branch structure. Figure
\ref{Fig-topological-transition}(b) verifies the numerical condition that
converts the indexed contribution into the sign \(\Lambda_{3}<0\): the
background contribution not weighted by the indices remains smaller than the
indexed contribution in the tested nonzero Euler characteristic samples. This
panel therefore contains only the even-dimensional sphere samples, because
they are the \(S^{D}\) cases with nonzero Euler characteristic in the
\(D=1,\ldots,8\) test set.
Figure \ref{Fig-topological-transition}(c) shows the resulting
\(\Lambda_{3}\) values. With nonzero Euler characteristic and the sign
condition, the computed coefficients are negative; with zero Euler
characteristic, both signs occur. Figure \ref{Fig-topological-transition}(d)
then tests the quantitative prediction for the hysteresis width by comparing
the value from Eq. (\ref{GEQ15}) with numerical continuation of
Eq. (\ref{TEQ1}). Figures \ref{Fig-topological-transition}(e) and
\ref{Fig-topological-transition}(f) show that the same computed coefficients
produce the transition classification over dimensions and in the
\(\left(\Lambda_{3},\Lambda_{5}\right)\) plane.
\section{Representative geometric and topological indicators}\label{Case-study-section}

In this section, we apply the theory presented in Sec. III to representative
homogeneous manifolds. We first
identify the geometric source coefficient $\kappa\left(M\right)$ that is determined by the
averaged tangent projection on the chosen embedding. This coefficient multiplies the coupling-induced source term in the
linearized kinetic equation. Together with the intrinsic-drift response, it
determines the critical coupling for the loss of stability of the incoherent
state. We then
state the Euler characteristic. This gives an unconditional prediction for the
net defect charge of the critical texture through the Poincar\'e-Hopf theorem.
When the locality conditions of Appendix \ref{Appendix-D} and the sign
condition of Appendix \ref{Appendix-E} are also imposed, the same Euler
characteristic gives the conditional cubic-sign information used in the
one-mode normal form. Under those additional conditions, a non-zero Euler
characteristic removes the generic continuous branch from the local normal
form and can lead to a discontinuous transition when $\Lambda_{5}>0$, whereas
for a zero Euler characteristic the local branch is determined by the
normal-form coefficients. In
Table \ref{Table-5}, the first four cases have nonzero Euler characteristic and
the last four have zero Euler characteristic. Appendix \ref{Appendix-J} collects
the explicit geometric and topological computations.

When the linearized problem is symmetry-degenerate, we interpret the one-mode
reduction branch by branch. In other words, we first select a critical
direction after symmetry breaking and then apply the local theory of
Subsecs. \ref{Nonlinear-response-subsection}-\ref{Scaling-scenarios-subsection}
to that branch.

\begin{table*}[t]
  \caption{Representative geometric and topological indicators. The first row is the hypersphere case;
  Rows 2--4 have $\chi\left(M\right)\neq 0$; Rows 5--8 have
  $\chi\left(M\right)=0$. In Column 4, $Q_{\mathrm{def}}$ and
  $N_{\mathrm{def}}^{\min}$ denote the net defect charge and the minimum number
  of simple defect cores. Column 5 is stated within the one-mode reduction,
  under the locality conditions of Appendix \ref{Appendix-D} and the sign
  condition of Appendix \ref{Appendix-E}; a discontinuous branch is marked as
  selected only when the additional stabilizing condition $\Lambda_{5}>0$
  holds. The status of these locality and sign assumptions for the listed
  families is summarized in Table \ref{Table-6}.}
  \label{Table-5}
  \begin{ruledtabular}
  \scriptsize
  \renewcommand{\arraystretch}{1.5}
  \begin{tabular}{@{}lllll@{}}
    Case &
    $\kappa\left(M\right)$ &
    $\chi\left(M\right)$ &
    Defect content &
    Conditional local scenario \\ \hline
    $S^{D-1}$ &
    $\left(D-1\right)/D$ &
    $1+(-1)^{D-1}$ &
    \begin{tabular}[t]{@{}l@{}}
    Odd $D$: $Q_{\mathrm{def}}=2$, $N_{\mathrm{def}}^{\min}=2$. \\
    Even $D$: $Q_{\mathrm{def}}=0$.
    \end{tabular} &
    \begin{tabular}[t]{@{}l@{}}
    Odd $D$: under Appendices \ref{Appendix-D}--\ref{Appendix-E}, \\
    continuous and tricritical branches \\
    are absent; a discontinuous branch occurs \\
    if $\Lambda_{5}>0$. \\
    Even $D$: the Euler characteristic \\
    gives no branch selection.
    \end{tabular} \\ \hline
    $S^{2m}\times S^{2m}$ &
    $\frac{2m}{2m+1}$ &
    $4$ &
    \begin{tabular}[t]{@{}l@{}}
    $Q_{\mathrm{def}}=4$, \\
    $N_{\mathrm{def}}^{\min}=4$.
    \end{tabular} &
    \begin{tabular}[t]{@{}l@{}}
    Under Appendices \ref{Appendix-D}--\ref{Appendix-E}, \\
    continuous and tricritical branches \\
    are absent. A discontinuous branch occurs \\
    if $\Lambda_{5}>0$.
    \end{tabular} \\ \hline
    \begin{tabular}[t]{@{}l@{}}
    $\mathrm{Gr}_{k}\left(\mathbb{C}^{n}\right)$, \\
    $2\leq k\leq n-2$
    \end{tabular} &
    $\frac{2k\left(n-k\right)}{n^{2}-1}$ &
    $\binom{n}{k}$ &
    \begin{tabular}[t]{@{}l@{}}
    $Q_{\mathrm{def}}=\binom{n}{k}$, \\
    $N_{\mathrm{def}}^{\min}=\binom{n}{k}$.
    \end{tabular} &
    \begin{tabular}[t]{@{}l@{}}
    Under Appendices \ref{Appendix-D}--\ref{Appendix-E}, \\
    continuous and tricritical branches \\
    are absent. A discontinuous branch occurs \\
    if $\Lambda_{5}>0$.
    \end{tabular} \\ \hline
    \begin{tabular}[t]{@{}l@{}}
    $\mathbb{CP}^{m}$, \\
    $m\geq 2$
    \end{tabular} &
    $\frac{2}{m+2}$ &
    $m+1$ &
    \begin{tabular}[t]{@{}l@{}}
    $Q_{\mathrm{def}}=m+1$, \\
    $N_{\mathrm{def}}^{\min}=m+1$.
    \end{tabular} &
    \begin{tabular}[t]{@{}l@{}}
    Under Appendices \ref{Appendix-D}--\ref{Appendix-E}, \\
    continuous and tricritical branches \\
    are absent. A discontinuous branch occurs \\
    if $\Lambda_{5}>0$.
    \end{tabular} \\ \hline
    $\mathbb{T}^{d}$ &
    $1/2$ &
    $0$ &
    \begin{tabular}[t]{@{}l@{}}
    $Q_{\mathrm{def}}=0$. \\
    Defect-free texture allowed.
    \end{tabular} &
    \begin{tabular}[t]{@{}l@{}}
    The Euler characteristic gives no \\
    branch selection. The normal-form \\
    coefficients decide \\
    the local scenario.
    \end{tabular} \\ \hline
    $\mathrm{St}\left(p,n\right),\ 2\leq p<n$ &
    $\left(2n-p-1\right)/\left(2n\right)$ &
    $0$ &
    \begin{tabular}[t]{@{}l@{}}
    $Q_{\mathrm{def}}=0$. \\
    Defect-free texture allowed.
    \end{tabular} &
    \begin{tabular}[t]{@{}l@{}}
    The Euler characteristic gives no \\
    branch selection. The normal-form \\
    coefficients decide \\
    the local scenario.
    \end{tabular} \\ \hline
    $\mathrm{SO}\left(n\right)$ &
    $\left(n-1\right)/\left(2n\right)$ &
    $0$ &
    \begin{tabular}[t]{@{}l@{}}
    $Q_{\mathrm{def}}=0$. \\
    Defect-free texture allowed.
    \end{tabular} &
    \begin{tabular}[t]{@{}l@{}}
    The Euler characteristic gives no \\
    branch selection. The normal-form \\
    coefficients decide \\
    the local scenario.
    \end{tabular} \\ \hline
    $\mathrm{U}\left(d\right)$ &
    $1/2$ &
    $0$ &
    \begin{tabular}[t]{@{}l@{}}
    $Q_{\mathrm{def}}=0$. \\
    Defect-free texture allowed.
    \end{tabular} &
    \begin{tabular}[t]{@{}l@{}}
    The Euler characteristic gives no \\
    branch selection. The normal-form \\
    coefficients decide \\
    the local scenario.
    \end{tabular}
  \end{tabular}
  \end{ruledtabular}
\end{table*}

The entries in Table \ref{Table-5} should be read as geometric and topological
indicator data. The coefficient $\kappa\left(M\right)$ specifies how strongly
the geometry of $M$ enters the coupling-induced source term of the linearized
kinetic equation, and it therefore directly affects the coupling strength
required for the incoherent state to lose stability once the intrinsic-drift
response has been fixed. The Euler characteristic gives the net topological
charge that any critical texture must carry. The last column is conditional: it
uses the local one-mode reduction only under the hypotheses stated in
Appendices \ref{Appendix-D} and \ref{Appendix-E}.

We give the explicit averaged-projector, Euler-characteristic, and
defect-counting calculations in Appendix \ref{Appendix-J}. Appendix
\ref{Appendix-K} provides a concise family-by-family interpretation of the
entries in Table \ref{Table-5}.

\begin{figure*}[!t]
\includegraphics[width=\textwidth]{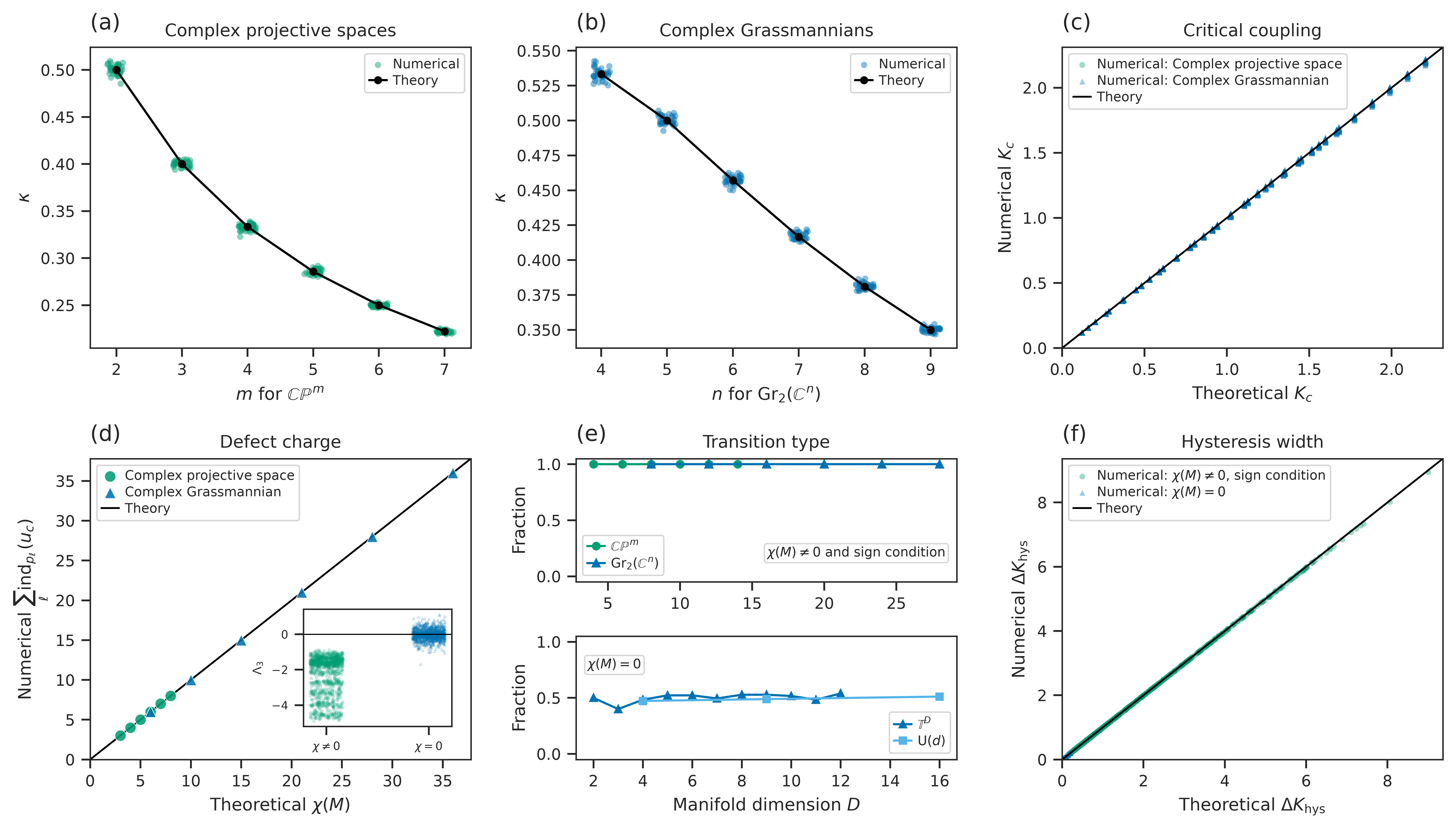}
\caption[Numerical validation on non-spherical homogeneous manifolds]{Numerical validation on non-spherical homogeneous manifolds. The
tested manifolds are complex projective spaces \(\mathbb{CP}^{m}\) and
complex Grassmannians \(\mathrm{Gr}_{2}\left(\mathbb{C}^{n}\right)\),
represented by Hermitian projectors. (a) Monte Carlo estimate of the averaged
tangent-projection factor on \(\mathbb{CP}^{m}\). The numerical points follow
the black curve
\(\kappa\left(\mathbb{CP}^{m}\right)=2/\left(m+2\right)\). (b) Monte Carlo
estimate of the averaged tangent-projection factor on
\(\mathrm{Gr}_{2}\left(\mathbb{C}^{n}\right)\). The numerical points follow
the black curve
\(\kappa\left(\mathrm{Gr}_{2}\left(\mathbb{C}^{n}\right)\right)
=4\left(n-2\right)/\left(n^{2}-1\right)\). (c) Critical coupling computed
with a Lorentzian intrinsic-drift distribution. The numerical value uses the
sampled \(\kappa\left(M\right)\), while the theoretical value uses the
corresponding expression in Table \ref{Table-5}. (d) Defect charge for the
diagonal height-gradient field on the projector manifolds. The numerical
index sum equals \(\chi\left(M\right)\), and the inset shows the cubic
coefficient computed from the corresponding local density. For nonzero Euler
characteristic together with the sign condition, the computed coefficients
satisfy \(\Lambda_{3}<0\); for zero Euler characteristic, the topology does
not impose a fixed sign. (e) Fraction of computed local response equations
that give a discontinuous transition as the manifold dimension is varied. In
the upper part, \(\mathbb{CP}^{m}\) has real dimension \(D=2m\), and
\(\mathrm{Gr}_{2}\left(\mathbb{C}^{n}\right)\) has real dimension
\(D=4\left(n-2\right)\). These families with nonzero Euler characteristic and the sign condition
select the discontinuous branch throughout the tested range. In the lower
part, \(\mathbb{T}^{D}\) has dimension \(D\), and \(\mathrm{U}\left(d\right)\)
has real dimension \(D=d^{2}\). These reference families with zero Euler characteristic do not
have a topologically fixed transition type. (f) Hysteresis width obtained by
numerical continuation of the perturbed quintic response equation. The
horizontal coordinate is Eq. (\ref{GEQ15}), the vertical coordinate is the
numerical continuation result, and the black line is the theoretical
diagonal.}
\label{Fig-nonspherical-validation}
\end{figure*}

Figure \ref{Fig-nonspherical-validation} applies the same numerical procedure
to non-spherical representative manifolds. For the geometric calculation, we
represent \(\mathbb{CP}^{m}\) and
\(\mathrm{Gr}_{2}\left(\mathbb{C}^{n}\right)\) as Hermitian projector
manifolds. We sample Haar-distributed projectors and isotropic traceless
Hermitian ambient directions, project those directions onto the tangent space,
and average the squared projection ratio to estimate \(\kappa\left(M\right)\).
We then insert the sampled \(\kappa\left(M\right)\) into the Lorentzian
intrinsic-drift response to compute \(K_{c}\). For the topological calculation,
we use the diagonal height-gradient field on the same projector manifolds. Its
zeros are the coordinate projectors. We compute their local indices from the
linearized tangent field in local charts, use the same local-density construction as in
Figure \ref{Fig-topological-transition} to obtain \(\Lambda_{3}\), and continue
the response equation.

Figure \ref{Fig-nonspherical-validation}(a) and Figure
\ref{Fig-nonspherical-validation}(b) compare the sampled averaged tangent
projection with the analytic values of \(\kappa\left(M\right)\) for two
non-spherical families. Figure \ref{Fig-nonspherical-validation}(c) shows how
this geometric coefficient determines the critical coupling after the
Lorentzian intrinsic-drift response is fixed. Figure
\ref{Fig-nonspherical-validation}(d) uses the diagonal height-gradient field to
compute the index data on the same projector manifolds and then evaluates the
corresponding local cubic coefficient. Figure
\ref{Fig-nonspherical-validation}(e) displays the phase transition type
obtained from these computed coefficients. The families with nonzero Euler
characteristic and the sign condition remain on the discontinuous branch,
whereas the reference families with zero Euler characteristic show both local
transition types because the Euler characteristic does not fix the sign of
\(\Lambda_{3}\). Figure \ref{Fig-nonspherical-validation}(f) compares the
hysteresis width from numerical continuation with Eq. (\ref{GEQ15}) for the
samples with \(\Lambda_{3}<0\) and \(\Lambda_{5}>0\).

\begin{figure*}[!t]
\includegraphics[width=\textwidth]{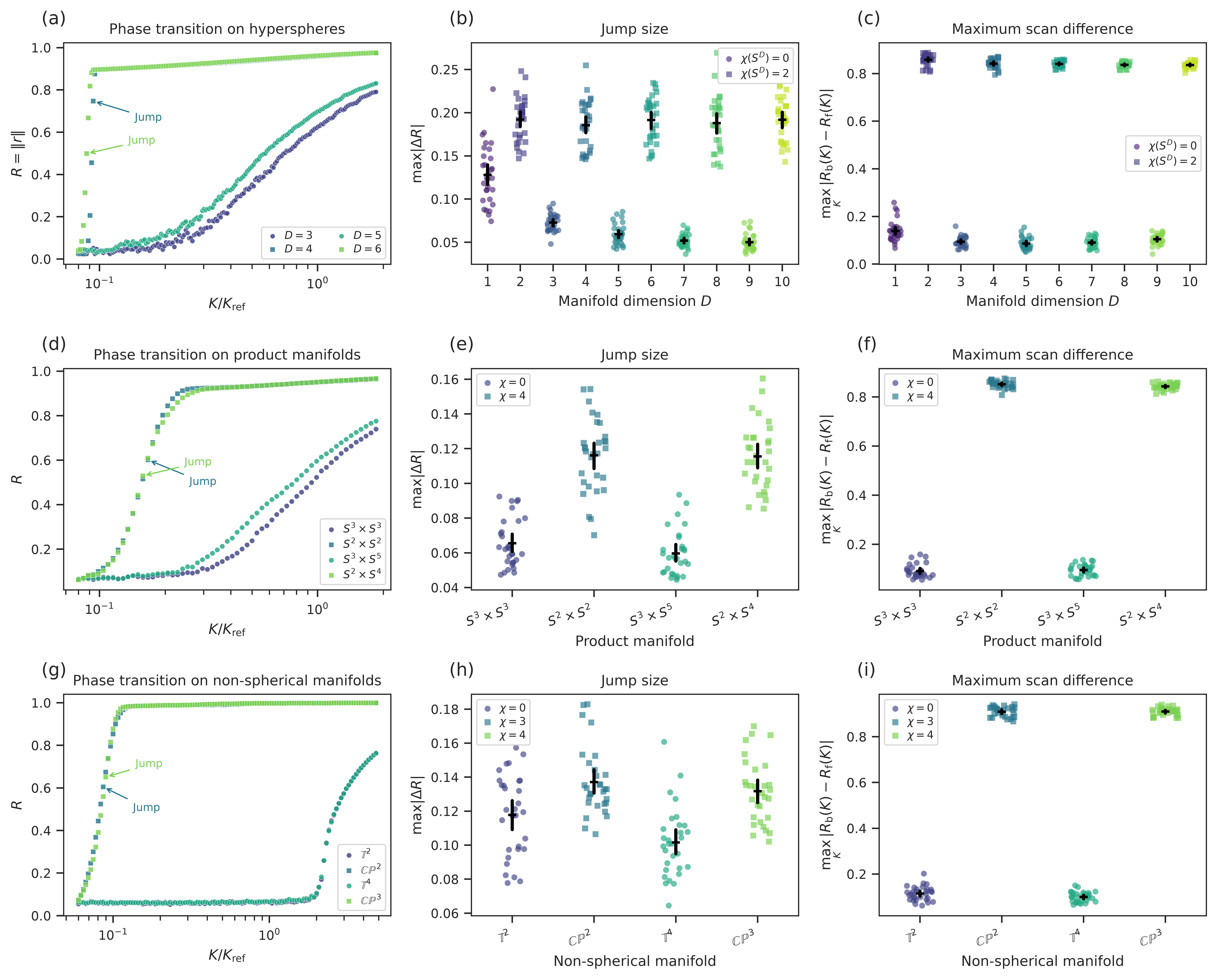}
\caption[Direct finite-N dynamical tests of the phase transition scenarios]{Direct finite-\(N\) dynamical tests of the phase transition scenarios.
The figure uses three groups of manifolds: hyperspheres \(S^{D}\), product
manifolds \(S^{p}\times S^{q}\), and non-spherical representatives
\(\mathbb{T}^{D}\) and \(\mathbb{CP}^{m}\). In all panels, \(R\) is the
computed order parameter and \(K_{\mathrm{ref}}=\Delta\), where \(\Delta\) is
the width parameter of the intrinsic-drift distribution. The zero Euler
characteristic cases use ordinary drift ensembles, whereas the nonzero Euler
characteristic cases use structurally prescribed sign-conditioned drift
ensembles. All generated samples are retained. (a) Forward \(K\)-scan for
hyperspheres. The odd-dimensional cases have \(\chi\left(S^{D}\right)=0\) and
show a continuous increase of \(R\), whereas the even-dimensional cases have
\(\chi\left(S^{D}\right)=2\) and show a finite jump. (b) Jump size
\(\max\left|\Delta R\right|\) over \(D=1,\ldots,10\). The square markers denote
the even-dimensional nonzero Euler characteristic cases. (c) Maximum
difference between the backward and forward scans,
\(\max_{K}\left|R_{\mathrm{b}}\left(K\right)-R_{\mathrm{f}}\left(K\right)\right|\),
for the same hypersphere data. (d) Forward \(K\)-scan for product manifolds.
The zero Euler characteristic controls \(S^{3}\times S^{3}\) and
\(S^{3}\times S^{5}\) increase continuously, whereas the sign-conditioned
nonzero Euler characteristic cases \(S^{2}\times S^{2}\) and
\(S^{2}\times S^{4}\) show jumps. (e) Jump size for the product manifolds.
(f) Maximum scan-direction difference for the product manifolds. (g) Forward
\(K\)-scan for the non-spherical representatives \(\mathbb{T}^{2}\),
\(\mathbb{T}^{4}\), \(\mathbb{CP}^{2}\), and \(\mathbb{CP}^{3}\). The torus
cases have \(\chi=0\), whereas \(\mathbb{CP}^{2}\) and \(\mathbb{CP}^{3}\)
have nonzero Euler characteristic and use the sign-conditioned Hermitian drift
ensemble. (h) Jump size for these non-spherical representatives. (i) Maximum
scan-direction difference for the same data. In the statistical panels, the
colored markers are independent finite-\(N\) samples and the black markers show
the sample mean with a 95\% confidence interval.}
\label{Fig-direct-dynamics}
\end{figure*}

Finally, Figure \ref{Fig-direct-dynamics} checks the phase transition scenarios
by direct integration of finite oscillator systems. Figures
\ref{Fig-direct-dynamics}(a)--\ref{Fig-direct-dynamics}(c) show the
hypersphere results. Figure \ref{Fig-direct-dynamics}(a) shows that the
odd-dimensional zero Euler characteristic cases increase continuously in the
forward scan, while the even-dimensional nonzero Euler characteristic cases
show a finite jump. Figures \ref{Fig-direct-dynamics}(b) and
\ref{Fig-direct-dynamics}(c) quantify the same distinction through the jump
size and the maximum difference between backward and forward scans.

Figure \ref{Fig-direct-dynamics}(d) extends the same finite-\(N\) test to
product manifolds. The zero Euler characteristic controls
\(S^{3}\times S^{3}\) and \(S^{3}\times S^{5}\) show continuous growth of the
order parameter, whereas the sign-conditioned nonzero Euler characteristic
cases \(S^{2}\times S^{2}\) and \(S^{2}\times S^{4}\) show finite jumps and
large scan-direction differences, as quantified in Figures
\ref{Fig-direct-dynamics}(e) and \ref{Fig-direct-dynamics}(f). Figures
\ref{Fig-direct-dynamics}(g)--\ref{Fig-direct-dynamics}(i) repeat the same
test on non-spherical representatives from Table \ref{Table-5}. The torus
cases \(\mathbb{T}^{2}\) and \(\mathbb{T}^{4}\) provide zero Euler
characteristic controls, while \(\mathbb{CP}^{2}\) and \(\mathbb{CP}^{3}\)
provide nonzero Euler characteristic tests under a sign-conditioned Hermitian
drift ensemble. The finite-\(N\) data therefore give a dynamical verification
of the conditional topological prediction across hyperspheres, product
manifolds, and non-spherical homogeneous manifolds.

\section{Conclusion}\label{Conclusion-section}

In this work, we extended the Kuramoto-Sakaguchi framework from spheres to
compact, connected, orientable, and homogeneous manifolds of arbitrary
dimension. We derived the continuum equation on the manifold, identified the
linear and nonlinear terms in the order parameter response equation near the
incoherent state, and connected those terms with the geometry and topology of
the state manifold. For the chosen embedding, the
averaged tangent projection determines the coefficient
$\kappa\left(M\right)$. This
coefficient multiplies the coupling-induced source term in the linearized
kinetic equation. The critical coupling $K_{c}$ is determined by this
geometric coefficient together with the intrinsic-drift response. The manifold
topology affects the nonlinear response through the Euler characteristic. The unconditional consequence is the
Poincar\'e-Hopf defect constraint on the critical tangent field. The
phase transition consequence is conditional: within the one-mode local
reduction of Appendices \ref{Appendix-D} and \ref{Appendix-E}, a nonzero Euler
characteristic gives the cubic sign used in the reduced normal form, and the
additional stabilization condition $\Lambda_{5}>0$ gives a discontinuous local
transition. When $\chi\left(M\right)=0$, the Euler characteristic gives no
sign selection for the cubic coefficient, and the local branch is determined
by the normal-form coefficients.

We then evaluated these indicators on representative homogeneous manifolds.
The hypersphere case recovers the topological part of the classical parity
distinction and shows how the odd and even cases fall on the two
Euler-characteristic sides of the theory. The family
$S^{2m}\times S^{2m}$ shows that the same nonzero Euler characteristic defect constraint
extends beyond a single sphere while the geometric projection factor varies.
The complex Grassmannian and complex projective families show that the
nonzero Euler characteristic defect constraint also applies to non-spherical complex
homogeneous spaces. The torus, real Stiefel, rotation-group, and unitary cases
give the complementary zero Euler characteristic class: the Euler characteristic imposes no
nonzero net defect charge, and the local branch is determined by the
normal-form coefficients. These representative manifolds illustrate how the
averaged tangent projection and the Euler characteristic enter the analysis:
geometry fixes the geometric contribution to the coupling strength carried by
the linear source term, while topology supplies an unconditional defect
constraint and a conditional contribution to local branch selection. The numerical data and the experiment code used to generate our results are available in an open repository \cite{DoloMingGithub2026}.

Several concrete statistical-physics directions now become accessible within
the present framework:
\begin{enumerate}
\item Finite-size scaling near the loss of stability of the incoherent state. One can study how
the finite-$N$ order parameter fluctuations, Binder-like cumulants \cite{Binder1981ZPhysB}, and
pseudocritical couplings depend on whether $\chi\left(M\right)$ vanishes. The
most direct comparison is between odd hyperspheres or $\mathbb{CP}^{m}$, where
the conditional normal form predicts a subcritical local branch under
Appendices \ref{Appendix-D} and \ref{Appendix-E}, and zero Euler characteristic cases such as
$\mathbb{T}^{d}$ or $\mathrm{U}\left(d\right)$, where the branch is determined
by the normal-form coefficients
\cite{Buendia2025PRL}.
\item Noise-induced switching and metastability on manifolds with nonzero Euler characteristic. By
adding intrinsic noise to the kinetic equation, one can derive the stationary
order parameter distribution near the subcritical branch and compute switching
times between incoherent and synchronized states. The present quintic normal
form gives a direct starting point for studying how a conditionally selected
discontinuous transition affects activation barriers and residence times
\cite{Buendia2025PRL}.
\item Defect kinetics near the loss of stability of the incoherent state. The theory predicts a net
defect charge and a lower bound on the number of simple defect cores when
$\chi\left(M\right)\neq 0$. A natural next step is to simulate the critical
texture on manifolds such as $S^{2m}\times S^{2m}$, $\mathbb{CP}^{m}$, and
$\mathrm{Gr}_{k}\left(\mathbb{C}^{n}\right)$, and then measure defect
creation, motion, annihilation, and coarsening as $K$ crosses the threshold
\cite{SarkarGupte2021PRE,RouzaireLevis2022FrontPhys,Rouzaire2021PRL}.
\item Quenched disorder in the intrinsic drifts. The present theory already
contains the drift field through the measure $h\left(V\right)$. One can study
how the width and symmetry of that disorder modify the critical coupling, the
cubic coefficient, and the accessibility of the continuous branch. This is
especially relevant for matrix-valued models on $\mathrm{St}\left(p,n\right)$,
$\mathrm{SO}\left(n\right)$, and $\mathrm{U}\left(d\right)$
\cite{Petkoski2012PRE,Barabash2014PRE,Kundu2017PRE}.
\item Topology versus interaction-network structure. The present analysis is
mean-field in character. A concrete next problem is to place the same manifold
state spaces on sparse random graphs, modular graphs, or low-dimensional
lattices, and ask whether the conditional local scenarios survive when
the coupling network itself is no longer complete
\cite{Battiston2020PhysRep,SkardalArenas2020CommunPhys,Millan2022CommunPhys,Nurisso2024Chaos,Gambuzza2021NatCommun,Gallo2022CommunPhys,Neuhauser2020PRE,Millan2025NatPhys,Berner2023PhysRep}.
\item Nonequilibrium hysteresis under ramps and external forcing. Appendix
\ref{Appendix-F} and Appendix \ref{Appendix-G} give local static scaling and
hysteresis scales. One can now study slow and fast parameter ramps, periodic
driving, and weak external fields to determine how the loop area, jump point,
and lag depend on the manifold topology. This direction is especially natural
for the cases with nonzero Euler characteristic, where the conditional normal form can select the
discontinuous side of the local bifurcation structure \cite{Leyva2018DCDSB,Dai2020Chaos,Bayani2023Chaos,Lu2016Chaos}.
\end{enumerate}

 \section*{Acknowledgements}
 Author appreciates Dr. Hedong Hou at the Institut de Math{\'e}matiques d'Orsay for their inspiring discussions. This project is supported by the High Performance Scientific Computing Research Program at Infplane Computing Technologies Ltd.

\appendix
\section{Gradient flow structure for $\alpha=0$}\label{Appendix-A}

In this appendix, we show that the coupling term in Eq. (\ref{II-B-EQ001}) admits a
gradient-flow structure when the phase-lag parameter $\alpha$ vanishes. 

When $\alpha=0$, the coupling in Eq. (\ref{II-B-EQ003}) simplifies to
\begin{align}
\Gamma\left(\sigma_{i},\sigma_{j};0\right)
    =\mathsf{P}^{\perp}_{\sigma_{i}}F\left(\sigma_{j}\right), \label{AEQ1}
\end{align}
such that the discrete dynamics in Eq. (\ref{II-B-EQ001}) becomes
\begin{align}
    \frac{\mathsf{d}}{\mathsf{d}t}\sigma_{i}
    = V_{i}\left(\sigma_{i}\right)
    + \frac{K}{N}\sum_{j=1}^{N}
\mathsf{P}^{\perp}_{\sigma_{i}}F\left(\sigma_{j}\right).
    \label{AEQ2}
\end{align}
Then, we construct a potential whose Riemannian gradient reproduces the
coupling term in Eq. (\ref{AEQ2}). We consider the
mean-field potential of the oscillator system
\begin{align}
    U(\sigma_1,\dots,\sigma_N)
    = -\frac{K}{2N}\sum_{i,j=1}^{N}
      \big\langle F\left(\sigma_{i}\right),F\left(\sigma_{j}\right)\big\rangle.
    \label{AEQ3}
\end{align}
Under the standing assumption that
the Riemannian metric on $M$ is induced by the embedding
$F:M\hookrightarrow\mathbb{R}^D$, the Riemannian gradient of
$U$ with respect to $\sigma_{i}$ is given by
\begin{align}
    \nabla_{\sigma_i}U
    = -\frac{K}{N}\sum_{j=1}^{N}
      \mathsf{P}^{\perp}_{\sigma_{i}}F\left(\sigma_{j}\right)
    = -\frac{K}{N}\sum_{j=1}^{N}
      \Gamma\left(\sigma_{i},\sigma_{j};0\right).\label{AEQ4}
\end{align}
Hence, the dynamics can be written compactly as
\begin{align}
    \frac{\mathsf{d}}{\mathsf{d}t}\sigma_{i}
    = V_{i}\left(\sigma_{i}\right)
      - \nabla_{\sigma_i}U,
    \;\;\;i=1,\dots,N
    \label{AEQ5}
\end{align}
when $\alpha=0$. As described by this result, the coupling part of the manifold Kuramoto model is a
gradient flow of the potential in Eq. (\ref{AEQ3}),
perturbed by the intrinsic drift fields $V_i$. This generalizes the
well-known gradient structure of the classic Kuramoto model on the
circle and of its high-dimensional extensions on spheres or other related
homogeneous manifolds \cite{Rodrigues2016PhysRep,Pietras2019PhysRep,Chandra2019PRX,Lipton2021Chaos,Markdahl2021CommunPhys,Zou2023PRL}.

\section{Special cases of the general model}
\label{Appendix-B}

In this appendix, we briefly explain how several classic synchronization models arise as the special cases of the general
framework defined in Eqs. (\ref{II-B-EQ001}, \ref{II-B-EQ003}). In all the cases, the state of
oscillator $i$ is denoted by $\sigma_{i}\in M$, the embedding is a smooth map $F:M\hookrightarrow\mathbb{R}^{D_{a}}$, and the coupling
is defined by $\Gamma\left(\sigma_{i},\sigma_{j};\alpha\right)
    =
    \mathsf{P}^{\perp}_{\sigma_{i}}
    \left[\left(\mathbf{I}-\alpha A\right)F\left(\sigma_{j}\right)\right]$ with a fixed skew-symmetric matrix
$A\in\mathfrak{so}\left(D_{a}\right)$ and a scalar phase-lag parameter
$\alpha$.

\paragraph{Kuramoto model on $S^{1}$.}
We take the state manifold
$M = S^{1} \cong \mathbb{R}/2\pi\mathbb{Z}$. 
Each oscillator state $\sigma_{i}\in S^{1}$ is actually an angular
coordinate $\sigma_{i}\in\mathbb{R}/2\pi\mathbb{Z}$.
With this identification, we write the embedding as $F\left(\sigma_{i}\right)
 = \left(\cos\sigma_{i},\sin\sigma_{i}\right)$, while the phase-lag is turned off by setting $A=\mathbf{0}$ and $\alpha=0$. The
tangent space at $\sigma_{i}$ is one-dimensional and spanned by
$JF\left(\sigma_{i}\right)
 = \left(-\sin\sigma_{i},\cos\sigma_{i}\right)$, where $J$ is the
$90^\circ$ counterclockwise rotation matrix $J
    =
    \left(\begin{smallmatrix}
        0 & -1 \\
        1 & \phantom{-}0
    \end{smallmatrix}\right)$. With these
definitions, the projection $\mathsf{P}^{\perp}_{\sigma_{i}}$
extracts the component of $F\left(\sigma_{j}\right)$ along
$JF\left(\sigma_{i}\right)$, such that Eq. (\ref{II-B-EQ001}) reduces to
the classic Kuramoto model with coupling
$\sin\left(\sigma_{j}-\sigma_{i}\right)$ \cite{Rodrigues2016PhysRep,Pietras2019PhysRep}.

\paragraph{Kuramoto-Sakaguchi model on $S^{1}$.}
We keep the same manifold $M=S^{1}$ and embedding
$F\left(\sigma\right)=\left(\cos\sigma,\sin\sigma\right)$. Additionally, we
introduce the phase frustration by choosing
$A=J$
and a small non-zero $\alpha$. The matrix $J$ generates rotations
in $\mathbb{R}^{2}$, and the factor
$\left(\mathbf{I}-\alpha A\right)F\left(\sigma_{j}\right)$ may be viewed as
a first-order approximation to the finite phase-lag action
$\exp\left(-\alpha A\right)F\left(\sigma_{j}\right)$. Projecting onto the tangent direction at $\sigma_{i}$ again yields a scalar coupling along $JF\left(\sigma_{i}\right)$, which has a form of $\sin\left(\sigma_{j}-\sigma_{i}\right)
 - \alpha\cos\left(\sigma_{j}-\sigma_{i}\right)$ (i.e., the
first-order expansion of the Kuramoto-Sakaguchi interaction
$\sin\left(\sigma_{j}-\sigma_{i}-\alpha\right)$ for small
$\vert\alpha\vert$) \cite{Rodrigues2016PhysRep,Pietras2019PhysRep}.

\paragraph{Generalized Kuramoto model on $S^{D-1}$.}
In this case, we define the state manifold as $M=S^{D-1}
 = \left\{x\in\mathbb{R}^{D}:
          \left\|x\right\|=1\right\}$ and let the embedding be the
canonical inclusion $F\left(\sigma\right)=\sigma$. Thus, each state
$\sigma_{i}$ is a unit vector in $\mathbb{R}^{D}$ and the tangent
space at $\sigma_{i}$ is
$T_{\sigma_{i}}S^{D-1}
 = \left\{v\in\mathbb{R}^{D}:
          \left\langle v,\sigma_{i}\right\rangle=0\right\}$. Given with
$A=0$ and $\alpha=0$, the coupling term becomes
$\Gamma\left(\sigma_{i},\sigma_{j};0\right)
 = \mathsf{P}^{\perp}_{\sigma_{i}}\sigma_{j}
 = \sigma_{j}
   - \left\langle\sigma_{i},\sigma_{j}\right\rangle\sigma_{i}$,
which coincides with the standard high-dimensional coupling term on the unit sphere used in generalized Kuramoto models on
$S^{D-1}$ \cite{Chandra2019PRX,Lipton2021Chaos,Markdahl2021CommunPhys,Zou2023PRL}.

\paragraph{Frustrated Kuramoto model on $S^{D-1}$.}
Again, we take $M=S^{D-1}$ and $F\left(\sigma\right)=\sigma$. Additionally, we allow a non-trivial phase-lag operator
$A\in\mathfrak{so}\left(D\right)$ together with a small nonzero
$\alpha$. The factor
$\left(\mathbf{I}-\alpha A\right)\sigma_{j}
 = \sigma_{j}-\alpha A\sigma_{j}$ is the first-order truncation of the
corresponding finite phase-lag action generated by $A$, such that the
interaction $\Gamma\left(\sigma_{i},\sigma_{j};\alpha\right)$
implements a multi-dimensional frustration of the Kuramoto coupling term on the sphere. Such matrix-valued phase lags generalize the
scalar phase shift of the Kuramoto-Sakaguchi model and recover frustrated high-dimensional Kuramoto systems studied on $S^{D-1}$ \cite{Buzanello2022Chaos,DeAguiar2023PRE}.

\paragraph{Vector-phase Kuramoto model on $\mathbb{T}^{d}$.}
To represent systems defined with $d$ angular degrees of freedom per
oscillator, we let the state manifold be the flat torus
$\mathbb{T}^{d}
  = \left(\mathbb{R}/2\pi\mathbb{Z}\right)^{d}$ and write
$\sigma_{i}
 = \left(\sigma_{i}^{1},\dots,\sigma_{i}^{d}\right)$, where each
$\sigma_{i}^{k}\in\mathbb{R}/2\pi\mathbb{Z}$. A natural embedding
is obtained as the product of $d$ circle embeddings
\begin{equation}
    F\left(\sigma_{i}\right)
    =
    \left(
        \cos\sigma_{i}^{1},\sin\sigma_{i}^{1},\dots,
        \cos\sigma_{i}^{d},\sin\sigma_{i}^{d}
    \right)
    \in\mathbb{R}^{2d}.
\end{equation}
We define $A$ to be block-diagonal with $d$ copies of $J$ on the
diagonal, which yields a phase-lag operator that acts independently on
each angular component. The projected coupling then gives a system
of $d$ coupled Kuramoto-type equations for the vector
$\left(\sigma_{i}^{1},\dots,\sigma_{i}^{d}\right)$, which can be
interpreted as a vector-phase or multi-frequency Kuramoto model
on $\mathbb{T}^{d}$.

\paragraph{Kuramoto models on Stiefel and rotation manifolds.}
For synchronization of matrices with orthogonality constraints, we
define $M$ as a matrix manifold such as the real Stiefel manifold
\cite{EdelmanAriasSmith1998SIAM}
$\mathrm{St}\left(p,n\right)
 = \left\{X\in\mathbb{R}^{n\times p}:
          X^{\mathsf{T}}X=\mathbf{I}_{p}\right\}$ or the special orthogonal
group $\mathrm{SO}\left(n\right)$ \cite{Hall2015LieGroups}. The embedding is the canonical
matrix inclusion $F\left(X\right)=X$ into
$\mathbb{R}^{n\times p}$ (or into $\mathbb{R}^{n\times n}$ for
$\mathrm{SO}\left(n\right)$). A direct choice of phase-lag
operator is $A\left(X\right)=BX$ with
$B\in\mathfrak{so}\left(n\right)$ fixed, such that
$\left(\mathbf{I}-\alpha A\right)X_{j}
 = X_{j}-\alpha BX_{j}$ realizes a linearized matrix-valued phase shift
along the group direction generated by $B$. The projection
$\mathsf{P}^{\perp}_{\sigma_{i}}$ enforces the orthogonality
constraint and Eq. (\ref{II-B-EQ001}) recovers Kuramoto-type consensus
dynamics on Stiefel or rotation manifolds \cite{Markdahl2019Automatica,Markdahl2017TAC,Tron2012CDC}.

\paragraph{Lohe and unitary Kuramoto models.}
To describe synchronization of unitary matrices, we define $M$ as a unitary matrix group (e.g., $\mathrm{U}\left(d\right)$ or
$\mathrm{SU}\left(d\right)$ \cite{Hall2015LieGroups}) and embed each unitary
$U\in\mathrm{U}\left(d\right)$ as
$F\left(U\right)=U$ in $\mathbb{C}^{d\times d}
 \cong\mathbb{R}^{2d^{2}}$. The Lie algebra
$\mathfrak{u}\left(d\right)$ consists of skew-Hermitian matrices \cite{Hall2015LieGroups} and we let $A\in\mathfrak{u}\left(d\right)$ act on $U$ by
left or right multiplication. The factor
$\left(\mathbf{I}-\alpha A\right)U_{j}$ then implements a linearized phase-lag along the group direction generated by $A$. The projected
coupling reproduces Lohe-type matrix models and quantum Kuramoto ensembles on unitary groups \cite{GolseHa2019ARMA,HaKoRyoo2017JSP,HaKoRyoo2018JSP,KimKim2023KRM,KimKim2025EJAM,HaPark2020SIADS,HaKangPark2021JMP,HaKangPark2021CPAA,HaPark2021JSP,Ryoo2025CMS,Lohe2019JMP,HaKimPark2020PhysicaD}.

\paragraph{Network-distributed Kuramoto models on $M$.}
Finally, the same manifold framework applies when oscillators are coupled through a network instead of a global mean field. In this case, the global scalar coupling strength $K$ in Eq. (\ref{II-B-EQ001}) is
replaced by a weighted adjacency matrix
$K_{ij}\in\mathbb{R}$ and the mean-field sum is replaced by
$\sum_{j=1}^{N}K_{ij}\Gamma\left(\sigma_{i},\sigma_{j};\alpha\right)$.
For any of the choices of $M$, $F$ and $A$ described above, this definition creates the Kuramoto- or Kuramoto-Sakaguchi-type dynamics on complex
networks, with the state manifold $M$ encoding the internal symmetry and geometry of individual oscillators \cite{Rodrigues2016PhysRep,Pietras2019PhysRep,Coutinho2013PRE,Ji2013PRL,Ji2014PRE,Boccaletti2016PhysRep,Maistrenko2014PRE,Wu2018SciRep,Kundu2017PRE,Leyva2018DCDSB,Dai2020Chaos,Bayani2023Chaos,Millan2018SciRep,Battiston2020PhysRep,SkardalArenas2020CommunPhys,Millan2022CommunPhys,Nurisso2024Chaos,Gambuzza2021NatCommun,Gallo2022CommunPhys,Neuhauser2020PRE,Millan2025NatPhys,Berner2023PhysRep}. 

\section{Continuum limit and kinetic equation}
\label{Appendix-C}

In this appendix, we provide a formal derivation of the continuum limit for the general manifold-based Kuramoto-Sakaguchi model in Eqs. (\ref{II-B-EQ001}, \ref{II-B-EQ003}). The strategy follows the standard mean-field limit for the classic Kuramoto model on $S^{1}$ and its generalizations \cite{Sznitman1991PropagationChaos,GolseHa2019ARMA}, which can be adapted to the present geometric setting.

To keep the derivation compact, we work in this appendix with the joint density on $TM$. The main-text notation is recovered by the factorization $\varrho\left(\sigma,V,t\right)=\rho\left(\sigma,V,t\right)h\left(V\right)$, where $\rho$ is the conditional spatial density for a fixed intrinsic drift field and $h\left(V\right)$ is the time-independent distribution of $V$.

\paragraph{Empirical measure and weak formulation}

For a fixed number $N$ of oscillators, the dynamics is defined by Eq. (\ref{II-B-EQ001}), which we recall here for convenience
\begin{align}
    \frac{\mathsf{d}\sigma_{i}}{\mathsf{d}t}
    &=
    V\left(\sigma_{i}\right)
    +
    \frac{K}{N}\sum_{j=1}^{N}
    \Gamma\left(\sigma_{i},\sigma_{j};\alpha\right),
    \label{CEQ1}
\end{align}
where $\sigma_{i}\left(t\right)\in M$ denotes the oscillator state. The coupling is
\begin{align}
    \Gamma\left(\sigma_{i},\sigma_{j};\alpha\right)
    &=
    \mathsf{P}^{\perp}_{\sigma_{i}}
    \left[
        \left(\mathbf{I}-\alpha A\right)
        F\left(\sigma_{j}\right)
    \right],
    \label{CEQ2}
\end{align}
where $F:M\hookrightarrow\mathbb{R}^{D_{a}}$ is a fixed smooth embedding,
$\mathsf{P}^{\perp}_{\sigma_{i}}:\mathbb{R}^{D_{a}}\to T_{\sigma_{i}}M$
is the orthogonal projection, $A\in\mathfrak{so}\left(D_{a}\right)$ is a
skew-symmetric matrix, and $\alpha\in\mathbb{R}$ is the phase-lag parameter.

The empirical measure on $TM$ associated with the configuration
$\left\{\left(\sigma_{i}\left(t\right),V_{i}\right)\right\}_{i=1}^{N}$
is defined by
\begin{align}
    \nu_{t}^{N}
    &=
    \frac{1}{N}\sum_{i=1}^{N}
    \delta_{\left(\sigma_{i}\left(t\right),V_{i}\right)},
    \label{CEQ3}
\end{align}
where $\delta_{\left(\sigma,V\right)}$ is the Dirac mass on $TM$. By its construction, $\nu_{t}^{N}$ is a probability measure on $TM$. The corresponding
configuration-space empirical measure on $M$ is the marginal $\mu_{t}^{N}
    =
    \frac{1}{N}\sum_{i=1}^{N}
    \delta_{\sigma_{i}\left(t\right)}$, which is obtained from $\nu_{t}^{N}$ by integrating out $V$.

To derive an evolution equation for $\nu_{t}^{N}$, we consider it in a weak form. Given a smooth function
$\Phi\in C^{\infty}\left(TM\right)$, we define
\begin{align}
    \left\langle\Phi,\nu_{t}^{N}\right\rangle
    &=
    \int_{TM}
    \Phi\left(\sigma,V\right)\,
    \nu_{t}^{N}\left(\mathsf{d}\sigma\mathsf{d}V\right)\notag\\
    &=
    \frac{1}{N}\sum_{i=1}^{N}
    \Phi\left(\sigma_{i}\left(t\right),V_{i}\right)
    \label{CEQ4}
\end{align}
Differentiating Eq. (\ref{CEQ4}) with respect to $t$ and using the chain rule on the manifold $M$, we obtain
\begin{align}
    \frac{\mathsf{d}}{\mathsf{d}t}
    \left\langle\Phi,\nu_{t}^{N}\right\rangle
    &=
    \frac{1}{N}\sum_{i=1}^{N}
    \left\langle
        \nabla_{M}\Phi\left(\sigma_{i},V_{i}\right),
        \frac{\mathsf{d}\sigma_{i}}{\mathsf{d}t}
    \right\rangle_{g},
    \label{CEQ5}
\end{align}
where $\nabla_{M}\Phi$ denotes the Riemannian gradient of
$\Phi$ and
$\left\langle\cdot,\cdot\right\rangle_{g}$ is the Riemannian
inner product on $T_{\sigma}M$ induced by the metric $g$.

Substituting Eq. (\ref{CEQ1}) into Eq. (\ref{CEQ5}) and separating
intrinsic and coupling contributions gives
\begin{align}
    &\frac{\mathsf{d}}{\mathsf{d}t}
    \left\langle\Phi,\nu_{t}^{N}\right\rangle=
    \frac{1}{N}\sum_{i=1}^{N}
    \left\langle
        \nabla_{M}\Phi\left(\sigma_{i},V_{i}\right),
        V\left(\sigma_{i}\right)
    \right\rangle_{g}\notag\\
    &+
    \frac{K}{N^{2}}
    \sum_{i=1}^{N}\sum_{j=1}^{N}
    \left\langle
        \nabla_{M}\Phi\left(\sigma_{i},V_{i}\right),
        \Gamma\left(\sigma_{i},\sigma_{j};\alpha\right)
    \right\rangle_{g}.
    \label{CEQ6}
\end{align}
Based on the empirical measure defined in Eq. (\ref{CEQ3}), the
right side of Eq. (\ref{CEQ6}) can be written as
\begin{align}
    \frac{\mathsf{d}}{\mathsf{d}t}
    \left\langle\Phi,\nu_{t}^{N}\right\rangle
    &=
    \int_{TM}
    \left\langle
        \nabla_{M}\Phi\left(\sigma,V\right),
        V\left(\sigma\right)
    \right\rangle_{g}
    \nu_{t}^{N}\left(\mathsf{d}\sigma\mathsf{d}V\right)\notag\\
    &\quad+
    K\int_{TM}\int_{TM}
    \left\langle
        \nabla_{M}\Phi\left(\sigma,V\right),
        \Gamma\left(\sigma,\sigma^{\prime};\alpha\right)
    \right\rangle_{g}\notag\\
    &\qquad\qquad\qquad\qquad
    \nu_{t}^{N}\left(\mathsf{d}\sigma\mathsf{d}V\right)\,
    \nu_{t}^{N}\left(\mathsf{d}\sigma^{\prime}\mathsf{d}V^{\prime}\right).
    \label{CEQ7}
\end{align}
Eq. (\ref{CEQ7}) is the weak formulation of the finite-$N$ dynamics in terms of the empirical measure $\nu_{t}^{N}$.

\paragraph{Mean-field limit and continuity equation}

We now formally take the mean-field limit $N\to\infty$. We assume that the empirical measures
$\nu_{t}^{N}$ converge weakly to a Borel probability measure, i.e,
$\nu_{t}$ on $M$ for each $t\geq 0$
\begin{align}
    \nu_{t}^{N}
    &\xrightarrow[N\to\infty]{\ \mathrm{weakly}\ }
    \nu_{t}.
    \label{CEQ8}
\end{align}
Eq. (\ref{CEQ8}) holds in the sense that
$\left\langle\Phi,\nu_{t}^{N}\right\rangle
 \to\left\langle\Phi,\nu_{t}\right\rangle$ for all
$\Phi\in C^{\infty}\left(TM\right)$. Based on Eq. (\ref{CEQ8}), we can formally obtain the limit in Eq. (\ref{CEQ7})
\begin{align}
    &\frac{\mathsf{d}}{\mathsf{d}t}
    \left\langle\Phi,\nu_{t}\right\rangle=
    \int_{TM}
    \left\langle
        \nabla_{M}\Phi\left(\sigma,V\right),
        V\left(\sigma\right)
    \right\rangle_{g}
    \nu_{t}\left(\mathsf{d}\sigma\mathsf{d}V\right)
    \notag\\
    &+
    K\int_{TM}\int_{TM}
    \left\langle
        \nabla_{M}\Phi\left(\sigma,V\right),
        \Gamma\left(\sigma,\sigma^{\prime};\alpha\right)
    \right\rangle_{g}\notag\\
    &\qquad\qquad\qquad\qquad
    \nu_{t}\left(\mathsf{d}\sigma\mathsf{d}V\right)
    \nu_{t}\left(\mathsf{d}\sigma^{\prime}\mathsf{d}V^{\prime}\right).
    \label{CEQ9}
\end{align}
This is the weak formulation of the mean-field equation for the limit measure $\mu_{t}$.

On the tangent bundle $TM$, it is natural to assume that $\nu_{t}$ admits a density $\rho\left(\sigma,V,t\right)$ with respect to the
product of the Riemannian volume measure
$\mu\left(\mathsf{d}\sigma\right)$ on $M$ and a reference measure
$\mathsf{d}V$ on each fiber $T_{\sigma}M$, i.e.,
\begin{align}
    \nu_{t}\left(\mathsf{d}\sigma\mathsf{d}V\right)
    &=
    \rho\left(\sigma,V,t\right)\,
    \mu\left(\mathsf{d}\sigma\right)\mathsf{d}V,
    \label{CEQ10}
\end{align}
with the normalization
\begin{align}
    \int_{TM}
    \rho\left(\sigma,V,t\right)\,
    \mu\left(\mathsf{d}\sigma\right)\mathsf{d}V
    &=
    1.
    \label{CEQ11}
\end{align}
In terms of this density, the continuum order parameter takes the form
\begin{align}
    r\left(t\right)
    &=
    \int_{TM}\int_{M}
    F\left(\sigma^{\prime}\right)\,
    \rho\left(\sigma^{\prime},V^{\prime},t\right)\,
    \mathsf{d}\mu\left(\sigma^{\prime}\right)\mathsf{d}V^{\prime}.
    \label{CEQ12}
\end{align}

Using the explicit form of the coupling in Eq. (\ref{CEQ2}), the interaction contribution in Eq. (\ref{CEQ9}) can be re-written as
\begin{widetext}
\begin{align}
    &\int_{TM}\int_{TM}
    \left\langle
        \nabla_{M}\Phi\left(\sigma,V\right),
        \Gamma\left(\sigma,\sigma^{\prime};\alpha\right)
    \right\rangle_{g}
    \nu_{t}\left(\mathsf{d}\sigma\mathsf{d}V\right)\,
    \nu_{t}\left(\mathsf{d}\sigma^{\prime}\mathsf{d}V^{\prime}\right)
    \notag\\
    =&
    \int_{TM}
    \left\langle
        \nabla_{M}\Phi\left(\sigma,V\right),
        \int_{TM}
        \Gamma\left(\sigma,\sigma^{\prime};\alpha\right)\,
        \nu_{t}\left(\mathsf{d}\sigma^{\prime}\mathsf{d}V^{\prime}\right)
    \right\rangle_{g}
    \notag\\
    =&
    \int_{TM}
    \left\langle
        \nabla_{M}\Phi\left(\sigma,V\right),
        \int_{TM}
        \Gamma\left(\sigma,\sigma^{\prime};\alpha\right)\,
        \rho\left(\sigma^{\prime},V^{\prime},t\right)\,
        \mathsf{d}\mu\left(\sigma^{\prime}\right)\mathsf{d}V^{\prime}
    \right\rangle_{g}
    \mathsf{d}\nu_{t}\left(\sigma,V\right),
    \label{CEQ13}
\end{align}
\end{widetext}
where we have used Eq. (\ref{CEQ10}) in the last line.

By substituting Eq. (\ref{CEQ2}) into the inner integral in
Eq. (\ref{CEQ13}) and using the linearity of
$\mathsf{P}_{\sigma}^{\perp}$ and $\left(\mathbf{I}-\alpha A\right)$ with
respect to $F\left(\sigma^{\prime}\right)$, we obtain
\begin{widetext}
\begin{align}
    \int_{TM}
    \Gamma\left(\sigma,\sigma^{\prime};\alpha\right)\,
    \rho\left(\sigma^{\prime},V^{\prime},t\right)\,
    \mathsf{d}\mu\left(\sigma^{\prime}\right)\mathsf{d}V^{\prime}
    &=
    \int_{TM}
    \mathsf{P}_{\sigma}^{\perp}
    \left[
        \left(\mathbf{I}-\alpha A\right)
        F\left(\sigma^{\prime}\right)
    \right]
    \rho\left(\sigma^{\prime},V^{\prime},t\right)\,
    \mathsf{d}\mu\left(\sigma^{\prime}\right)\mathsf{d}V^{\prime}
    \notag\\
    &=
    \mathsf{P}_{\sigma}^{\perp}
    \left[
        \left(\mathbf{I}-\alpha A\right)
        \int_{TM}
        F\left(\sigma^{\prime}\right)\,
        \rho\left(\sigma^{\prime},V^{\prime},t\right)\,
        \mathsf{d}\mu\left(\sigma^{\prime}\right)\mathsf{d}V^{\prime}
    \right]
    \notag\\
    &=
    \mathsf{P}_{\sigma}^{\perp}
    \left(
        r\left(t\right)
        -\alpha A r\left(t\right)
    \right),
    \label{CEQ14}
\end{align}
\end{widetext}
where we have used the definition of the
order parameter in Eq. (\ref{CEQ12}) in the last step. This shows explicitly that, in the mean–field
limit, the averaged coupling can be expressed in terms of the
order parameter as the projected field
$\mathsf{P}_{\sigma}^{\perp}\left(r\left(t\right)-\alpha A r\left(t\right)\right)$,
which is precisely the coupling structure used in Eq. (\ref{II-D-EQ003}) in our main text.

After substituting Eq. (\ref{CEQ10}) and Eqs. (\ref{CEQ13}-\ref{CEQ14}) into Eq. (\ref{CEQ9}) and integrating
by parts on $M$ with respect to the Riemannian volume
$\mu\left(\mathsf{d}\sigma\right)$ (based on that $M$ is compact and
without boundary), we obtain the weak formulation
\begin{align}
    &\frac{\mathsf{d}}{\mathsf{d}t}
    \int_{TM}
    \Phi\left(\sigma,V\right)\,
    \rho\left(\sigma,V,t\right)\,
    \mathsf{d}\mu\left(\sigma\right)\mathsf{d}V\notag\\
    =&
    -\int_{TM}
    \Phi\left(\sigma,V\right)\,
    \nabla_{M}\cdot
    \left(
        \rho\left(\sigma,V,t\right)\,
        \mathbf{v}\left(\sigma,V,t\right)
    \right)
    \mathsf{d}\mu\left(\sigma\right)\mathsf{d}V,
    \label{CEQ15}
\end{align}
where the mean–field velocity field on $TM$ is given by
\begin{align}
    \mathbf{v}\left(\sigma,V,t\right)
    &=
    V\left(\sigma\right)
    +
    K\,\mathsf{P}_{\sigma}^{\perp}
    \left(
        r\left(t\right)
        -\alpha A r\left(t\right)
    \right)
    \label{CEQ16}
\end{align}
according to our analysis of the averaged coupling.

Because Eq. (\ref{CEQ15}) holds for all smooth functions
$\Phi\in C^{\infty}\left(TM\right)$, it is equivalent to the continuity equation on $TM$
\begin{align}
    \frac{\partial}{\partial t}
    \rho\left(\sigma,V,t\right)
    +
    \nabla_{M}\cdot
    \left(
        \rho\left(\sigma,V,t\right)\,
        \mathbf{v}\left(\sigma,V,t\right)
    \right)
    &=
    0,
    \label{CEQ17}
\end{align}
where $\mathbf{v}\left(\sigma,V,t\right)$ is given by Eq. (\ref{CEQ16}). Eq. (\ref{CEQ17}) serves as the kinetic 
equation underlying the macroscopic model
Eqs. (\ref{II-D-EQ004}-\ref{II-D-EQ006}). The associated normalization $\int_{TM}
    \rho\left(\sigma,V,t\right)\,
    \mathsf{d}\mu\left(\sigma\right)\mathsf{d}V
    =
    1$ is preserved for all $t$.

\section{Local bifurcation reduction of the order parameter response equation}
\label{Appendix-D}

In this appendix, we provide the technical reduction used in
Subsec. \ref{Nonlinear-response-subsection}. This
appendix has two purposes. The first purpose is to derive the scalar amplitude
equation near the incoherent state. The second purpose is to state explicit
conditions under which the cubic coefficient becomes a local integral over
$M$. We assume that the steady order parameter response near the incoherent
state is described by the equation
\begin{align}
G\left(K,r\right)
=
\mathcal{L}\left(K\right)r+\mathcal{N}\left(K,r\right)
=0,\label{DEQ1}
\end{align}
on a Banach space $X$ of admissible order parameter perturbations, with
$G\in C^{4}$ in a neighborhood of $\left(K_{c},0\right)$ and
$G\left(K,0\right)=0$. Let
\begin{align}
L_{c}
=
\mathrm{D}_{r}G\left(K_{c},0\right),\label{DEQ2}
\end{align}
and assume that $L_{c}$ is Fredholm of index $0$. We assume that
$\ker L_{c}=\operatorname{span}\left\{r_{c}\right\}$ is one-dimensional. We
assume that the adjoint kernel is generated by $r_{c}^{\dagger}$. We normalize
the dual pairing by $\langle r_{c}^{\dagger},r_{c}\rangle=1$. We also assume
that the corresponding simple eigenvalue satisfies the transversality condition
\begin{align}
\lambda\left(K_{c}\right)=0,\qquad
\lambda^{\prime}\!\left(K_{c}\right)\neq 0.\label{DEQ3}
\end{align}
These assumptions describe a standard codimension-one instability. The
critical mode $r_{c}$ is the only neutral collective direction at $K_{c}$. The
remaining directions are determined by this mode near threshold. The transversality
condition states that the instability is crossed at a nonzero rate when $K$
passes through $K_{c}$.
We introduce the spectral projectors
$P\left[f\right]=r_{c}\langle r_{c}^{\dagger},f\rangle$ and $Q=\mathbf{I}-P$, and decompose
\begin{align}
r
=
a\,r_{c}+w,\qquad
\langle r_{c}^{\dagger},w\rangle=0.\label{DEQ4}
\end{align}
Expanding the nonlinear part at $K=K_{c}$ gives
\begin{align}
\mathcal{N}\left(K_{c},r\right)
=
\mathcal{N}_{2}\left(r,r\right)
+
\mathcal{N}_{3}\left(r,r,r\right)
+
O\left(\|r\|_{X}^{4}\right),\label{DEQ5}
\end{align}
where $\mathcal{N}_{2}$ and $\mathcal{N}_{3}$ are the symmetric multilinear
Taylor coefficients. To exclude a transcritical-type correction in the scalar
normal form, we further impose the condition
\begin{align}
\Big\langle r_{c}^{\dagger},\mathcal{N}_{2}\left(r_{c},r_{c}\right)\Big\rangle
=
0.\label{DEQ6}
\end{align}
Eq. (\ref{DEQ6}) is automatic whenever the steady response equation is odd
under $r\mapsto -r$. In more general settings, we impose Eq. (\ref{DEQ6}) as
an additional condition. Eq. (\ref{DEQ6}) states that the incoherent state has
no preferred sign along the critical mode at quadratic order.

We give a concrete example in which Eq. (\ref{DEQ6}) follows from symmetry.
Take the classical circle case $M=S^{1}$, write
$F\left(\theta\right)=e^{i\theta}$, set $\alpha=0$, and take
$V=\omega\partial_{\theta}$ with an even frequency density
$g\left(\omega\right)$. The half-period shift
$\theta\mapsto\theta+\pi$ preserves the volume measure and the intrinsic
transport, while it sends $F\left(\theta\right)$ to
$-F\left(\theta\right)$. Therefore a steady density with order parameter $r$
is mapped to a steady density with order parameter $-r$. The corresponding
steady response map satisfies
\begin{align}
G\left(K,-r\right)
=
-G\left(K,r\right)
\label{DEQ6a}
\end{align}
on the first Fourier mode. Expanding Eq. (\ref{DEQ6a}) in the critical
coordinate eliminates all even scalar terms in the kernel equation. In
particular, the projected quadratic coefficient in Eq. (\ref{DEQ6}) vanishes.
This example shows the mechanism behind the symmetry condition; outside such
an explicit sign-reversing equivariance, Eq. (\ref{DEQ6}) must be verified or
assumed for the chosen model.
The range
equation then yields
\begin{align}
w\left(a;K\right)
=&
a^{2}w_{2}+O\left(a^{3}\right)+O\left(\left|K-K_{c}\right|a\right),\notag\\
w_{2}
=&
-\left(QL_{c}Q\right)^{-1}
Q\,\mathcal{N}_{2}\left(r_{c},r_{c}\right).\label{DEQ7}
\end{align}
The correction $w_{2}$ is the stable component generated by the quadratic
self-interaction of the critical mode. Eq. (\ref{DEQ7}) shows that $w_{2}$ is
not an independent degree of freedom. The stable part is determined by the
critical amplitude. Eq. (\ref{DEQ7}) then enters the kernel equation and yields
the reduced amplitude equation
\begin{align}
\Psi\left(a;K\right)
&=
\lambda\left(K\right)a+\Lambda_{3}a^{3}\notag\\
&\quad
+O\left(a^{4}\right)
+O\left(\left|K-K_{c}\right|a^{2}\right)
=0,\label{DEQ8}
\end{align}
with
\begin{align}
\Lambda_{3}
=
\Big\langle r_{c}^{\dagger},\,
\mathcal{N}_{3}\left(r_{c},r_{c},r_{c}\right)
+
2\mathcal{N}_{2}\left(r_{c},w_{2}\right)
\Big\rangle.\label{DEQ9}
\end{align}
Eq. (\ref{DEQ8}) is the technical statement used in
Subsec. \ref{Nonlinear-response-subsection}. Eq. (\ref{DEQ9})
shows that the cubic coefficient contains a direct cubic term and a mixed term
produced by the stable correction.
Moreover, Eq. (\ref{DEQ3}) implies the local expansion
\begin{align}
\lambda\left(K\right)
=
\lambda^{\prime}\!\left(K_{c}\right)\left(K-K_{c}\right)
+
O\left(\left(K-K_{c}\right)^{2}\right),\label{DEQ10}
\end{align}
so every nontrivial small-amplitude branch with $\Lambda_{3}\neq 0$ satisfies
\begin{align}
\left|a\right|
\sim
\sqrt{-\frac{\lambda\left(K\right)}{\Lambda_{3}}}
=
\sqrt{-\frac{\lambda^{\prime}\!\left(K_{c}\right)\left(K-K_{c}\right)}{\Lambda_{3}}}.\label{DEQ11}
\end{align}
Eq. (\ref{DEQ11}) is the usual square-root scaling near a cubic bifurcation.

We next connect Eq. (\ref{DEQ9}) with the topology of $M$. We introduce the induced
critical tangent field
\begin{align}
u_{c}\left(\sigma\right)
&=
\mathsf{P}_{\sigma}^{\perp}
\left(r_{c}-\alpha A r_{c}\right).\label{DEQ12}
\end{align}
This field is the relevant geometric object for the nonlinear transport. The
microscopic velocity in Eq. (\ref{II-D-EQ003}) depends on the order parameter
through the projected field
$\mathsf{P}_{\sigma}^{\perp}\left(r-\alpha A r\right)$. For this reason, the
topological content of the critical response is carried by $u_{c}$.
For any stable correction $w\in QX$, we also define the induced tangent field
\begin{align}
\widetilde{u}_{w}\left(\sigma\right)
&=
\mathsf{P}_{\sigma}^{\perp}
\left(w-\alpha A w\right).\label{DEQ13}
\end{align}
We now state a sufficient set of assumptions under which the cubic coefficient
becomes pointwise local. These assumptions are not consequences of the
Poincar\'e-Hopf theorem, and they are not automatic consequences of
homogeneity. They are hypotheses on the nonlinear response operator and on the
stable inverse in Eq. (\ref{DEQ7}). In particular, the correction $w_{2}$
contains the inverse of the global operator $QL_{c}Q$, so its locality must be
checked for the chosen model and critical subspace.

\paragraph{Locality assumptions.-}Let
$\mathcal{S}_{2}\left(u_{c}\right)\subset QX$ denote the set of stable
corrections $w$ with the following property. The induced tangent field
$\widetilde{u}_{w}\left(\sigma\right)$ can be written pointwise as
$S\left(\sigma\right)\left[u_{c}\left(\sigma\right),u_{c}\left(\sigma\right)\right]$
for some $S\in \Gamma\left(\operatorname{Sym}^{2}T^{\ast}M\otimes TM\right)$.
This set collects the local quadratic tensor subspace generated by the critical
field. It is a specified subspace of stable corrections, not the full stable
space.

\emph{Assumption D1: local direct cubic term.} We first require the direct
cubic contribution to be local in the induced tangent field. This requirement
states that the direct cubic response is assembled from a pointwise cubic
density on $M$. Concretely, we assume that
there exists a smooth symmetric trilinear tensor field $T_{3,0}\left(\sigma\right)$
such that
\begin{align}
&\Big\langle r_{c}^{\dagger},\mathcal{N}_{3}\left(r_{c},r_{c},r_{c}\right)\Big\rangle\notag\\
=&
\int_{M}
T_{3,0}\left(\sigma\right)\!\left[
u_{c}\left(\sigma\right),
u_{c}\left(\sigma\right),
u_{c}\left(\sigma\right)
\right]\mathsf{d}\mu\left(\sigma\right),\label{DEQ14}
\end{align}
\emph{Assumption D2: closure of the quadratic stable correction.} We next
require the quadratic source to stay inside the local quadratic subspace. We
also require the stable inverse to preserve that subspace:
\begin{align}
Q\,\mathcal{N}_{2}\left(r_{c},r_{c}\right)\in \mathcal{S}_{2}\left(u_{c}\right).\label{DEQ15}
\end{align}
\begin{align}
\left(QL_{c}Q\right)^{-1}\mathcal{S}_{2}\left(u_{c}\right)\subseteq \mathcal{S}_{2}\left(u_{c}\right).\label{DEQ16}
\end{align}
The second inclusion is the main locality hypothesis. It says that solving the
stable range equation does not generate nonlocal dependence on the critical
field inside the selected subspace. Under Eq. (\ref{DEQ15}) and
Eq. (\ref{DEQ16}), the stable correction produced by the quadratic
self-interaction remains local and quadratic in $u_{c}$. Therefore there
exists a smooth symmetric bilinear bundle map $S_{2}\left(\sigma\right)$. This
map satisfies
\begin{align}
\widetilde{u}_{w_{2}}\left(\sigma\right)
&=
S_{2}\left(\sigma\right)
\left[u_{c}\left(\sigma\right),u_{c}\left(\sigma\right)\right].\label{DEQ17}
\end{align}
\emph{Assumption D3: local mixed term.} We finally require the mixed quadratic
contribution to be local in the pair
$\left(u_{c},\widetilde{u}_{w}\right)$. This requirement states that the
interaction between the critical mode and the stable correction is evaluated
pointwise on $M$ after the stable correction has been restricted to
$\mathcal{S}_{2}\left(u_{c}\right)$. We therefore assume that there exists a
smooth bilinear form $B\left(\sigma\right)$. This form satisfies the following
relation for every $w\in \mathcal{S}_{2}\left(u_{c}\right)$:
\begin{align}
\Big\langle r_{c}^{\dagger},\mathcal{N}_{2}\left(r_{c},w\right)\Big\rangle
=
\int_{M}
B\left(\sigma\right)\!\left[
u_{c}\left(\sigma\right),
\widetilde{u}_{w}\left(\sigma\right)
\right]
\,\mathsf{d}\mu\left(\sigma\right).\label{DEQ18}
\end{align}
Eq. (\ref{DEQ17}) turns Eq. (\ref{DEQ18}) into
\begin{align}
\Big\langle r_{c}^{\dagger},\mathcal{N}_{2}\left(r_{c},w_{2}\right)\Big\rangle
&=
\int_{M}
T_{3,1}\left(\sigma\right)\!\left[
u_{c}\left(\sigma\right),
u_{c}\left(\sigma\right),
u_{c}\left(\sigma\right)
\right]\notag\\
&\quad\,\mathsf{d}\mu\left(\sigma\right),\label{DEQ19}
\end{align}
The tensor field $T_{3,1}\left(\sigma\right)$ is obtained by symmetrizing the
contraction of $B\left(\sigma\right)$ with $S_{2}\left(\sigma\right)$.
Eqs. (\ref{DEQ9}, \ref{DEQ14}, and \ref{DEQ19}) then give
\begin{align}
\Lambda_{3}
&=
\int_{M}
T\left(\sigma\right)\!\left[
u_{c}\left(\sigma\right),
u_{c}\left(\sigma\right),
u_{c}\left(\sigma\right)
\right]\mathsf{d}\mu\left(\sigma\right),\label{DEQ20}
\end{align}
with $T=T_{3,0}+2T_{3,1}$. This is precisely the local tensor
representation used in Subsec. \ref{Nonlinear-response-subsection}. The three assumptions above state that the
direct cubic term, the quadratic correction, and their mixed contribution all
preserve the local tensor structure selected by the projected tangent field.

A useful way to verify Eq. (\ref{DEQ16}) is through symmetry, but the required
closure must still be checked. Suppose that the quadratic source generated by
$u_{c}\otimes u_{c}$ belongs to a finite-dimensional $G$-invariant harmonic
subspace for the transitive isometry group $G$ of $M$, and suppose that
$QL_{c}Q$ is $G$-equivariant and invertible on that same subspace. If the subspace
is irreducible, Schur's lemma \cite{Hall2015LieGroups} implies that $QL_{c}Q$
acts there as a scalar multiple of the identity. This verifies the preservation
condition in Eq. (\ref{DEQ16}) on that subspace. Without such a componentwise
verification, Eq. (\ref{DEQ20}) should be read as an additional local
reduction assumption rather than as a theorem following from homogeneity.
Under Assumptions D1--D3, Subsec. \ref{Nonlinear-response-subsection} reduces
the cubic coefficient to the topology of the induced tangent field. Appendix
\ref{Appendix-E} then supplies a separate sign condition needed to pass from
the forced zeros of $u_{c}$ to a definite sign of $\Lambda_{3}$.

\section{Sign condition for the cubic coefficient}
\label{Appendix-E}

In this appendix, we state the additional sign condition used in
Subsec. \ref{Nonlinear-response-subsection}. Under the locality assumptions of
Appendix \ref{Appendix-D}, the cubic coefficient can be written as a local
integral over $M$. The Poincar\'e-Hopf theorem fixes the total index of the
zeros of $u_{c}$, but it does not by itself fix the sign of this integral. The
sign statement therefore requires an additional model-dependent dominance
condition on the local cubic density. We write this density in
Eq. (\ref{III-C-EQ003}) as
\begin{align}
q\left(\sigma\right)
&=
T\left(\sigma\right)
\left[
 u_{c}\left(\sigma\right),
 u_{c}\left(\sigma\right),
 u_{c}\left(\sigma\right)
\right],\label{EEQ1}
\end{align}
The density $q\left(\sigma\right)$ measures the local contribution of the
critical tangent field to the cubic coefficient. The coefficient $\Lambda_{3}$
is the integral of this density:
\begin{align}
\Lambda_{3}
=
\int_{M}q\left(\sigma\right)\mathsf{d}\mu\left(\sigma\right).\label{EEQ2}
\end{align}
We assume that the zero set of $u_{c}$ consists of isolated points
$\left\{p_{1},\dots,p_{m}\right\}$. This situation is generic for a smooth
vector field after a small perturbation. For each zero $p_{\ell}$, we choose a
small geodesic neighborhood $U_{\ell}$ \cite{Lee2018Riemannian}. The sets
$U_{\ell}$ are disjoint and contain no other zeros. We decompose the local
contribution from each core as
\begin{align}
\int_{U_{\ell}}
q\left(\sigma\right)\mathsf{d}\mu\left(\sigma\right)
=
-\gamma_{\ell}\operatorname{ind}_{p_{\ell}}\left(u_{c}\right)
+\delta_{\ell},
\qquad
\gamma_{\ell}>0.
\label{EEQ3}
\end{align}
Here $\gamma_{\ell}$ measures the indexed part of the local contribution,
and $\delta_{\ell}$ records the local remainder not weighted by the index. The sign
convention in Eq. (\ref{EEQ3}) is the one used in the main text. We then define
the weighted indexed sum
\begin{align}
\Gamma\left(u_{c}\right)
=
\sum_{\ell=1}^{m}
\gamma_{\ell}\operatorname{ind}_{p_{\ell}}\left(u_{c}\right).
\label{EEQ4}
\end{align}
The Euler characteristic fixes the unweighted sum of the indices, not the
weighted sum in Eq. (\ref{EEQ4}). Thus the condition
$\chi\left(M\right)\neq0$ does not by itself imply
$\Gamma\left(u_{c}\right)>0$ when the local indices have mixed signs or when
the weights vary.

Let
$U=\bigcup_{\ell=1}^{m}U_{\ell}$ and define the remainder not weighted by the
indices by
\begin{align}
\mathcal{R}_{U}
=
\sum_{\ell=1}^{m}\delta_{\ell}
+
\int_{M\setminus U}
q\left(\sigma\right)\mathsf{d}\mu\left(\sigma\right).
\label{EEQ5}
\end{align}
The sign condition used in the main text is the following signed dominance
assumption:
\begin{align}
\Gamma\left(u_{c}\right)>0,
\qquad
\left|\mathcal{R}_{U}\right|
<
\Gamma\left(u_{c}\right).
\label{EEQ6}
\end{align}
Under Eq. (\ref{EEQ6}), Eq. (\ref{EEQ2}) becomes
\begin{align}
\Lambda_{3}
=
-\Gamma\left(u_{c}\right)
+
\mathcal{R}_{U}
<
0.
\label{EEQ7}
\end{align}
This is the additional sign condition used in the main text when we pass from
the topological defect constraint to a definite subcritical sign of the cubic
normal-form coefficient. The quantities $\gamma_{\ell}$, $\delta_{\ell}$, and
$\mathcal{R}_{U}$ depend on the nonlinear response operator, the chosen
critical branch, and the drift ensemble. They are not determined by the Euler
characteristic alone.

When $\chi\left(M\right)=0$, the above sign conclusion is unavailable. In that
case the sign of $\Lambda_{3}$ must be determined from the analytic structure of
$T\left(\sigma\right)$ and from the detailed nonlinear reduction.

\section{Dynamical normal form and weak-field scaling}
\label{Appendix-F}

In this appendix, we record two direct consequences of the reduction derived in
Appendix \ref{Appendix-D}. The first consequence is a scalar dynamical normal
form near the critical coupling. The second consequence is the standard
weak-field scaling law near a continuous branch. These statements do not add a
new topological mechanism. They use the same one-dimensional reduction that
already underlies Subsec. \ref{Nonlinear-response-subsection}.

\paragraph{Dynamical normal form.-}We assume that the local order parameter
dynamics near the incoherent state is generated by the same response map as in
Appendix \ref{Appendix-D}. More precisely, we assume that
\begin{align}
\frac{\mathsf{d}}{\mathsf{d}t}r
=
-G\left(K,r\right),\label{FEQ1}
\end{align}
where $G\left(K,r\right)$ is defined by Eq. (\ref{DEQ1}). We also assume that
Eq. (\ref{FEQ1}) admits a $C^{3}$ local center manifold \cite{HaragusIooss2011} at
$\left(K_{c},0\right)$. The center manifold is tangent to
$\operatorname{span}\left\{r_{c}\right\}$ and uses the same critical mode as
the steady reduction in Appendix \ref{Appendix-D}. Therefore the reduced center
variable $a$ satisfies
\begin{align}
\frac{\mathsf{d}}{\mathsf{d}t}a
&=
-\Psi\left(a;K\right)\notag\\
&=
-\lambda\left(K\right)a-\Lambda_{3}a^{3}\notag\\
&\quad
+O\left(a^{4}\right)
+O\left(\left|K-K_{c}\right|a^{2}\right),\label{FEQ2}
\end{align}
where $\Psi\left(a;K\right)$ is the steady reduced equation in
Eq. (\ref{DEQ8}). Hence the equilibria of Eq. (\ref{FEQ2}) are exactly the
small-amplitude steady branches already studied in
Subsec. \ref{Nonlinear-response-subsection}.

Let $a_{\ast}\left(K\right)$ be a stable equilibrium of Eq. (\ref{FEQ2}).
Linearizing Eq. (\ref{FEQ2}) around $a_{\ast}\left(K\right)$ gives the local
relaxation rate
\begin{align}
\tau^{-1}\left(K\right)
=
\left|\partial_{a}\Psi\left(a_{\ast}\left(K\right);K\right)\right|.\label{FEQ3}
\end{align}
For the incoherent branch $a_{\ast}=0$, Eq. (\ref{FEQ3}) gives
\begin{align}
\tau^{-1}\left(K\right)
=
\left|\lambda\left(K\right)\right|
+o\left(\left|K-K_{c}\right|\right),\label{FEQ4}
\end{align}
and therefore
\begin{align}
\tau\left(K\right)
\sim
\left|\lambda\left(K\right)\right|^{-1}
\sim
\left|K-K_{c}\right|^{-1}.\label{FEQ5}
\end{align}
This is the usual critical slowing-down law. When a continuous branch exists,
Eq. (\ref{DEQ11}) gives
$a_{\ast}^{2}\left(K\right)=-\lambda\left(K\right)/\Lambda_{3}+o\left(\lambda\right)$.
Substituting this relation into Eq. (\ref{FEQ3}) yields
\begin{align}
\tau^{-1}\left(K\right)
&=
\left|\lambda\left(K\right)+3\Lambda_{3}a_{\ast}^{2}\left(K\right)\right|\notag\\
&=
2\left|\lambda\left(K\right)\right|
+o\left(\left|K-K_{c}\right|\right),\label{FEQ6}
\end{align}
so the same scaling $\tau\sim \left|K-K_{c}\right|^{-1}$ holds on the
continuous branch.

\paragraph{Weak-field scaling.-}We next add a weak external bias in the
critical direction. We assume that the full response equation takes the form
\begin{align}
\frac{\mathsf{d}}{\mathsf{d}t}r
=
-G\left(K,r\right)+h\,\eta,\label{FEQ7}
\end{align}
where $\eta\in X$ is a fixed bias direction with
$\langle r_{c}^{\dagger},\eta\rangle\neq 0$. After rescaling the field
amplitude $h$, the reduced steady equation on the center manifold becomes
\begin{align}
\Psi_{h}\left(a;K\right)
&=
\lambda\left(K\right)a+\Lambda_{3}a^{3}-h\notag\\
&\quad
+O\left(a^{4}\right)
+O\left(\left|K-K_{c}\right|a^{2}\right)
+O\left(\left|h\right|\left|a\right|\right)\notag\\
&=
0.\label{FEQ8}
\end{align}
At the critical coupling $K=K_{c}$, Eq. (\ref{FEQ8}) gives
\begin{align}
\left|a\right|
\sim
\left|\frac{h}{\Lambda_{3}}\right|^{1/3},\label{FEQ9}
\end{align}
provided $\Lambda_{3}\neq 0$. This is the standard mean-field field-response
law, so the exponent is $\delta=3$.

When $h=0$ and a continuous branch exists, Eq. (\ref{DEQ11}) already gives
\begin{align}
\left|a_{\ast}\left(K\right)\right|
\sim
\left|K-K_{c}\right|^{1/2},\label{FEQ10}
\end{align}
so the order parameter exponent is $\beta=1/2$. To obtain the differential
susceptibility, we differentiate Eq. (\ref{FEQ8}) with respect to $h$ along a
steady branch. This gives
\begin{align}
\frac{\partial}{\partial h}a_{\ast}
=
\left(
\lambda\left(K\right)
+3\Lambda_{3}a_{\ast}^{2}\left(K\right)
\right)^{-1}
+o\left(\left|K-K_{c}\right|^{-1}\right).\label{FEQ11}
\end{align}
On the incoherent side, where $a_{\ast}=0$, Eq. (\ref{FEQ11}) gives
\begin{align}
\frac{\partial}{\partial h}a_{\ast}
\sim
\lambda\left(K\right)^{-1}.\label{FEQ12}
\end{align}
On a continuous branch, Eqs. (\ref{DEQ11}, \ref{FEQ11}) give
\begin{align}
\frac{\partial}{\partial h}a_{\ast}
\sim
\left(-2\lambda\left(K\right)\right)^{-1}.\label{FEQ13}
\end{align}
Therefore, on both sides of a continuous transition,
\begin{align}
\left|\frac{\partial}{\partial h}a_{\ast}\right|
\sim
\left|K-K_{c}\right|^{-1},\label{FEQ14}
\end{align}
which is the standard mean-field susceptibility law $\gamma=1$.

The present one-dimensional reduction therefore gives the usual mean-field
triplet
$\beta=\frac{1}{2}$, $\gamma=1$, and $\delta=3$ whenever a continuous branch
exists. In the present theory, topology does not change these exponents at the
generic codimension-one level. Instead, topology affects the availability of the
continuous branch described by Eqs. (\ref{FEQ9}-\ref{FEQ14}). Under the sign
condition in Appendix \ref{Appendix-E}, a manifold with
$\chi\left(M\right)\neq 0$ gives $\Lambda_{3}<0$, so that branch is absent from
the conditional one-mode normal form. A manifold with
$\chi\left(M\right)=0$ leaves the sign of $\Lambda_{3}$ to the analytic
coefficients.

\section{Quintic normal form and hysteresis scales}
\label{Appendix-G}

In this appendix, we extend the local reduction to quintic order.
This extension is useful when the cubic coefficient is negative and the
continuous branch is therefore absent in the conditional normal form. The cubic
normal form then does not
determine the stable large-amplitude branch by itself. The hysteresis analysis
only needs one additional property beyond Appendix \ref{Appendix-D}. The
reduced scalar equation must be odd up to quintic order:
\begin{align}
\Psi\left(-a;K\right)
=
-\Psi\left(a;K\right)
+
O\left(a^{7}\right)
+
O\left(\left|K-K_{c}\right|a^{3}\right).\label{GEQ1}
\end{align}
This condition is natural when the reduced order parameter has no preferred
sign and no external bias is present. The amplitudes $a$ and $-a$ describe the
same critical mode with opposite sign. In the absence of an external field or
built-in chirality, the reduced dynamics should not distinguish them.

A weaker sufficient condition at the model level is a local
$\mathbb{Z}_{2}$ symmetry of the response map near the incoherent state.
Assume that there exists a bounded linear involution
$\mathcal{R}:X\to X$ such that
\begin{align*}
\mathcal{R}^{2}
=
\mathbf{I},\qquad
G\left(K,\mathcal{R}r\right)
=
\mathcal{R}G\left(K,r\right),\qquad
\mathcal{R}r_{c}
=
-r_{c}.
\end{align*}
Then $L_{c}$ commutes with $\mathcal{R}$. The local center manifold can be
chosen $\mathcal{R}$-invariant. The amplitude coordinate changes from $a$ to
$-a$. The reduced scalar equation is therefore odd. Eq. (\ref{GEQ1}) follows.
This condition is weaker than $G\left(K,-r\right)=-G\left(K,r\right)$. This
condition is also weaker than assuming a specific involutive isometry of $M$.
A manifold isometry is only one concrete realization of this local
$\mathbb{Z}_{2}$ symmetry.

If Eq. (\ref{GEQ1}) fails, the reduced scalar equation contains even terms.
The two signs of $a$ are then not equivalent. The local bifurcation becomes
imperfect or asymmetric. The symmetric branch formulas derived below do not
hold in the same form.

Eq. (\ref{GEQ1}) is enough for the quintic normal form and for the hysteresis
scalings derived below. We now keep the general Taylor expansion of
$\mathcal{N}\left(K_{c},r\right)$ up to quintic order. We also assume that
$G\in C^{7}$ in a neighborhood of $\left(K_{c},0\right)$, so the reduced
equation admits a controlled quintic expansion. The range correction then has
the form
\begin{align}
w\left(a;K\right)
&=
a^{2}w_{2}+a^{3}w_{3}+a^{4}w_{4}\notag\\
&\qquad
+O\left(a^{5}\right)
+O\left(\left|K-K_{c}\right|a\right).\label{GEQ2}
\end{align}
Here $w_{2}$ is given by Eq. (\ref{DEQ7}). The coefficients
$w_{3},w_{4}\in QX$ satisfy
\begin{align}
w_{3}
=
-\left(QL_{c}Q\right)^{-1}
Q\left(
2\mathcal{N}_{2}\left(r_{c},w_{2}\right)
+
\mathcal{N}_{3}\left(r_{c},r_{c},r_{c}\right)
\right),\label{GEQ3}
\end{align}
and
\begin{align}
w_{4}
&=
-\left(QL_{c}Q\right)^{-1}
Q\Big(
\mathcal{N}_{2}\left(w_{2},w_{2}\right)
+
2\mathcal{N}_{2}\left(r_{c},w_{3}\right)\notag\\
&\qquad
+3\mathcal{N}_{3}\left(r_{c},r_{c},w_{2}\right)\notag\\
&\qquad
+\mathcal{N}_{4}\left(r_{c},r_{c},r_{c},r_{c}\right)
\Big).\label{GEQ4}
\end{align}
Substituting Eqs. (\ref{GEQ2}-\ref{GEQ4}) into the kernel equation yields the
quintic normal form
\begin{align}
\Psi_{5}\left(a;K\right)
&=
\lambda\left(K\right)a+\Lambda_{3}a^{3}+\Lambda_{5}a^{5}\notag\\
&\quad
+O\left(a^{7}\right)
+O\left(\left|K-K_{c}\right|a^{3}\right)
=
0,\label{GEQ5}
\end{align}
where $\Lambda_{3}$ is still given by Eq. (\ref{DEQ9}), and
\begin{align}
\Lambda_{5}
&=
\Big\langle r_{c}^{\dagger},
\mathcal{N}_{5}\left(r_{c},r_{c},r_{c},r_{c},r_{c}\right)\notag\\
&\qquad
+4\mathcal{N}_{4}\left(r_{c},r_{c},r_{c},w_{2}\right)
+3\mathcal{N}_{3}\left(r_{c},r_{c},w_{3}\right)\notag\\
&\qquad
+3\mathcal{N}_{3}\left(r_{c},w_{2},w_{2}\right)
+2\mathcal{N}_{2}\left(r_{c},w_{4}\right)\notag\\
&\qquad
+2\mathcal{N}_{2}\left(w_{2},w_{3}\right)
\Big\rangle.\label{GEQ6}
\end{align}
Eq. (\ref{GEQ1}) removes the quadratic and quartic terms from the reduced
scalar equation. Eq. (\ref{GEQ1}) does not require $w_{2}$ or $w_{4}$ to
vanish. Therefore the hysteresis analysis only needs Eq. (\ref{GEQ5}) together
with the sign condition $\Lambda_{5}>0$. The explicit form of $\Lambda_{5}$
changes, but the later formulas do not.

Eq. (\ref{GEQ6}) also shows why a universal topological sign rule for
$\Lambda_{5}$ is unavailable in the present framework. Unlike $\Lambda_{3}$,
the quintic coefficient depends on the slaved corrections $w_{2},w_{3},w_{4}$
and on the higher multilinear forms $\mathcal{N}_{2}$-$\mathcal{N}_{5}$. These
objects are fixed by the analytic structure of the transport law. They are not
fixed by the zero set of $u_{c}$ alone. One may, for example, add a local
quintic perturbation proportional to
$\langle r_{c}^{\dagger},r\rangle^{5}r_{c}$. Such a perturbation changes
$\Lambda_{5}$. It does not change $L_{c}$, the topology of $M$, or the cubic
sign statement in Appendix \ref{Appendix-E}. Therefore topology alone does not
determine the sign of the general coefficient $\Lambda_{5}$.

To obtain a shorter coefficient formula, we now add one more simplification. We
assume that the nonlinear Taylor expansion along the critical branch contains
no even terms up to quintic order:
\begin{align}
\mathcal{N}\left(K_{c},r\right)
=
\mathcal{N}_{3}\left(r,r,r\right)
+
\mathcal{N}_{5}\left(r,r,r,r,r\right)
+
O\left(\|r\|_{X}^{7}\right).\label{GEQ7}
\end{align}
Under Eq. (\ref{GEQ7}), Eq. (\ref{DEQ9}) simplifies because $w_{2}=0$.
Eqs. (\ref{GEQ2}-\ref{GEQ4}) then reduce to
\begin{align*}
w\left(a;K\right)
=
a^{3}w_{3}
+
O\left(a^{5}\right)
+
O\left(\left|K-K_{c}\right|a\right),\\
w_{3}
=
-\left(QL_{c}Q\right)^{-1}
Q\,\mathcal{N}_{3}\left(r_{c},r_{c},r_{c}\right).
\end{align*}
In that case, Eq. (\ref{GEQ6}) becomes
\begin{align}
\Lambda_{5}
=
\Big\langle r_{c}^{\dagger},
\mathcal{N}_{5}\left(r_{c},r_{c},r_{c},r_{c},r_{c}\right)
+3\mathcal{N}_{3}\left(r_{c},r_{c},w_{3}\right)
\Big\rangle.\label{GEQ8}
\end{align}
Eq. (\ref{GEQ5}) is the quintic extension of Eq. (\ref{DEQ8}).
Eq. (\ref{GEQ8}) is a compact special case of the general coefficient
formula in Eq. (\ref{GEQ6}). Eq. (\ref{GEQ5}) is the appropriate local normal
form when the cubic term is not sufficient to describe the stable
large-amplitude branch.

We now assume $\lambda^{\prime}\!\left(K_{c}\right)>0$ and
$\Lambda_{3}<0<\Lambda_{5}$. This is the subcritical situation selected by
Appendix \ref{Appendix-E} when $\chi\left(M\right)\neq 0$ and the quintic term
stabilizes the large-amplitude branch. For nontrivial equilibria,
Eq. (\ref{GEQ5}) reduces to
\begin{align}
\lambda\left(K\right)+\Lambda_{3}x+\Lambda_{5}x^{2}
=
0,\qquad x:=a^{2}>0.\label{GEQ9}
\end{align}
Therefore the two nonzero branches are
\begin{align}
x_{\pm}\left(K\right)
=
\frac{-\Lambda_{3}\pm
\sqrt{\Lambda_{3}^{2}-4\Lambda_{5}\lambda\left(K\right)}}{2\Lambda_{5}}.\label{GEQ10}
\end{align}
The reduced dynamics in Appendix \ref{Appendix-F} gives
\begin{align}
\frac{\mathsf{d}}{\mathsf{d}t}a
=
-\Psi_{5}\left(a;K\right)
+
O\left(a^{7}\right)
+
O\left(\left|K-K_{c}\right|a^{3}\right).\label{GEQ11}
\end{align}
Linearizing Eq. (\ref{GEQ11}) along a nonzero branch gives
\begin{align}
\partial_{a}\Psi_{5}\left(a;K\right)
=
2a^{2}\left(\Lambda_{3}+2\Lambda_{5}a^{2}\right)
+
O\left(a^{6}\right).\label{GEQ12}
\end{align}
Hence the branch $x_{+}\left(K\right)$ is locally stable and the branch
$x_{-}\left(K\right)$ is locally unstable at leading order.

The saddle-node point is determined by
$\Lambda_{3}+2\Lambda_{5}a^{2}=0$. Therefore
\begin{align}
x_{\mathrm{sn}}
=
-\frac{\Lambda_{3}}{2\Lambda_{5}},\qquad
\lambda_{\mathrm{sn}}
=
\frac{\Lambda_{3}^{2}}{4\Lambda_{5}}.\label{GEQ13}
\end{align}
Using Eq. (\ref{DEQ10}), we obtain the leading spinodal location
\begin{align}
K_{\mathrm{sn}}-K_{c}
\sim
\frac{\Lambda_{3}^{2}}
{4\lambda^{\prime}\!\left(K_{c}\right)\Lambda_{5}}.\label{GEQ14}
\end{align}
This quantity is the local hysteresis width:
\begin{align}
\Delta K_{\mathrm{hys}}
:=
K_{\mathrm{sn}}-K_{c}
\sim
\frac{\Lambda_{3}^{2}}
{4\lambda^{\prime}\!\left(K_{c}\right)\Lambda_{5}}.\label{GEQ15}
\end{align}

At $K=K_{c}$, the stable large-amplitude branch satisfies
\begin{align}
\left|a_{\mathrm{jump}}\right|
=
\sqrt{-\frac{\Lambda_{3}}{\Lambda_{5}}}.\label{GEQ16}
\end{align}
This is the jump amplitude when the incoherent branch loses stability at
$K=K_{c}$. At the saddle-node point, the synchronized branch disappears with
the smaller amplitude
\begin{align}
\left|a_{\mathrm{sn}}\right|
=
\sqrt{-\frac{\Lambda_{3}}{2\Lambda_{5}}}.\label{GEQ17}
\end{align}
Therefore, under the additional condition $\Lambda_{5}>0$, the conditional sign
statement for $\Lambda_{3}$ obtained in Appendix \ref{Appendix-E} extends to a
local description of the discontinuous regime: the sign condition gives
$\Lambda_{3}<0$, while the positive quintic coefficient fixes the
width of the hysteresis window and the size of the order parameter jump.

The same quintic normal form also refines the continuous regime. If
$\Lambda_{3}>0$ and $\lambda\left(K\right)<0$, the small supercritical branch
of Eq. (\ref{GEQ9}) satisfies
\begin{align}
x_{\mathrm{cont}}\left(K\right)
=
-\frac{\lambda\left(K\right)}{\Lambda_{3}}
-\frac{\Lambda_{5}}{\Lambda_{3}^{3}}\lambda\left(K\right)^{2}
+O\left(\lambda\left(K\right)^{3}\right).\label{GEQ18}
\end{align}
Therefore $\Lambda_{5}$ gives the first higher-order correction to the
continuous branch. It does not change the generic mean-field exponents derived
in Appendix \ref{Appendix-F}. Those exponents are still controlled by the cubic
term. The role of $\Lambda_{5}$ becomes leading only when $\Lambda_{3}=0$.
If that codimension-two situation occurs and $\Lambda_{5}>0$, the transition
is still continuous, but the scaling changes to
\begin{align}
\left|a_{\ast}\right|
\sim
\left(-\frac{\lambda\left(K\right)}{\Lambda_{5}}\right)^{1/4},
\qquad
\left|a\right|
\sim
\left|\frac{h}{\Lambda_{5}}\right|^{1/5}.\label{GEQ19}
\end{align}
This is the usual tricritical scaling. In the present theory, topology does
not force this tricritical point. Such a point can only arise when the cubic
coefficient is not already fixed to a definite nonzero sign.

\section{Critical defect count and charge balance}\label{Appendix-H}

In this appendix, we record the counting argument used in
Subsec. \ref{Defect-structure-subsection}. In the main text, the field
$u_{c}$ is read as the critical texture, its isolated zeros are the defect cores,
and the local index is the corresponding defect charge. Let
$p_{1},\ldots,p_{N_{\mathrm{def}}}$ be the isolated zeros of the critical
tangent field $u_{c}$. The Poincar\'e-Hopf theorem \cite{Milnor1997DifferentiableViewpoint} gives
\begin{align}
\sum_{\ell=1}^{N_{\mathrm{def}}}\operatorname{ind}_{p_{\ell}}\left(u_{c}\right)
=
\chi\left(M\right).\label{HEQ1}
\end{align}
Assume that each defect core is nondegenerate. Then each local charge is equal
to $+1$ or $-1$. Let $N_{+}$ and $N_{-}$ denote the numbers of positive and
negative defect cores. Eq. (\ref{HEQ1}) then becomes
\begin{align}
N_{+}-N_{-}
=
\chi\left(M\right).\label{HEQ2}
\end{align}
Therefore the total number of simple critical defects satisfies
\begin{align}
N_{\mathrm{def}}
=
N_{+}+N_{-}
\geq
\left|\chi\left(M\right)\right|.\label{HEQ3}
\end{align}
This is the generic counting statement used in the main text.

When $\chi\left(M\right)=0$, Eq. (\ref{HEQ2}) gives $N_{+}=N_{-}$. Simple
critical defects must therefore be charge neutral as a set. Under smooth
parameter variation, the total index remains fixed, so simple defects can only
be created or annihilated through events with zero net topological charge.
When $\chi\left(M\right)\neq 0$, Eq. (\ref{HEQ2}) shows that a nonzero
background charge remains in the critical texture. That net charge cannot be
removed by local charge-neutral defect events.

\section{Conditional accessibility of local phase transition scenarios}\label{Appendix-I}

In this appendix, we collect the logical consequences of
Appendices \ref{Appendix-E}-\ref{Appendix-G}. The statements below concern only
the local phase transition scenarios captured by the one-mode cubic and quintic
reductions stated there. They do not claim a classification of the full global
dynamics away from threshold. When symmetry produces a multidimensional
critical space, these statements apply after selecting an individual
symmetry-broken critical branch. The generic continuous branch in Appendix
\ref{Appendix-F} requires
\begin{align}
\Lambda_{3}>0.\label{IEQ1}
\end{align}
The tricritical continuous branch in Appendix \ref{Appendix-G} requires
\begin{align}
\Lambda_{3}=0,\qquad
\Lambda_{5}>0.\label{IEQ2}
\end{align}
The discontinuous branch with local hysteresis requires
\begin{align}
\Lambda_{3}<0,\qquad
\Lambda_{5}>0.\label{IEQ3}
\end{align}

Suppose that the sign condition of Appendix \ref{Appendix-E} holds. If
$\chi\left(M\right)\neq 0$, then Appendix \ref{Appendix-E} gives
\begin{align}
\Lambda_{3}<0.\label{IEQ4}
\end{align}
Eq. (\ref{IEQ4}) is incompatible with Eqs. (\ref{IEQ1}, \ref{IEQ2}).
Therefore, within the conditional one-mode normal form, a non-zero Euler
characteristic removes the generic continuous and tricritical continuous
scenarios. Under the
additional analytic condition $\Lambda_{5}>0$, Eq. (\ref{IEQ3}) remains as the
local scenario selected near threshold.

If $\chi\left(M\right)=0$, the sign of $\Lambda_{3}$ is not fixed by topology.
In that case topology gives no sign selection for $\Lambda_{3}$. The
continuous, tricritical continuous, and discontinuous scenarios are then
distinguished by Eq. (\ref{IEQ1}), Eq. (\ref{IEQ2}), and Eq. (\ref{IEQ3}),
respectively. Their actual realization is decided by the analytic coefficients
of the reduced equation. This is the sense in which the topology of $M$ enters
only conditionally in the local phase transition classification.

\section{Geometric and topological indicators for representative manifolds}\label{Appendix-J}

In this appendix, we collect the representative-manifold calculations used in
Sec. \ref{Case-study-section}. We first compute the averaged
tangent-projection factor $\kappa\left(M\right)$, which gives the geometric
coefficient multiplying the coupling-induced source term in the linearized
kinetic equation. We then collect the Euler characteristics and the
defect-counting consequences obtained from Appendix \ref{Appendix-H}.

\subsection{Averaged tangent-projection factors}

\paragraph{Hypersphere.} Let $M=S^{D-1}\subset\mathbb{R}^{D}$ with the standard
embedding. The tangent projector at $\sigma\in S^{D-1}$ is
\begin{align}
\mathsf{P}_{\sigma}^{\perp}
=
\mathbf{I}_{D}-\sigma\sigma^{\top}.\label{JEQ1}
\end{align}
Its average over the sphere is
\begin{align}
\overline{\mathsf{P}}_{S^{D-1}}
:=
\frac{1}{\textsf{Vol}\left(S^{D-1}\right)}
\int_{S^{D-1}}\mathsf{P}_{\sigma}^{\perp}\,
\mathsf{d}\mu\left(\sigma\right).\label{JEQ2}
\end{align}
Rotational invariance implies
$\overline{\mathsf{P}}_{S^{D-1}}=c\,\mathbf{I}_{D}$ for some scalar $c$. Taking the trace
of Eq. (\ref{JEQ2}) gives
\begin{align}
cD
=
\operatorname{tr}\left(\overline{\mathsf{P}}_{S^{D-1}}\right)
=
D-1,\label{JEQ3}
\end{align}
so
\begin{align}
\overline{\mathsf{P}}_{S^{D-1}}
=
\frac{D-1}{D}\mathbf{I}_{D},\qquad
\kappa\left(S^{D-1}\right)
=
\frac{D-1}{D}.\label{JEQ4}
\end{align}

\paragraph{Flat torus.} Let
$M=\mathbb{T}^{d}=\left(S^{1}\right)^{d}$ with the standard product embedding
\begin{align}
F\left(\theta_{1},\ldots,\theta_{d}\right)
=
\left(
\cos\theta_{1},\sin\theta_{1},
\ldots,
\cos\theta_{d},\sin\theta_{d}
\right).\label{JEQ5}
\end{align}
The tangent projector is block diagonal,
\begin{align}
\mathsf{P}_{\left(\theta_{1},\ldots,\theta_{d}\right)}^{\perp}
=
\operatorname{diag}\left(
\mathsf{P}\left(\theta_{1}\right),
\ldots,
\mathsf{P}\left(\theta_{d}\right)
\right).\label{JEQ6}
\end{align}
where
\begin{align}
\mathsf{P}\left(\theta\right)
=
\begin{pmatrix}
\sin^{2}\theta & -\sin\theta\cos\theta\\
-\sin\theta\cos\theta & \cos^{2}\theta
\end{pmatrix}.\label{JEQ7}
\end{align}
Averaging Eq. (\ref{JEQ7}) over one circle gives
\begin{align}
\frac{1}{2\pi}\int_{0}^{2\pi}\mathsf{P}\left(\theta\right)\,
\mathsf{d}\theta
=
\frac{1}{2}\mathbf{I}_{2}.\label{JEQ8}
\end{align}
Therefore
\begin{align}
\overline{\mathsf{P}}_{\mathbb{T}^{d}}
=
\frac{1}{2}\mathbf{I}_{2d},\qquad
\kappa\left(\mathbb{T}^{d}\right)
=
\frac{1}{2}.\label{JEQ9}
\end{align}

\paragraph{Real Stiefel manifold.} Let
$M=\mathrm{St}\left(p,n\right)\subset\mathbb{R}^{n\times p}$ with
$2\leq p<n$, equipped with the canonical Frobenius metric. For
$X\in\mathrm{St}\left(p,n\right)$ and $Z\in\mathbb{R}^{n\times p}$, the
orthogonal projection onto
$T_{X}\mathrm{St}\left(p,n\right)
=
\left\{\Xi:X^{\mathsf{T}}\Xi+\Xi^{\mathsf{T}}X=0\right\}$ is
\begin{align}
\mathsf{P}_{X}^{\perp}\left(Z\right)
&=
Z-X\,\operatorname{sym}\left(X^{\mathsf{T}}Z\right)
\notag\\
&=
Z-\frac{1}{2}X\left(X^{\mathsf{T}}Z+Z^{\mathsf{T}}X\right).\label{JEQ10}
\end{align}
Let
\begin{align}
\overline{\mathsf{P}}_{\mathrm{St}\left(p,n\right)}
:=
\frac{1}{\textsf{Vol}\left(\mathrm{St}\left(p,n\right)\right)}
\int_{\mathrm{St}\left(p,n\right)}\mathsf{P}_{X}^{\perp}\,
\mathsf{d}\mu\left(X\right).\label{JEQ11}
\end{align}
The canonical left-right action of
$\mathrm{SO}\left(n\right)\times\mathrm{SO}\left(p\right)$ on
$\mathbb{R}^{n\times p}$ implies
\begin{align}
\overline{\mathsf{P}}_{\mathrm{St}\left(p,n\right)}
\left(Q_{1}ZR\right)
=
Q_{1}\,
\overline{\mathsf{P}}_{\mathrm{St}\left(p,n\right)}
\left(Z\right)\,
R\label{JEQ12}
\end{align}
for all $Q_{1}\in\mathrm{SO}\left(n\right)$ and
$R\in\mathrm{SO}\left(p\right)$. Therefore
$\overline{\mathsf{P}}_{\mathrm{St}\left(p,n\right)}$ commutes with the standard
tensor-product action of $\mathrm{SO}\left(n\right)\times\mathrm{SO}\left(p\right)$
on $\mathbb{R}^{n\times p}$. By Schur's lemma, it is therefore a scalar
multiple of the identity on $\mathbb{R}^{n\times p}$. Since each tangent projector has
trace
$np-\frac{p\left(p+1\right)}{2}$, we obtain
\begin{align}
\overline{\mathsf{P}}_{\mathrm{St}\left(p,n\right)}
=
\frac{2n-p-1}{2n}\mathbf{I}_{np},\qquad
\kappa\left(\mathrm{St}\left(p,n\right)\right)
=
\frac{2n-p-1}{2n}.\label{JEQ13}
\end{align}

\paragraph{Rotation group.} Let $M=\mathrm{SO}\left(n\right)$ with the
canonical embedding in $\mathbb{R}^{n\times n}$ equipped with the Frobenius
inner product. For $R\in\mathrm{SO}\left(n\right)$ and
$Z\in\mathbb{R}^{n\times n}$, the orthogonal projection onto
$T_{R}\mathrm{SO}\left(n\right)=R\mathfrak{so}\left(n\right)$ is
\begin{align}
\mathsf{P}_{R}^{\perp}\left(Z\right)
=
R\,\operatorname{skew}\left(R^{\mathsf{T}}Z\right)
=
\frac{1}{2}\left(Z-RZ^{\mathsf{T}}R\right).\label{JEQ14}
\end{align}
Let
\begin{align}
\overline{\mathsf{P}}_{\mathrm{SO}\left(n\right)}
:=
\frac{1}{\textsf{Vol}\left(\mathrm{SO}\left(n\right)\right)}
\int_{\mathrm{SO}\left(n\right)}\mathsf{P}_{R}^{\perp}\,
\mathsf{d}\mu\left(R\right).\label{JEQ15}
\end{align}
Bi-invariance of the Haar measure gives
\begin{align}
\overline{\mathsf{P}}_{\mathrm{SO}\left(n\right)}
\left(Q_{1}ZQ_{2}\right)
=
Q_{1}\,
\overline{\mathsf{P}}_{\mathrm{SO}\left(n\right)}
\left(Z\right)\,
Q_{2}\label{JEQ16}
\end{align}
for all $Q_{1},Q_{2}\in\mathrm{SO}\left(n\right)$. Therefore
$\overline{\mathsf{P}}_{\mathrm{SO}\left(n\right)}$ commutes with the left-right
action of $\mathrm{SO}\left(n\right)\times\mathrm{SO}\left(n\right)$ on
$\mathbb{R}^{n\times n}$. By the real form of Schur's lemma,
$\overline{\mathsf{P}}_{\mathrm{SO}\left(n\right)}$ is therefore a scalar
multiple of the identity on $\mathbb{R}^{n\times n}$. Since each tangent projector has trace
$n\left(n-1\right)/2$, we obtain
\begin{align}
\overline{\mathsf{P}}_{\mathrm{SO}\left(n\right)}
=
\frac{n-1}{2n}\mathbf{I}_{n^{2}},\qquad
\kappa\left(\mathrm{SO}\left(n\right)\right)
=
\frac{n-1}{2n}.\label{JEQ17}
\end{align}

\paragraph{Unitary group.} Let $M=\mathrm{U}\left(d\right)$ with the canonical
embedding in $\mathbb{C}^{d\times d}\cong\mathbb{R}^{2d^{2}}$, equipped with
the real Frobenius inner product
$\langle X,Y\rangle=\operatorname{Re}\operatorname{tr}\left(X^{\dagger}Y\right)$.
For $U\in\mathrm{U}\left(d\right)$ and $Z\in\mathbb{C}^{d\times d}$, the
orthogonal projection onto
$T_{U}\mathrm{U}\left(d\right)=U\mathfrak{u}\left(d\right)$ is
\begin{align}
\mathsf{P}_{U}^{\perp}\left(Z\right)
=
U\,\operatorname{skew}\left(U^{\dagger}Z\right)
=
\frac{1}{2}\left(Z-UZ^{\dagger}U\right).\label{JEQ18}
\end{align}
Let
\begin{align}
\overline{\mathsf{P}}_{\mathrm{U}\left(d\right)}
:=
\frac{1}{\textsf{Vol}\left(\mathrm{U}\left(d\right)\right)}
\int_{\mathrm{U}\left(d\right)}\mathsf{P}_{U}^{\perp}\,
\mathsf{d}\mu\left(U\right).\label{JEQ19}
\end{align}
For any phase $\lambda\in\mathrm{U}\left(1\right)\subset\mathrm{U}\left(d\right)$,
left invariance of the Haar measure gives
\begin{align}
\int_{\mathrm{U}\left(d\right)}UZ^{\dagger}U\,\mathsf{d}\mu\left(U\right)
=
\lambda^{2}
\int_{\mathrm{U}\left(d\right)}UZ^{\dagger}U\,\mathsf{d}\mu\left(U\right).
\label{JEQ20}
\end{align}
Choosing $\lambda$ with $\lambda^{2}\neq 1$ shows that the integral in
Eq. (\ref{JEQ20}) vanishes. Therefore
\begin{align}
\overline{\mathsf{P}}_{\mathrm{U}\left(d\right)}
=
\frac{1}{2}\mathbf{I}_{2d^{2}},\qquad
\kappa\left(\mathrm{U}\left(d\right)\right)
=
\frac{1}{2}.\label{JEQ21}
\end{align}

\paragraph{Special unitary group.} For $M=\mathrm{SU}\left(d\right)$, the
tangent space consists of traceless skew-Hermitian directions. The canonical
projection is therefore obtained from Eq. (\ref{JEQ18}) by removing the trace
component of $U^{\dagger}Z$ before applying the left translation by $U$. In the
canonical embedding into $\mathbb{C}^{d\times d}$, the averaged tangent
projector then separates the trace direction from its orthogonal complement.
For this reason, the linear response is described by a representation-dependent
operator rather than by a single scalar coefficient on the full ambient space.
The special case $\mathrm{SU}\left(2\right)\cong S^{3}$ is already covered by
the hypersphere case.

\paragraph{Complex Grassmannian.} Let
$M=\mathrm{Gr}_{k}\left(\mathbb{C}^{n}\right)$ with $1\leq k\leq n-1$. We
represent a point of $M$ by the rank-$k$ Hermitian projector $P$ onto the
corresponding $k$-plane, and we use the centered projector embedding
$F\left(P\right)=P-\frac{k}{n}\mathbf{I}_{n}$ into the real vector space of traceless
Hermitian matrices. For a traceless Hermitian matrix $Z$, the orthogonal
projection onto the tangent space of the conjugacy orbit through $P$ is
\begin{align}
\mathsf{P}_{P}^{\perp}\left(Z\right)
=
PZ+ZP-2PZP.\label{JEQ25}
\end{align}
Let
\begin{align}
\overline{\mathsf{P}}_{\mathrm{Gr}_{k}\left(\mathbb{C}^{n}\right)}
:=
\frac{1}{\textsf{Vol}\left(\mathrm{Gr}_{k}\left(\mathbb{C}^{n}\right)\right)}
\int_{\mathrm{Gr}_{k}\left(\mathbb{C}^{n}\right)}\mathsf{P}_{P}^{\perp}\,
\mathsf{d}\mu\left(P\right).\label{JEQ26}
\end{align}
The $\mathrm{U}\left(n\right)$-invariance of the Haar measure implies that
$\overline{\mathsf{P}}_{\mathrm{Gr}_{k}\left(\mathbb{C}^{n}\right)}$ commutes
with conjugation on the traceless Hermitian subspace. Therefore it is a scalar
multiple of the identity on that subspace. Since the real dimension of
$\mathrm{Gr}_{k}\left(\mathbb{C}^{n}\right)$ is $2k\left(n-k\right)$ and the
ambient real dimension is $n^{2}-1$, we obtain
\begin{align}
\overline{\mathsf{P}}_{\mathrm{Gr}_{k}\left(\mathbb{C}^{n}\right)}
&=
\frac{2k\left(n-k\right)}{n^{2}-1}\mathbf{I}_{n^{2}-1},\qquad
\notag\\
\kappa\left(\mathrm{Gr}_{k}\left(\mathbb{C}^{n}\right)\right)
&=
\frac{2k\left(n-k\right)}{n^{2}-1}.\label{JEQ27}
\end{align}

\paragraph{Complex projective space.} Since
$\mathbb{CP}^{m}=\mathrm{Gr}_{1}\left(\mathbb{C}^{m+1}\right)$,
Eq. (\ref{JEQ27}) gives
\begin{align}
\kappa\left(\mathbb{CP}^{m}\right)
=
\frac{2}{m+2}.\label{JEQ28}
\end{align}

\paragraph{Sphere product.} Let
$M=\left(S^{m}\right)^{q}\subset\left(\mathbb{R}^{m+1}\right)^{q}
=\mathbb{R}^{q\left(m+1\right)}$ with the product embedding
$F\left(x_{1},\ldots,x_{q}\right)=\left(x_{1},\ldots,x_{q}\right)$. The tangent
projector at $\left(x_{1},\ldots,x_{q}\right)$ is
\begin{align}
\mathsf{P}_{\left(x_{1},\ldots,x_{q}\right)}^{\perp}
=
\operatorname{diag}\left(
\mathbf{I}_{m+1}-x_{1}x_{1}^{\top},
\ldots,
\mathbf{I}_{m+1}-x_{q}x_{q}^{\top}
\right).\label{JEQ22}
\end{align}
Using the sphere computation from Eqs. (\ref{JEQ1}-\ref{JEQ4}) on each factor,
we obtain
\begin{align}
\overline{\mathsf{P}}_{\left(S^{m}\right)^{q}}
=
\frac{m}{m+1}\mathbf{I}_{q\left(m+1\right)},\label{JEQ23}
\end{align}
and therefore
\begin{align}
\kappa\left(\left(S^{m}\right)^{q}\right)
=
\frac{m}{m+1}.\label{JEQ24}
\end{align}
In particular,
\begin{align}
\kappa\left(S^{2m}\times S^{2m}\right)
=
\frac{2m}{2m+1}.\label{JEQ29}
\end{align}

\subsection{Euler characteristics and defect consequences}

We next collect the Euler characteristics and the corresponding
defect-counting consequences used in Sec. \ref{Case-study-section}. These
statements combine the standard Euler-characteristic identities with the
counting result of Appendix \ref{Appendix-H}
\cite{Milnor1997DifferentiableViewpoint,Hatcher2002AlgebraicTopology,Harris1992AlgebraicGeometry}.

\begin{table*}[!t]
  \caption{Topological defect constraint and additional assumptions for the
  representative manifolds. The defect constraint is unconditional and follows
  from the Poincar\'e-Hopf theorem together with the Euler-characteristic
  calculations collected in this appendix. The locality and sign columns state
  the extra hypotheses needed before the last column of Table \ref{Table-5} can
  be read as a local phase transition conclusion.}
  \label{Table-6}
  \begingroup
  \scriptsize
  \renewcommand{\arraystretch}{1.5}
  \newcommand{\TableVICell}[2]{\begin{minipage}[t]{#1}\raggedright #2\end{minipage}}
  \newcommand{\TableVIHead}[2]{\begin{minipage}[t]{#1}#2\end{minipage}}
  \newcommand{\TableVIHeadSep}{\\\noalign{\vskip 0.55em}\hline\noalign{\vskip 1.05em}}
  \newcommand{\TableVISep}{\\\noalign{\vskip 1.35em}}
  \newcommand{\TableVIEnd}{\\\noalign{\vskip 2.2em}}
  \begin{ruledtabular}
  \begin{tabular}{@{}llll@{}}
    \TableVIHead{0.13\textwidth}{Manifold family} &
    \TableVIHead{0.25\textwidth}{Unconditional topological defect constraint} &
    \TableVIHead{0.26\textwidth}{Locality condition for the one-mode reduction} &
    \TableVIHead{0.27\textwidth}{Sign condition for the cubic coefficient}
    \TableVIHeadSep
    \TableVICell{0.13\textwidth}{$S^{D-1}$} &
    \TableVICell{0.25\textwidth}{If $D$ is odd, Poincar\'e-Hopf gives
    $Q_{\mathrm{def}}=2$ and
    $N_{\mathrm{def}}^{\min}=2$ for simple zeros. If $D$ is even, the net
    defect charge is zero.} &
    \TableVICell{0.26\textwidth}{The locality hypotheses of Appendix
    \ref{Appendix-D} must be verified for the chosen rotation-field ensemble
    and for the selected critical branch.} &
    \TableVICell{0.27\textwidth}{For odd $D$, the signed dominance condition in
    Eq. (\ref{EEQ6}) must be verified before inferring the sign of
    $\Lambda_{3}$. For even $D$, topology gives no sign constraint on
    $\Lambda_{3}$.}
    \TableVISep
    \TableVICell{0.13\textwidth}{$S^{2m}\times S^{2m}$} &
    \TableVICell{0.25\textwidth}{Poincar\'e-Hopf gives $Q_{\mathrm{def}}=4$ and
    $N_{\mathrm{def}}^{\min}=4$ for simple zeros.} &
    \TableVICell{0.26\textwidth}{The locality hypotheses must be verified for
    the product embedding and for the selected order parameter branch.} &
    \TableVICell{0.27\textwidth}{The Euler characteristic fixes the net defect
    charge, but it does not by itself imply Eq. (\ref{EEQ6}). The signed
    dominance condition must be checked separately.}
    \TableVISep
    \TableVICell{0.13\textwidth}{$\mathrm{Gr}_{k}\left(\mathbb{C}^{n}\right)$} &
    \TableVICell{0.25\textwidth}{Poincar\'e-Hopf gives
    $Q_{\mathrm{def}}=\binom{n}{k}$ and
    $N_{\mathrm{def}}^{\min}=\binom{n}{k}$ for simple zeros.} &
    \TableVICell{0.26\textwidth}{The locality hypotheses must be verified for
    the Pl\"ucker embedding subspace and for the selected critical branch.} &
    \TableVICell{0.27\textwidth}{Schubert-cell topology determines
    $\chi\left(M\right)$, but the local weights $\gamma_{\ell}$ and the
    remainder $\mathcal{R}_{U}$ in Eq. (\ref{EEQ6}) must be controlled
    separately.}
    \TableVISep
    \TableVICell{0.13\textwidth}{$\mathbb{CP}^{m}$} &
    \TableVICell{0.25\textwidth}{Poincar\'e-Hopf gives
    $Q_{\mathrm{def}}=m+1$ and $N_{\mathrm{def}}^{\min}=m+1$ for simple zeros.} &
    \TableVICell{0.26\textwidth}{The locality hypotheses must be verified for
    the projective embedding subspace and for the selected critical branch.} &
    \TableVICell{0.27\textwidth}{The Euler characteristic fixes the net defect
    charge, but Eq. (\ref{EEQ6}) remains an additional signed dominance
    condition.}
    \TableVISep
    \TableVICell{0.13\textwidth}{$\mathbb{T}^{d}$} &
    \TableVICell{0.25\textwidth}{The Euler characteristic is zero. A
    defect-free critical texture is topologically allowed, and any simple
    defects must have zero total index.} &
    \TableVICell{0.26\textwidth}{The locality condition is required if one performs a
    local normal-form calculation for a specified drift ensemble and selected
    critical branch.} &
    \TableVICell{0.27\textwidth}{There is no topological sign constraint on
    $\Lambda_{3}$. The cubic coefficient is determined by the analytic
    normal-form calculation.}
    \TableVISep
    \TableVICell{0.13\textwidth}{$\mathrm{St}\left(p,n\right)$} &
    \TableVICell{0.25\textwidth}{The Euler characteristic is zero. A
    defect-free critical texture is topologically allowed, and any simple
    defects must have zero total index.} &
    \TableVICell{0.26\textwidth}{The locality condition is model dependent and must
    be checked after a specific matrix-valued model and a specific critical branch have been chosen.} &
    \TableVICell{0.27\textwidth}{There is no topological sign constraint on
    $\Lambda_{3}$. The sign must be obtained from the normal-form coefficients
    of the specified model.}
    \TableVISep
    \TableVICell{0.13\textwidth}{$\mathrm{SO}\left(n\right)$} &
    \TableVICell{0.25\textwidth}{The Euler characteristic is zero. A
    defect-free critical texture is topologically allowed, and any simple
    defects must have zero total index.} &
    \TableVICell{0.26\textwidth}{The locality condition is model dependent and must
    be checked after a specific rotation-group model and a specific critical branch have been chosen.} &
    \TableVICell{0.27\textwidth}{There is no topological sign constraint on
    $\Lambda_{3}$. The sign must be obtained from the normal-form coefficients
    of the specified model.}
    \TableVISep
    \TableVICell{0.13\textwidth}{$\mathrm{U}\left(d\right)$} &
    \TableVICell{0.25\textwidth}{The Euler characteristic is zero. A
    defect-free critical texture is topologically allowed, and any simple
    defects must have zero total index.} &
    \TableVICell{0.26\textwidth}{The locality condition is model dependent and must
    be checked after a specific unitary-group model and a specific critical branch have been chosen.} &
    \TableVICell{0.27\textwidth}{There is no topological sign constraint on
    $\Lambda_{3}$. The sign must be obtained from the normal-form coefficients
    of the specified model.}
    \TableVIEnd
  \end{tabular}
  \end{ruledtabular}
  \endgroup
\end{table*}

\paragraph{Hypersphere.} For $M=S^{D-1}$, one has \cite{Hatcher2002AlgebraicTopology}
\begin{align}
\chi\left(S^{D-1}\right)
=
1+(-1)^{D-1}.\label{KEQ1}
\end{align}
Hence
\begin{align}
\chi\left(S^{D-1}\right)
=
\begin{cases}
2,& D\;\text{odd},\\
0,& D\;\text{even}.
\end{cases}\label{KEQ2}
\end{align}
When $D$ is odd, Appendix \ref{Appendix-H} gives net defect charge $2$ and
$N_{\mathrm{def}}\geq 2$ on a generic branch. When $D$ is even, a smooth
defect-free critical texture is topologically allowed, and any simple defects must
be charge neutral as a set.

\paragraph{Flat torus.} For $M=\mathbb{T}^{d}=\left(S^{1}\right)^{d}$, one has \cite{Hatcher2002AlgebraicTopology}
\begin{align}
\chi\left(\mathbb{T}^{d}\right)
=
\chi\left(S^{1}\right)^{d}
=
0.\label{KEQ3}
\end{align}
Therefore a smooth defect-free critical texture is topologically allowed. If
simple defects are present, Appendix \ref{Appendix-H} gives zero net charge.

\paragraph{Real Stiefel manifold.} For
$M=\mathrm{St}\left(p,n\right)$ with $2\leq p<n$, the projection onto the first
vector gives a fiber bundle \cite{Hatcher2002AlgebraicTopology,EdelmanAriasSmith1998SIAM}
\begin{align}
\mathrm{St}\left(p-1,n-1\right)
\longrightarrow
\mathrm{St}\left(p,n\right)
\longrightarrow
S^{n-1}.\label{KEQ4}
\end{align}
The multiplicativity of the Euler characteristic for compact fiber bundles \cite{Hatcher2002AlgebraicTopology}
therefore gives
\begin{align}
\chi\left(\mathrm{St}\left(p,n\right)\right)
=
\prod_{j=0}^{p-1}\left(1+(-1)^{n-1-j}\right).\label{KEQ5}
\end{align}
For every $2\leq p<n$, this product contains a zero factor. Hence
\begin{align}
\chi\left(\mathrm{St}\left(p,n\right)\right)
=
0,\qquad 2\leq p<n.\label{KEQ6}
\end{align}
Therefore a smooth defect-free critical texture is topologically allowed on every
such real Stiefel manifold. If simple defects are present, Appendix
\ref{Appendix-H} gives zero net charge.

\paragraph{Rotation and unitary groups.} Every positive-dimensional compact
connected Lie group admits a nowhere-vanishing left-invariant vector field.
Therefore Poincar\'e-Hopf gives \cite{Hall2015LieGroups,Milnor1997DifferentiableViewpoint}
\begin{align}
\chi\left(\mathrm{SO}\left(n\right)\right)
=
0,\qquad
\chi\left(\mathrm{U}\left(d\right)\right)
=
0,\qquad
\chi\left(\mathrm{SU}\left(d\right)\right)
=
0.\label{KEQ7}
\end{align}
Hence both the rotation-group and unitary-group cases allow a smooth
defect-free critical texture. The same is true for $\mathrm{SU}\left(d\right)$.
If simple defects are present, Appendix \ref{Appendix-H} gives zero net charge
in all three cases.

\paragraph{Sphere product.} For $M=\left(S^{m}\right)^{q}$, one has \cite{Hatcher2002AlgebraicTopology}
\begin{align}
\chi\left(\left(S^{m}\right)^{q}\right)
=
\left(1+(-1)^{m}\right)^{q}.\label{KEQ8}
\end{align}
If $m$ is odd, then $\chi\left(\left(S^{m}\right)^{q}\right)=0$. A smooth
defect-free critical texture is then topologically allowed, and any simple
defects must be charge neutral as a set. If $m$ is even, then
$\chi\left(\left(S^{m}\right)^{q}\right)=2^{q}$. Appendix \ref{Appendix-H}
then gives
\begin{align}
N_{\mathrm{def}}
\geq
2^{q}.\label{KEQ9}
\end{align}
In particular, every manifold of the form $S^{2m}\times S^{2m}$ has $\chi=4$
and therefore carries at least four simple defect cores on a generic branch.

\paragraph{Complex Grassmannian.} The complex Grassmannian
$\mathrm{Gr}_{k}\left(\mathbb{C}^{n}\right)$ admits a Schubert-cell
decomposition with one cell for each $k$-element subset of \cite{Harris1992AlgebraicGeometry}
$\left\{1,\ldots,n\right\}$. Every Schubert cell has even real dimension.
Therefore the Euler characteristic equals the number of Schubert cells:
\begin{align}
\chi\left(\mathrm{Gr}_{k}\left(\mathbb{C}^{n}\right)\right)
=
\binom{n}{k}.\label{KEQ10}
\end{align}
Appendix \ref{Appendix-H} then gives
\begin{align}
N_{\mathrm{def}}
\geq
\binom{n}{k}.\label{KEQ11}
\end{align}
Hence the critical texture on $\mathrm{Gr}_{k}\left(\mathbb{C}^{n}\right)$ must
carry net defect charge $\binom{n}{k}$ and, on a generic branch, at least
$\binom{n}{k}$ simple defect cores.

\paragraph{Complex projective space.} Since \cite{Harris1992AlgebraicGeometry}
$\mathbb{CP}^{m}=\mathrm{Gr}_{1}\left(\mathbb{C}^{m+1}\right)$,
Eq. (\ref{KEQ10}) gives
\begin{align}
\chi\left(\mathbb{CP}^{m}\right)
=
m+1.\label{KEQ12}
\end{align}
Appendix \ref{Appendix-H} then gives
\begin{align}
N_{\mathrm{def}}
\geq
m+1.\label{KEQ13}
\end{align}
Hence the critical texture on $\mathbb{CP}^{m}$ must carry net defect charge
$m+1$ and, on a generic branch, at least $m+1$ simple defect cores.
\section{Interpretation of the representative manifold families}\label{Appendix-K}

In this appendix, we give a family-by-family reading of Table \ref{Table-5}.
The purpose is not to replace the linear stability condition or the nonlinear
normal-form calculation. Instead, we separate the two indicators that enter the
argument: the averaged geometric projection, which fixes the geometric strength
of the coupling-induced source term, and the Euler characteristic, which fixes
the net topological charge of the critical texture.

Table \ref{Table-6} should be read together with the last column of
Table \ref{Table-5}. The entries with nonzero Euler characteristic have an unconditional defect
content, but their discontinuous local branch is conditional on the locality
assumptions of Appendix \ref{Appendix-D}, the signed dominance condition of
Appendix \ref{Appendix-E}, and the stabilizing quintic condition
$\Lambda_{5}>0$. The entries with zero Euler characteristic have no topological sign constraint on
$\Lambda_{3}$; their local branch must be obtained from the normal-form
coefficients of the specified model.

\paragraph{Hyperspheres and even-sphere products.} For the standard
hypersphere $S^{D-1}\subset\mathbb{R}^{D}$, Appendix \ref{Appendix-J} gives
\begin{align}
\kappa\left(S^{D-1}\right)
=
\frac{D-1}{D}.
\end{align}
This coefficient is the averaged tangent-projection factor carried by the
coupling term. The complete critical coupling is still obtained from the
spectral self-consistency condition after the intrinsic rotation fields have
been specified. The same appendix gives
\begin{align}
\chi\left(S^{D-1}\right)
=
1+\left(-1\right)^{D-1}.
\end{align}
Thus odd ambient dimension $D$ gives net defect charge $2$, whereas even
ambient dimension gives zero net defect charge. Under the local hypotheses used
for the one-mode reduction, this reproduces the topological part of the known
parity distinction for hyperspherical synchronization.

The product $S^{2m}\times S^{2m}$ separates the geometric coefficient from the
Euler characteristic more clearly. We obtain
\begin{align}
\kappa\left(S^{2m}\times S^{2m}\right)
=
\frac{2m}{2m+1},
\qquad
\chi\left(S^{2m}\times S^{2m}\right)
=
4.
\end{align}
The geometry therefore fixes the strength with which the coupling drives the
linearized perturbation, while the topology fixes a nonzero net defect charge.
Under the same local hypotheses, the local normal form does not contain a
generic continuous branch, and the discontinuous branch requires the
stabilizing quintic condition $\Lambda_{5}>0$.

\paragraph{Complex Grassmannians and complex projective spaces.} For the
complex Grassmannian $\mathrm{Gr}_{k}\left(\mathbb{C}^{n}\right)$, the
Pl\"ucker embedding gives
\begin{align}
\kappa\left(\mathrm{Gr}_{k}\left(\mathbb{C}^{n}\right)\right)
=
\frac{2k\left(n-k\right)}{n^{2}-1}.
\end{align}
The Schubert-cell decomposition gives
\begin{align}
\chi\left(\mathrm{Gr}_{k}\left(\mathbb{C}^{n}\right)\right)
=
\binom{n}{k}.
\end{align}
Hence the critical texture carries net defect charge $\binom{n}{k}$ and, on a
generic branch with simple defects, at least $\binom{n}{k}$ defect cores. The
complex projective space is the special case
$\mathbb{CP}^{m}=\mathrm{Gr}_{1}\left(\mathbb{C}^{m+1}\right)$. Its geometric
and topological indicators reduce to
\begin{align}
\kappa\left(\mathbb{CP}^{m}\right)
=
\frac{2}{m+2},
\qquad
\chi\left(\mathbb{CP}^{m}\right)
=
m+1.
\end{align}
These two examples show that the same framework applies to homogeneous phase
spaces that are not hyperspheres.

\paragraph{Zero-Euler toroidal and matrix manifolds.} The remaining examples
in Table \ref{Table-5} have zero Euler characteristic. For the vector-phase
torus $\mathbb{T}^{d}$ and the unitary group $\mathrm{U}\left(d\right)$, the
averaged tangent-projection coefficient is $1/2$. For the real Stiefel manifold
and the rotation group, Appendix \ref{Appendix-J} gives
\begin{align}
\kappa\left(\mathrm{St}\left(p,n\right)\right)
=
\frac{2n-p-1}{2n},
\qquad
\kappa\left(\mathrm{SO}\left(n\right)\right)
=
\frac{n-1}{2n}.
\end{align}
The Euler-characteristic calculation in Appendix \ref{Appendix-J} gives
\begin{align}
\chi\left(\mathbb{T}^{d}\right)
=
\chi\left(\mathrm{St}\left(p,n\right)\right)
=
\chi\left(\mathrm{SO}\left(n\right)\right)
=
\chi\left(\mathrm{U}\left(d\right)\right)
=
0.
\end{align}
Therefore these manifolds do not impose a nonzero net defect charge on the
critical texture. A smooth defect-free critical texture is topologically
allowed, and if simple defects appear their total charge must vanish. In these
zero Euler characteristic cases, the local transition branch is decided by the coefficients of
the reduced normal form rather than by the Euler characteristic alone.
  \nocite{*}
  \bibliography{Main}

\end{document}